\documentclass[10pt]{iopart}

\usepackage{graphicx}

\usepackage{bm}

\usepackage{color}
\usepackage{threeparttable}
\usepackage{dcolumn}
\usepackage{here}
\usepackage{ulem}

\usepackage{makecell}
\usepackage{xspace}

\usepackage{iopams}
\expandafter\let\csname equation*\endcsname\relax
\expandafter\let\csname endequation*\endcsname\relax
\usepackage{amsmath}
\usepackage{braket}

\usepackage{xcolor}
\usepackage[final]{changes}

\sethighlightmarkup{textwidth=columnwidth}
\reversemarginpar

\definechangesauthor[name = JeanPascal, color=orange]{JP}
\definechangesauthor[name = Jacques, color=blue]{JF}
\definechangesauthor[name = Georg, color=green]{GK}
\definechangesauthor[name = Dai, color=purple]{DA}

\newcommand{\UTe}{UTe$_2$\xspace}
\newcommand{\Hm}{H_{\rm m}}
\newcommand{\Tsc}{T_{\rm sc}}
\newcommand{\Hc}{H_{\rm c2}}
\newcommand{\pc}{p_{\rm c}}

\begin{document}

\title[]{Unconventional Superconductivity in  UTe$_2$}

\author{D.~Aoki$^1$, J.-P.~Brison$^2$, J.~Flouquet$^2$, K.~Ishida$^3$, G.~Knebel$^2$, Y.~Tokunaga$^4$, Y.~Yanase$^{5,6}$}

\address{$^1$IMR, Tohoku University, Oarai, Ibaraki, 311-1313, Japan }
\address{$^2$Univ. Grenoble Alpes, CEA, Grenoble INP, IRIG, PHELIQS, F-38000 Grenoble, France}
\address{$^3$Department of Physics, Kyoto University, Kyoto 606-8502, Japan}
\address{$^4$ASRC, Japan Atomic Energy Agency, Tokai, Ibaraki 319-1195, Japan}
\address{$^5$Department of Physics, Graduate School of Science, Kyoto University, Kyoto 606-8502, Japan}
\address{$^6$Institute for Molecular Science, Okazaki 444-8585, Japan}
\ead{aoki@imr.tohoku.ac.jp, jean-pascal.brison@cea.fr, jflouc@aol.com, kishida@scphys.kyoto-u.ac.jp, georg.knebel@cea.fr, tokunaga.yo@jaea.go.jp, yanase@scphys.kyoto-u.ac.jp}
\vspace{10pt}


\begin{abstract}

The novel spin-triplet superconductor candidate UTe$_2$ was discovered only recently at the end of 2018 and attracted enormous attention. We review key experimental and theoretical progress which has been achieved in different laboratories. UTe$_2$ is a heavy-fermion paramagnet, but right after the discovery of superconductivity it has been  expected to be close to a ferromagnetic instability showing many similarities to the U-based ferromagnetic superconductors, URhGe and UCoGe. The competition between different types of magnetic interactions and the duality between the local and itinerant character of the $5f$ Uranium electrons, as well as the shift of the U valence appear as key parameters in the rich phase diagrams discovered recently under extreme conditions like low temperature, high magnetic field, and pressure. We discuss macroscopic and microscopic experiments 
at low temperature to clarify the normal phase properties at ambient pressure for field applied along the three axis of this orthorhombic structure. Special attention will be given to the occurrence of a metamagnetic transition at $H_{\rm m} =35$~T for a magnetic field applied along the hard magnetic axis $b$. Adding external pressure leads to strong changes in the magnetic and electronic properties with a direct feedback on superconductivity. Attention will be given on the possible evolution of the Fermi surface as a function of magnetic field and pressure. 

Superconductivity in  UTe$_2$ is extremely rich exhibiting various unconventional behaviors which will be highlighted. It shows an exceptionally huge superconducting upper critical field with a  re-entrant behavior under magnetic field and  the occurrence of multiple superconducting phases in the temperature field, pressure phase diagram. There is evidence for spin-triplet pairing.  Experimental indications exist for chiral superconductivity and spontaneous time reversal symmetry breaking in the superconducting state. The different theoretical approaches will be described. Notably we discuss that UTe$_2$ is a possible example for the realization of a fascinating topological superconductor. The recent discovery of superconductivity in \UTe reemphasizes that U-based heavy fermion compounds give unique examples to study and understand the strong interplay between the normal and superconducting properties in strongly correlated electron systems.

\end{abstract}

%
%
%
\maketitle
%
\ioptwocol

\section{Introduction}

Superconductivity in UTe$_2$ has been discovered only in late 2018 and rapidly confirmed \cite{Ran2019, Aoki2019}. While UTe$_2$ is paramagnetic already in the first report \cite{Ran2019} the possibility of spin-triplet superconductivity and the importance of ferromagnetic fluctuations for the superconducting pairing has been emphasized. It has been proposed that  UTe$_2$ belongs as paramagnetic end member to the family of the ferromagnetic superconductors like URhGe and UCoGe. However, rapidly it turned out that key differences in the crystal structure and by consequence on the magnetic interactions and the electronic structure make that the studies on UTe$_2$ open new unique phenomena. 

The electronic structure of UTe$_2$ is very particular. In band structure calculations within the local density approximation (LDA) a small gap of $\Delta = 130$~K occurs at the Fermi level. Thus, neglecting the electronic correlations, UTe$_2$ would be a small gap insulator \cite{Aoki2019}. Remarkably, the value of $\Delta$ is comparable to the antiferromagnetic interactions (as determined from the value of the Curie Weiss temperature above 100~K). As a consequence, on cooling the convergence of these two mechanisms lead to a very different scenario with a metallic ground state, the possibility of supplementary ferromagnetic interactions at low temperatures interfering the high temperature antiferromagnetic ones. From thermoelectric power, Hall effect and from the link between the $\gamma$ coefficient of the specific heat and the upper critical field it has been shown that UTe$_2$ appears as a good metal with roughly one charge carrier/U atom \cite{Niu2020}. Modifying the electronic band structure through a  magnetic field induced Zeeman shift or damping the magnetic interactions by external pressure will lead to a metamagnetic transition with a change in the balance between ferromagnetic and antiferromagnetic interactions, crossing through valence transition and Lifshitz transition \cite{Lifshitz_note}. All these ingredients will have a direct impact on the Cooper pairing with the observation of the multiple superconducting phases. 

The aim of this review is to present the status (August 2021) of research on unconventional superconductivity in \UTe from our point of view. It is a rapidly emerging field, and we apologize if interesting experiments or developments are incompletely covered. In sections \ref{sectionCrystalStructure} and \ref{electronic_structure}, we present respectively what is known on the crystalline and electronic structure of \UTe. In section \ref{normal_state_properties} we present the normal state properties of \UTe with focus on electronic transport and thermodynamics, as well as magnetic fluctuations detected by nuclear spin resonance (NMR) and inelastic neutron scattering experiments. 
In section \ref{section_supra} we review the superconducting properties of \UTe and discuss the possible symmetry and topology of unconventional superconducting phases. Section \ref{section_Pressure} introduces the properties of \UTe under high pressure and we discuss the superconducting phase diagrams under pressure and the magnetic properties. Finally in section \ref{section_theoretical_perspectives} we present perspectives on the superconductivity in \UTe from a theoretical point of view.

\section{Crystal growth}
\begin{figure}
\begin{center}
\includegraphics[width=.98\columnwidth]{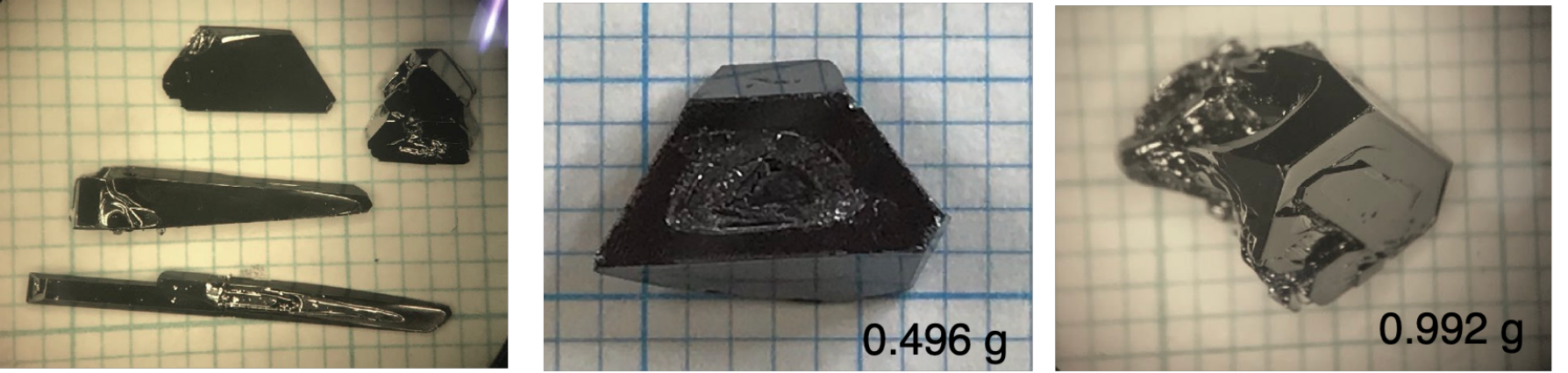}
\caption{Single crystals of UTe$_2$ grown using the CVT method.}
\label{photo}
\end{center}
\end{figure}

\added[id=DA]{High quality single crystals of UTe$_2$ are particularly important, because, for instance, the determination of the superconducting gap structure is strongly affected by the sample quality.
Furthermore the large residual density of state initially observed in the specific heat measurements at zero field by all groups, nearly $50\,{\%}$ raised the question of an intrinsic origin of ``ungapped electronic states''. They could have been the consequence of an "extreme" non-unitary triplet state, where only half the Fermi surface would have been paired.

As described later, the highest quality sample with $T_{\rm sc}\sim 2\,{\rm K}$ now display a residual $\gamma$-value as low as $10\,{\%}$, which of course excludes this intrinsic anomalous non unitary state.

Presently, most single crystals of UTe$_2$ are grown by the chemical vapor transport (CVT) method. Conditions for improved samples have been progressively discovered.
The first crystal growth were performed starting with stoichiometric amounts of U and Te, and iodine (density: a few ${\rm mg/cm^3}$) as the transport agent \cite{Ikeda2006}. They produced these single crystals with rather sharp specific jumps at the superconducting transition, but with large residual terms.
Higher quality single crystals are now grown from off-stoichiometric amounts with the ratio U : Te = 1 : 1.5--1.85 under a temperature gradient, 1060/1000 $^\circ$C in a horizontal furnace for 10-14 days\cite{Ran2019,Aoki2019,Aoki2019a}.
Optimising the growth condition, a large single crystal up to $\sim 1\,{\rm g}$ has been successfully obtained, as shown in Fig.~\ref{photo}.
Even higher quality samples seem to be produced by lowering the temperature of the synthesis, for instance, down to $800/710\,^\circ{\rm C}$ \cite{Rosa2021,Cairns2020}. 
}

\added[id=DA]{
Single crystals of UTe$_2$ can be also grown using the Te self-flux method\cite{Aoki2019,Ran2021crystals}. 
The quality is, however, not very good, with a residual resistivity ratio $\sim 3$, and no bulk superconductivity.
}
\section{Crystal structure of UTe$_2$}
\label{sectionCrystalStructure}

\begin{figure}
\begin{center}
\includegraphics[width=.98\columnwidth]{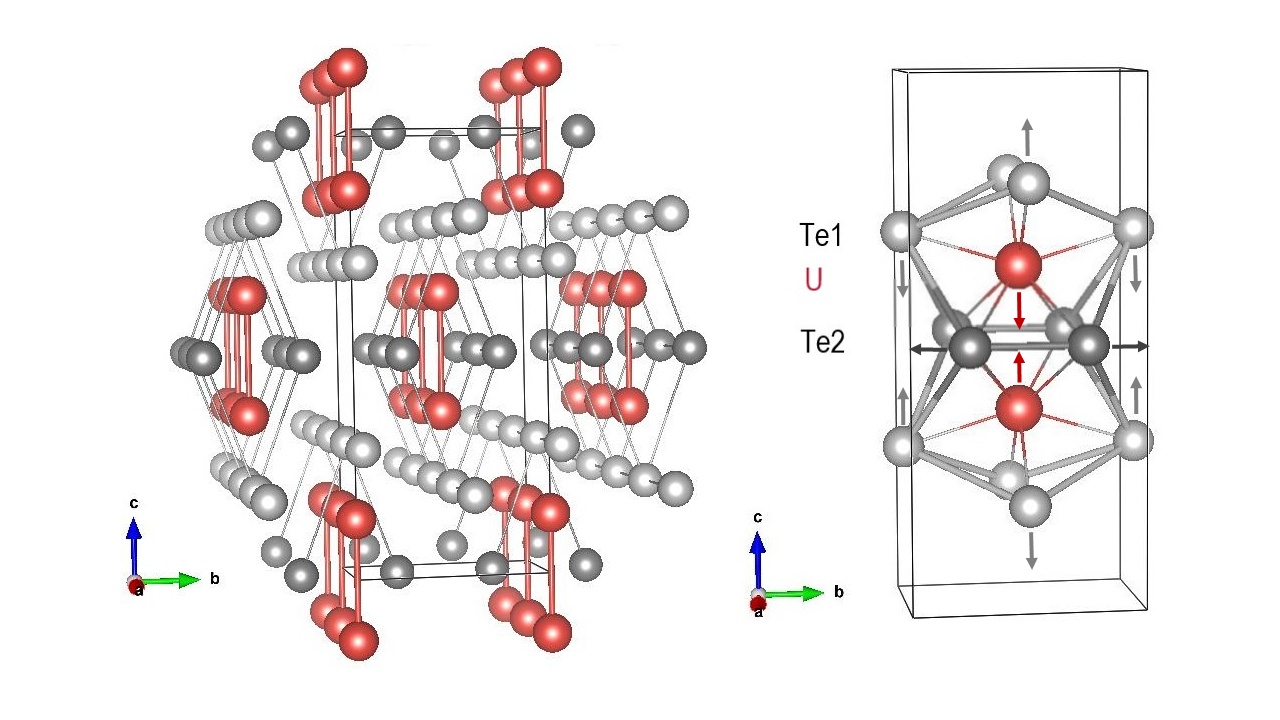}
\caption{(Left)Body-centered orthorhombic crystal structure of UTe$_2$. (Right) Zoom on the nearest neighbors of the U atoms, the arrows indicate the relative changes in inter-atomic distances when the temperature is lowered. The main effect of temperature on the structure is the reduction of the U - U distance on the ladder rungs as indicated by the red arrows in the right figure \cite{Hutanu2020,Stoewe1997}. }
\label{fig1}
\end{center}
\end{figure}

UTe$_2$  crystallizes in a body centered orthorhombic structure (see Fig.~\ref{fig1}) with the symmorphic space group $Immm$ (No. 71, $D_{2h}^{25}$).  The lattice parameters at ambient temperature are $a=4.159$~\AA, $b = 6.124$~\AA, and $c = 13.945$~\AA , respectively, and no change in the  crystal structure is found down to low temperatures \cite{Hutanu2020}.  There are four formula units per unit cell, and the 4 U atoms occupy $4i$ site. The 8 Te atoms are located at two different sites in the unit cell; $4j$ and $4h$ sites with point symmetries $mm2$ and $m2m$. Their respective sites are denoted Te(1) and Te(2).  The U atoms are surrounded by two Te(1) and four Te(2) atoms in form of a bi-trigonal prism. The Te(2) atoms form chains along the $b$ axis  with a distance of 3.05~\AA, which is the shortest distance between all atoms in the structure. The U atoms form a two-leg ladder structure along the $a$ axis with the shortest distance between the U atoms of about 3.78~\AA\  along the rung in $c$ direction. It is significantly smaller than distance of the leg  of 4.16~\AA\ along the $a$ axis. The distance between two such ladder chains is 4.91~\AA\ and  and next neighboring chains are displaced by $c/2$. In Fig.~\ref{fig1} the nearest neighbors of the U atoms are shown, the arrows indicate the relative changes in the inter-atomic distances on cooling \cite{Hutanu2020,Stoewe1997}. Remarkably, the distance between the two U atoms on the rung shrinks by 60\% more that the lattice parameter along the $c$ axis with decreasing temperature \cite{Hutanu2020}. 
Finally we note that the shortest U - U distance is larger than the so-called Hill limit \cite{Hill_note} $(\approx 3.5$\AA). We would expect that the U moments are localized and that at low temperature magnetic order occur. In the simple picture, unconventional superconductivity is not expected \cite{Hill1970}. 
Although the lattice constant of $c$-axis is rather long compared to those of $a$ and $b$-axes, the Brillouin zone is not flattened very much along $c$-axis, because of the ``body-centered'' orthorhombic structure, suggesting the formation of 3D Fermi surfaces rather than 2D Fermi surfaces.
The shape of Brillouin zone looks similar to that of the well-known ThCr$_2$Si$_2$-type structure, but it is elongated along $a$-axis.
We further want to emphasise that the $[011]$ direction, corresponds to the direction tilted by 23.7 deg from $b$ to $c$-axis, at which the field-reentrant superconductivity is observed above the metamagnetic transition field $H_{\rm m}$ (see section \ref{field_reinforcement} below) \cite{Ran2019a, Knafo2020}.
It should be stressed that the inversion center of the structure is not located at the U site, but exists at the middle of the U rung in spite of the body-centered orthorhombic structure with the global inversion symmetry,
meaning that the local inversion symmetry is broken.
The situation is similar to the case for the recently discovered heavy fermion superconductor CeRh$_2$As$_2$ with CaBe$_2$Ge$_2$-type structure (space group: $P4/nmm$) \cite{Khim2021},
where the inversion center is not located at the Ce atom with global inversion symmetry, leading to the broken inversion symmetry locally.

The substitution of Te by Se in UTe$_2$ leads to the drastic change of the crystal structure. 
At least up to U(Te$_{0.8}$Se$_{0.2}$)$_2$, the space group $Immm$ is retained, but further substituting with Se, the space group changes to $Pnma$, where ferromagnetic order occurs at relatively high temperatures with the semi-metallic electronic state.
For example, U(Te$_{0.36}$Se$_{0.64}$)$_2$ orders ferromagnetically below $T_{\rm Curie}\sim 69\,{\rm K}$ \cite{Noel1996}.
$\beta$-USe$_2$ is a ferromagnet with $T_{\rm Curie}=14\,{\rm K}$, showing the semiconducting behavior.
We note that $Pnma$ is a subgroup of $Pmmn$ (No.~59, $D^{13}_{2h}$), which is a subgroup of $Immm$.

\section{Electronic structure of UTe$_2$}
\label{electronic_structure}

\subsection{Theoretical predictions}
\label{section:electronicStructure_theory}

\begin{figure}[tbp]
\begin{center}
\includegraphics[width=1\columnwidth]{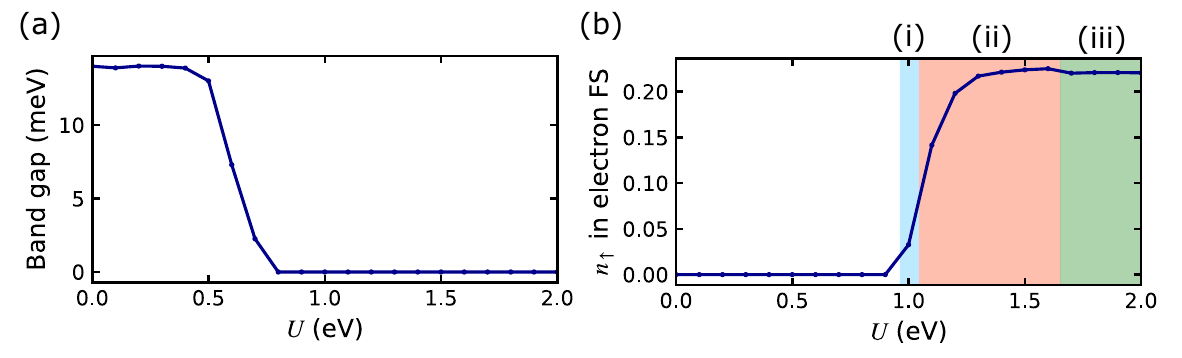}
\caption{The GGA+U calculations for the Coulomb interaction dependence of (a) the band gap at the Fermi level and (b) the electron carrier density per spin~\cite{IshizukaPRL2019}.
Insulator-metal transition occurs around $U = 1$~eV. Metallic states with different topology of FSs are labeled by (i)-(iii). Figures taken from Ref.~\cite{Ishizuka2019}.
}
\label{I-M_transition}
\end{center}
\end{figure}

As we mentioned in the introduction, a naive LDA band calculation for UTe$_2$ predicted an insulating state with a small gap of 13~meV~\cite{Aoki2019}. This is clearly incompatible with experiments showing the metallic conduction and superconducting instability. The discrepancy indicates that the electron correlation effects are crucially important in UTe$_2$. A way to take into account the electron correlations from first-principles is the GGA+U method, where GGA means ``generalized gradient approximation'' and the Hubbard $U$ is the on-site Coulomb repulsion, responsible for the development of strong electronic correlations. 
The GGA+U calculation for the paramagnetic state indeed revealed an insulator-metal transition~\cite{IshizukaPRL2019}. As shown in Fig.~\ref{I-M_transition}, the insulating gap appearing in the LDA calculations closes and conducting carriers appear for $U > 1$~eV. For a large $U$, the density of electron carriers reaches $n_{{\rm e}\uparrow} \simeq 0.2$ per spin. 
This leads to electron and hole carrier densities $n_{\rm e}=n_{\rm h} \simeq 0.4$ since UTe$_2$ is a compensated metal. The total density $n_{\rm e}+n_{\rm h} \simeq 0.8$ is close to unity and consistent with experiments~\cite{Niu2020}.
These results have been obtained by the relativistic full-potential linearized augmented plane wave + local orbitals method within the GGA approximation. The effect of Hubbard interaction $U$ is taken into account at the level of a static mean-field approximation. Another DFT+U calculation, however within the LSDA approximation and in a hypothetical ferromagnetic ground state~\cite{Shick_UTe2}, also predicts the metallic state for similar values of $U$. A long-range ferromagnetic order has not been observed in UTe$_2$, 
however it is argued~\cite{Shick_UTe2} that locally, the electronic structure could be considered as ferromagnetic due to slow long-range ferromagnetic fluctuations. Finally, the metal-insulator transition due to strong electronic correlations has also been confirmed by the density functional theory combined with the dynamical mean field theory (DFT+DMFT)~\cite{Xu_UTe2}, however, results are presented in this last case for much larger values of $U \simeq 7$~eV.

\begin{figure}[tbp]
\begin{center}
\includegraphics[width=1\columnwidth]{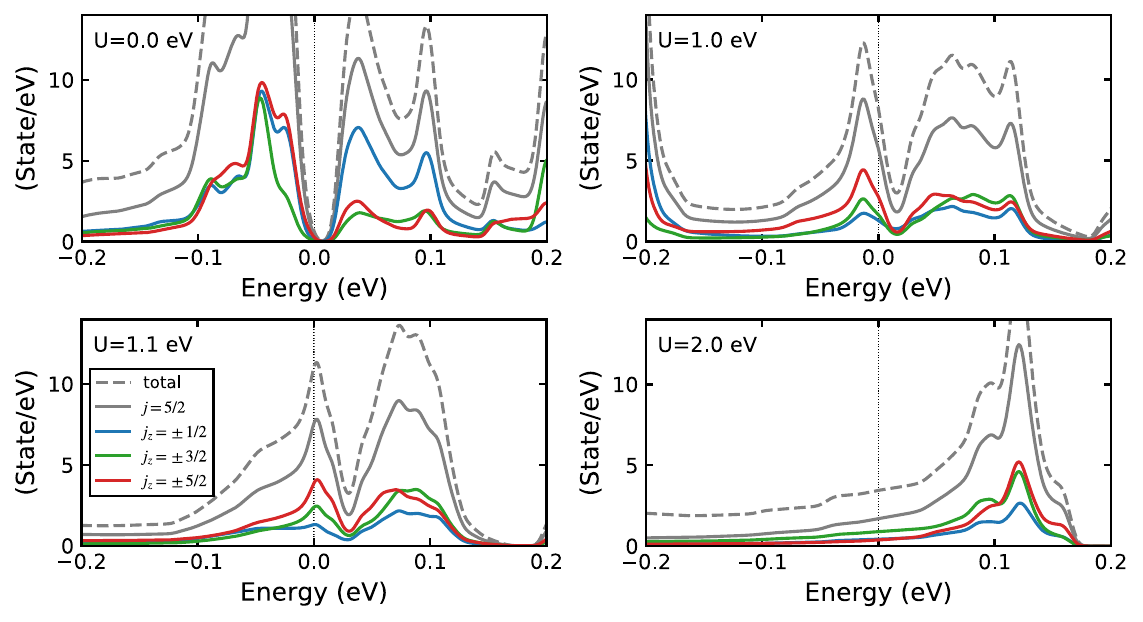}
\caption{The GGA+U calculations for the partial density of states for U-5f orbitals of $j = 5/2$, $j_z = \pm 1/2$, $\pm3/2$, and $\pm 5/2$ near the Fermi level~\cite{IshizukaPRL2019}. Dashed line represents the total density of states. Figures taken from Ref.~\cite{Ishizuka2019}.
}
\label{GGA+U-DOS}
\end{center}
\end{figure}

Discussion of the mechanism of insulator-metal transition by the Coulomb interaction, can be done with the results from the GGA approximation, giving the density of states for various values of $U$ (Fig.~\ref{GGA+U-DOS}). 
The density of states at $U=0$~eV reveals an hybridization gap at the Fermi energy. Because the density of states near the Fermi level mainly comes from the heavy $f$-electron bands, the result of the LDA calculation is regarded as a Kondo insulating state. 
Branching the Coulomb interaction shifts the hybridization gap to the high energy side. Such an energy shift of the hybridization gap should be prohibited in the single-orbital periodic Anderson model so as to preserve the charge neutrality. 
However, all the $J=5/2$ multiplets contribute to the low-energy electronic states in UTe$_2$. Hence the multi-orbital nature allows the energy shift, and the resulting insulator-metal transition.

\begin{figure}[htbp]
\begin{center}
\includegraphics[width=0.8\columnwidth]{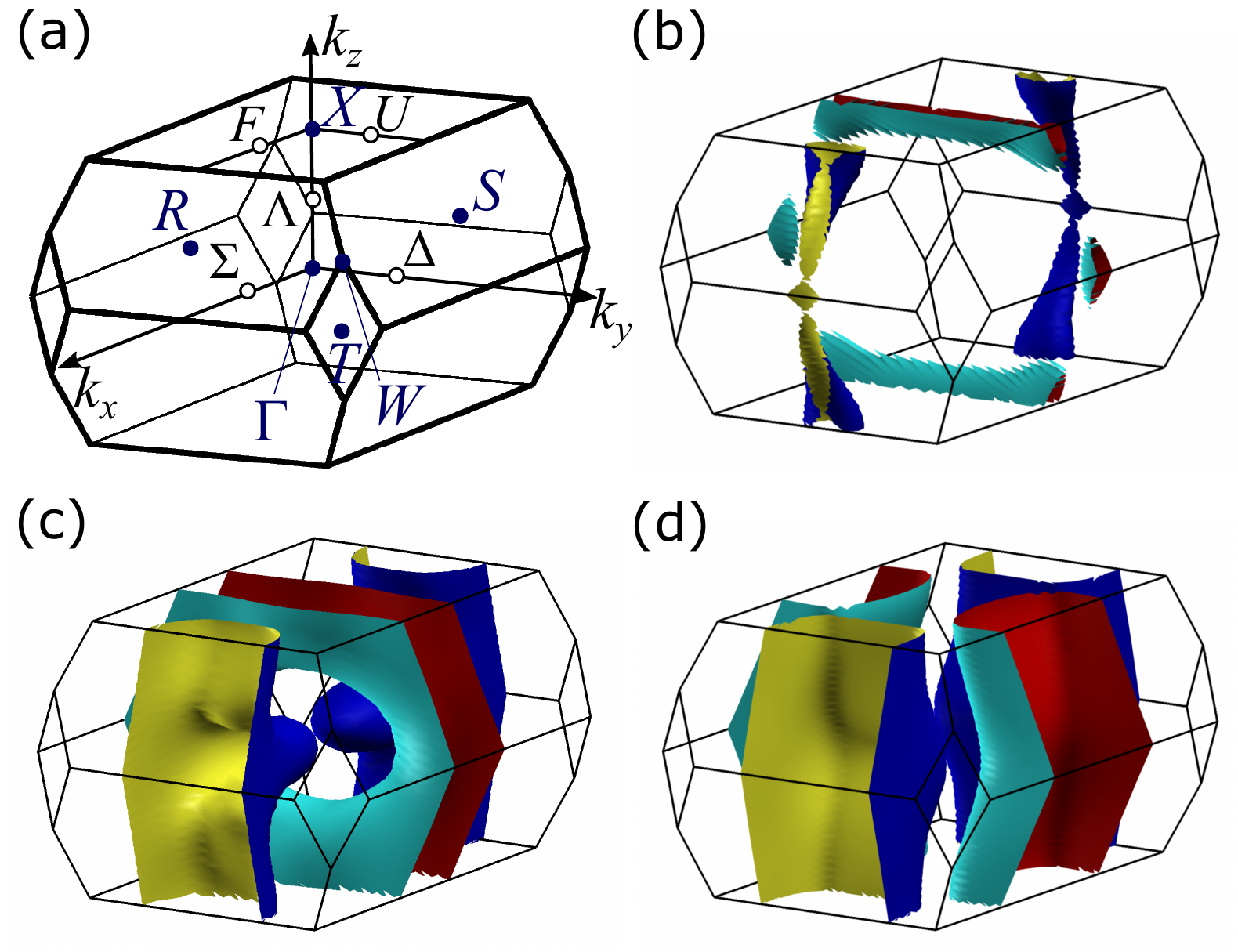}
\caption{(a) First Brillouin zone and symmetry points of body-centered orthorhombic crystals. (b)-(d) Fermi surfaces of UTe$_2$ by GGA+U calculations for (b) $U = 1.0$~eV [region (i)], (c) $U = 1.1$~eV [region (ii)], and (d) $U = 2.0$~eV [region (iii)]~\cite{IshizukaPRL2019}. 
The electron sheet (cyan and red) and the hole sheet (blue and yellow) are highlighted by colors. Figures taken from Ref.~\cite{Ishizuka2019}.
}
\label{GGA+U-FS}
\end{center}
\end{figure}

Fermi surfaces appear above the insulator-metal transition. Figure~\ref{GGA+U-FS} shows the Fermi surfaces for various values of $U$. We see Lifshitz transitions; the topology of Fermi surfaces changes with increasing $U$. We denote the regions (i), (ii), and (iii) in Fig.~\ref{I-M_transition} based on the different topology of the Fermi surfaces in Figs.~\ref{GGA+U-FS}(b), \ref{GGA+U-FS}(c), and \ref{GGA+U-FS}(d).

The Fermi surfaces in the large Coulomb interaction region [Fig.~\ref{GGA+U-FS}(d)] resemble the results of the DFT+DMFT calculation which assumes a large $U$~\cite{Xu_UTe2,Miao2020}. Thus, the rectangular Fermi surfaces are consistently obtained by the DFT+$U$ and DFT+DMFT calculations, and they are similar to the Fermi surfaces of ThTe$_2$ without $f$-electrons~\cite{Miao2020,Harima_UTe2,Nevidomskyy_UTe2}. Combining these results, we suppose that the $f$-electrons are almost localized in the large Coulomb interaction region. 
Indeed, the rectangular Fermi surfaces are formed by the conduction electrons mainly consisting of U-6$d$ electrons and Te(2)-5$p$ electrons. Interestingly, they have a quasi-one-dimensional character, and the conducting directions are orthogonal to each other. Thus, the perpendicular quasi-one-dimensional Fermi surfaces of 6$d$ and 5$p$ electrons hybridize and form rectangular Fermi surfaces in Fig.~\ref{GGA+U-FS}(d).
The electronic state in the large Coulomb interaction region is well understood in this way. 
However, such electronic structure is not compatible with expected properties of UTe$_2$. For instance, the large specific heat~\cite{Ran2019,Aoki2019} indicates the presence of an itinerant heavy $f$-electron band. A recent observation of moderate anisotropic electrical conductivity~\cite{Eo2021} requires three-dimensional Fermi surfaces. 
On the other hand, only light quasi-one-dimensional Fermi surfaces appear in the large $U$ region.

The expected properties, namely, the heavy, three-dimensional, and itinerant $f$-electron bands, are fulfilled in the intermediate Coulomb interaction region. Low-energy electronic states appear around the X point in the Brillouin zone, when we decrease the Coulomb interaction $U$. This heavy electron band hybridizes with the conduction electron band, and then, the ring-shaped Fermi surface of Fig.~\ref{GGA+U-FS}(c) is formed. This band structure is more realistic for UTe$_2$, although further verification is needed as other calculations predict qualitatively different band structure~\cite{Shick2021} which is in reasonable agreement with the soft X-ray ARPES experiment \cite{Fujimori2019}. Comparison with experiments is highly desirable. In this regard, an ARPES experiment~\cite{Miao2020} reported low-energy electronic states consistent with Fig.~\ref{GGA+U-FS}(c). 
In addition to the light rectangular Fermi surfaces, a signature of the heavy band around the X point has been observed. Note that the X point in the conventional notation is equivalent to the Z point in Ref.~\cite{Miao2020} (Fig.~\ref{ARPES}). 
Further ARPES studies clarifying the sensitivity to the bulk and surface states are desired. Quantum oscillation measurements are also awaited as the Fermi surfaces of heavy-fermion systems, such as UPt$_3$~\cite{McMullan_2008}, were precisely determined by them.

The strength of Coulomb interaction also affects the number of $f$-electrons per U ion. Generally speaking, Coulomb interaction decreases the $f$-electron number: $n_f = 2$ for the localized $5f^2$ state, while $n_f = 3$ for the itinerant $5f^3$ state. Theoretical values are reported as $n_f \simeq 2.3$~\cite{Miao2020}, $n_f \simeq 2.5$~\cite{Ishizuka2019}, $n_f \simeq 2.73$~\cite{Shick2021}. As mentioned in the next section, core level photo-emission experiments support an intermediate valence state \cite{Fujimori2021}.

\subsection{Experiment}
\label{Electronic_state_exp}

The electronic structure of UTe$_2$ has been studied by angle resolved photoemission spectroscopy (ARPES) using different photon energies. Experiments in Japan  have been performed on the soft X-ray beam line BL23SU at Spring 8 using photon energies between $h\nu = 560 - 800$~eV \cite{Fujimori2019, Fujimori2021}, and experiments in the US using synchrotron radiation at the Advanced Light Source (Berkeley) and an helium lamp light source (New York University) using photon energies $h\nu = 10 - 150$~eV \cite{Miao2020}. The results obtained in these experiments are very different and present, at first glance, contradicting views of the electronic structure. This may be related to the different inelastic mean free path of the photoelectrons which is a function of the  kinetic energy of the photon \cite{Seah1979} and in general, low energy photoemission can be very surface sensitive instead of probing the bulk. From the universal curve \cite{Seah1979, Fujimori_2016} one can expect that the soft X-ray experiments probe 15\AA\ depth, while for the low energy spectroscopy near 100~eV the probing depth is two times lower and thus more surface sensitive. 

\begin{figure}[htbp]
\begin{center}
\includegraphics[width=0.9\columnwidth]{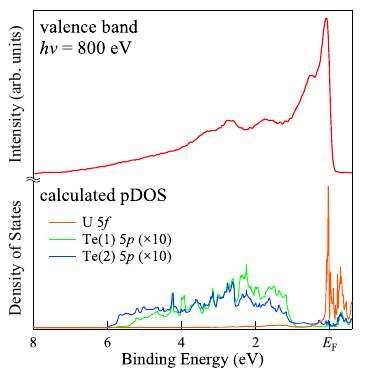}
\caption{AIPES spectrum measured at $h\nu = 800\,{\rm eV}$ and the theoretical partial DOS of U 5$f$ and Te 5$p$. Figure taken from Ref.~\cite{Fujimori2019}.}
\label{AIPES}
\end{center}
\end{figure}

Figure~\ref{AIPES} shows the valence band spectrum obtained by the angle-integrated photoelectron spectroscopy (AIPES) experiments performed at an energy of $h\nu=800$~eV \cite{Fujimori2019}.
The sharp peak below $E_{\rm F}$ is due to the U 5$f$ state with an itinerant character.
The long tail is indicative of an anomalous admixture of the U 5$f$ and Te 5$p$ bands observed at a higher binding energy. A similar peak is found for the partial U 5$f$ density of states obtained by resonant photoelectron spectroscopy at the  U $4d-5f$ absorption edge ($h\nu = 736$~eV) which suggests that the U $5f$ states in \UTe  are itinerant in nature, but with some hybridization with the Te 5$p$ states. The lower panel in Fig.~\ref{AIPES} indicates the theoretical U $5f$ and Te $5p$ partial density of states.  This calculation has been performed treating all the 5$f$ electrons as itinerant (see Ref.~\cite{Fujimori2019}). 

The core-level spectroscopy reveals that a mixed-valence state is realized and the dominant contribution is the itinerant 5$f^3$ configuration, by comparison to UB$_2$ with an itinerant 5$f^3$ configuration, and to UPd$_3$ with a localized 5$f^2$ configuration \cite{Fujimori2021}. This has been also concluded from the Uranium L3 X-ray absorption near-edge spectroscopy \cite{Thomas2020}. These itinerant 5$f$ states are in agreement with the electronic structure calculations presented in Refs.~\cite{Fujimori2019, Shick2021}.  
 
Contrary to the core-level spectroscopy, the spectrum at the U $5d-5f$ resonance ($h\nu = 98$~eV) shows no peak structure close to the Fermi energy, but clearly enhanced intensities at $-0.7$~eV binding energy \cite{Miao2020}. This results is in good agreement with the first principle-based dynamical mean field theory (DFT + DMFT) \cite{Miao2020}. The spectra are interpreted to originate from light Te bands and the U-$5f$ states occurs to be mainly localized, i.e. do not contribute to the Fermi surface. The conclusion of the experiment is that the the dominant configuration is U-5$f^2$. 

The overall band structure obtained from the different ARPES experiments is again very different. The soft X-ray experiments (at photon energies $h\nu \approx 600$~eV) are well explained by the band structure calculation based on the 5$f$-itinerant model.
However, the fine structure near the Fermi level was unresolved experimentally in the soft X-ray experiments, owing to insufficient resolution compared to the bandwidth of the flat bands. 

\begin{figure}[htbp]
\begin{center}
\includegraphics[width=0.9\columnwidth]{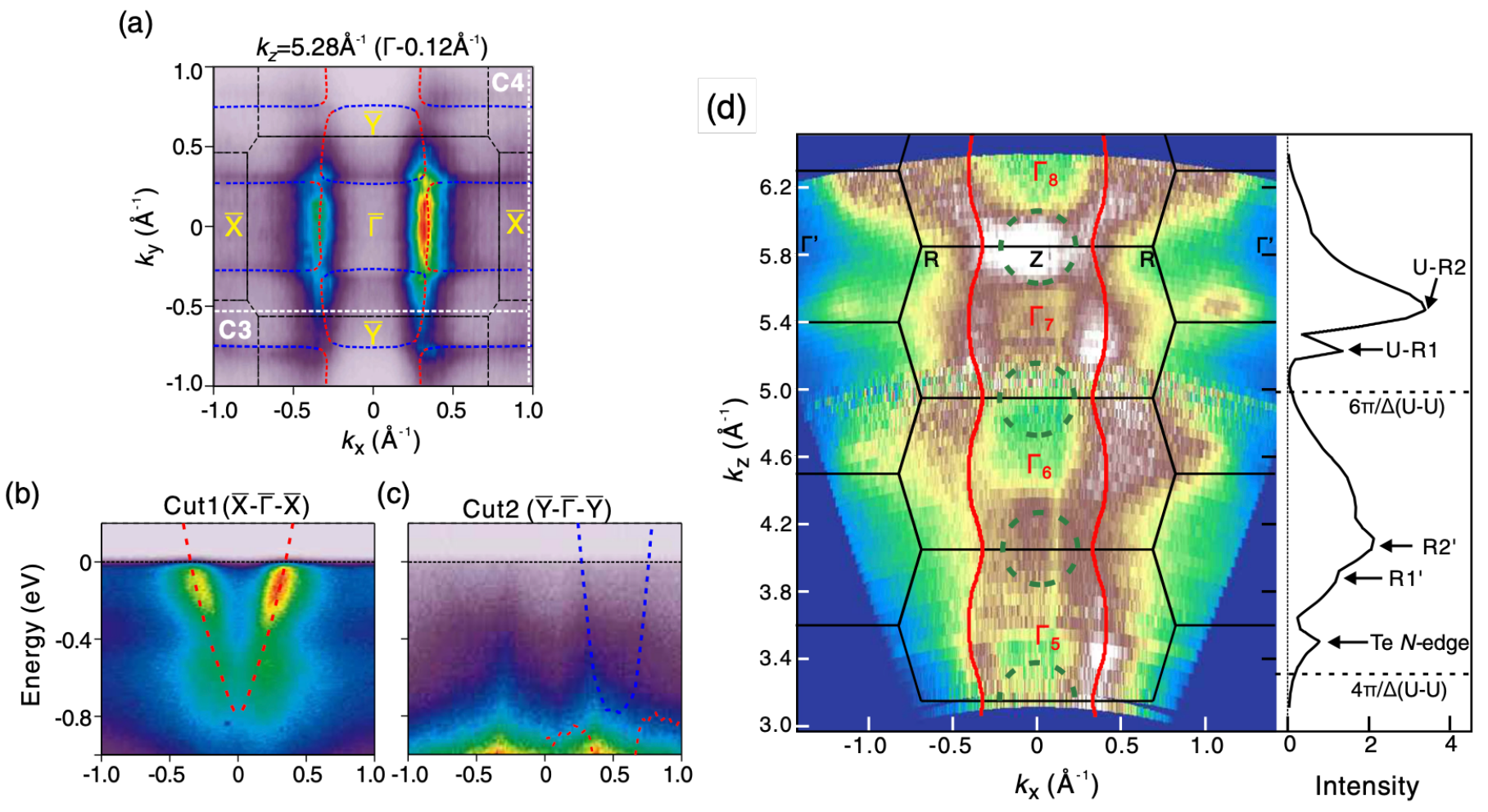}
\caption{Quasi-one-dimensional Fermi surfaces obtained by high resolution ARPES experiments~\cite{Miao2020}. (a)ARPES measurements at the $(001)$ surface. (b)(c) the band structure near the Fermi level along $k_x$ and $k_y$ directions. (d) Three dimensional dispersion and a heavy Z-point electron pocket. Figure taken from Ref.~\cite{Miao2020}.}
\label{ARPES}
\end{center}
\end{figure}

In difference, the high resolution ARPES experiments using a helium light source~\cite{Miao2020} revealed two light quasi-one-dimensional bands at the Fermi level, attributed to the U and Te chains.
Figures~\ref{ARPES}(a)-(c) show two light Fermi surfaces observed at $\Gamma$--$X$ and $\Gamma$--$Y$.
These Fermi surfaces coincide with the calculated Fermi surfaces for ThTe$_2$. 
Another heavy pocket is also detected at Z point, as shown in Fig.~\ref{ARPES}(d). 
This Fermi surface seems to be inconsistent with the results from the soft X-ray ARPES experiments~\cite{Shick2021}, which are more bulk sensitive,
than these ARPES measurements using a helium light source ($hv = 30 -150$~eV). 
Hence, no clear conclusion on the electronic structure on \UTe emerges from the photoemission experiments performed up to now.

No quantum oscillations have been detected, neither by de Haas-van Alphen (dHvA) nor by Shubnikov-de Haas (SdH) effect. 
Possible reasons why dHvA or SdH oscillations are not detected are: 1) the sample quality is insufficient, 2) the effective mass is large by comparison to the Fermi surface volume, 3) the curvature factor of the Fermi surfaces are unfavorable.
The residual resistivity ratio ($\mbox{RRR}\sim 40$) of the best available crystals of \UTe should be enough for detecting at least a part of the Fermi surfaces by dHvA or SdH experiments. 
Compared to the very large initial slope of the upper critical field, the moderately enhanced $\gamma$-value may point to the existence of rather small pockets (moderate $\gamma$) with large effective masses (large $\Hc$).
Hence, if the carrier number was small, and only these small Fermi surfaces with heavy quasi-particles existed, they would be very difficult to detect by quantum oscillations, as it is predicted in UBe$_{13}$.
However, in UTe$_2$, the carrier number indicated by Hall experiments \cite{Niu2020} seems not so small, revealing some volume of Fermi surfaces.
In addition, the 2D Fermi surfaces suggested by the low energy ARPES experiments and by the band calculations should be easily detected by dHvA or SdH experiments, because of the light effective mass of the associated carriers and the favourable curvature of the Fermi surface.

Naturally, the large $H_{\rm c2}$ also impedes to detect the dHvA or SdH signals in the conventional measurements using a superconducting magnet up to $15\,{\rm T}$. 
Nevertheless, the higher field experiments using a resistive magnet or a pulse field magnet also show no quantum oscillations up to now.
This failure of a detection of any quantum oscillations in \UTe is one of the deep puzzles of \UTe, calling for still improved sample quality. 
Progress in the knowledge of the Fermi surface topology is a clear key challenge; a next generation of experiments is highly demanded, both for ARPES and quantum oscillations measurements.

\section{Normal state properties}
\label{normal_state_properties}

\subsection{Basic properties of UTe$_2$ and behavior for field  $H\parallel a$, the easy magnetization axis at $p=0$} 
\label{subsectionHpara_a}

\begin{figure}
\begin{center}
\includegraphics[angle=-90, width=1\columnwidth]{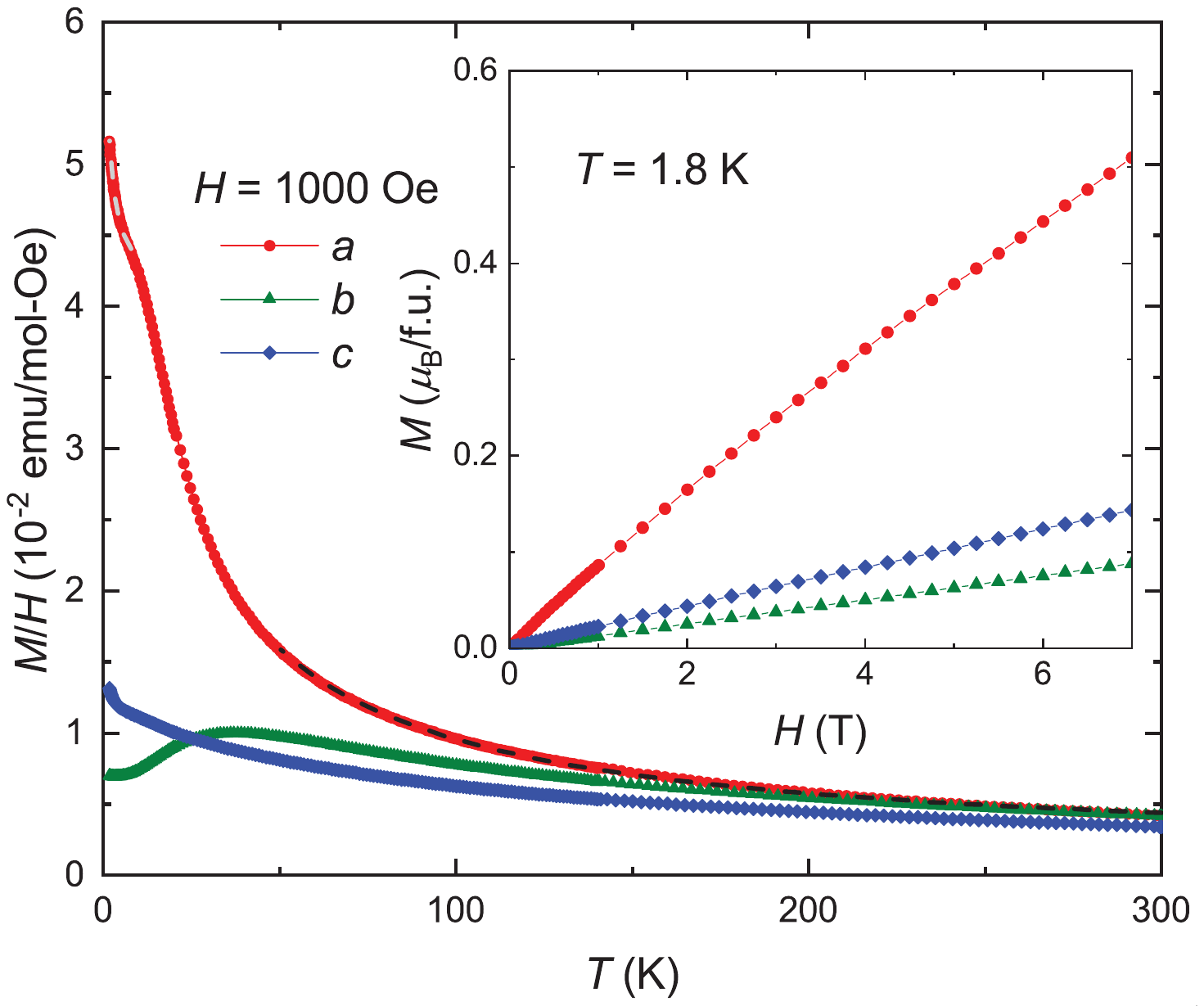}
\caption{Magnetization divided by magnetic field $m/H$ measured at 0.1~T for the three crystallographic axes. The black dashed line indicates the Curie-Weiss behavior for the $a$ axis. The gray dashed line is a fit to a power law temperature dependence below 10~K.  The inset shows the field dependence of the magnetization measured at 1.8~K. (taken from Ref. \cite{Ran2019})   }
\label{fig2}
\end{center}
\end{figure}

The magnetic susceptibility $\chi$ of UTe$_2$ is shown in  Fig.~\ref{fig2} for the three crystallographic axes $a$, $b$, $c$. For all directions the susceptibility follows a Curie Weiss behavior $\chi = C/(T - \Theta)$ for temperatures above 150~K with respective Curie-Weiss temperatures $\Theta = -60$, -110, and -135~K along the $a$, $b$, and $c$ axis \cite{Ikeda2006, Knafo2020}. The negative value of $\Theta$ may be, at first glance, indicative for a dominant antiferromagnetic exchange at high temperatures. The corresponding effective magnetic moments obtained from the fits above 150~K are $\mu_{\rm eff} = 3.44\,\mu_{\rm B}$,  $3.5 \,\mu_{\rm B}$, and $3.33\,  \mu_{\rm B}$ for the $a$, $b$, and $c$ axis, respectively.  These values are very  close to the free ion values 3.58 $\mu_{\rm B}$ for $5f^2$ (U$^{4+}$) or 3.62 $\mu_{\rm B}$ for $5f^3$  (U$^{3+}$) states. However, the crystal field splitting and the Kondo effect may have also influence on the temperature dependence of the susceptibility. On cooling, $\chi_a$ growths stronger than the Curie Weiss behavior below 100~K, starts to saturate around 20~K, but finally increases strongly in small fields below $T_a^\ast \approx 12$~K down to the onset of superconductivity. We note that on approaching the superconducting transition at $T_{\rm sc} \approx 1.6$~K, at low field no Fermi-liquid behavior $\chi_a = \chi_0 + a T^2$ is detected. A quantum critical scaling of $M_a/T$ versus $H/T^{1.5}$ at low temperature (the subtraction of a linear term in $M_a (H)$)  has been proposed to indicate the proximity to a ferromagnetic quantum critical point \cite{Ran2019}. However, this scaling rises two open questions namely the origin for the subtraction of the additional term and the scaling exponents apply for disordered metals  near ferromagnetic  quantum criticality \cite{Kirkpatrick2015}. Thus, in principle it should not be applicable for UTe$_2$, which shows unconventional superconductivity and is in the clean limit. The increase of the $a$ axis susceptibility below 10~K remains to be clarified. 

In difference to the $a$ axis, the susceptibility $\chi_b$ follows the Curie Weiss behavior down to 50~K, and it shows a broad maximum at $T_{\rm max} \sim 35$~K. Such a maximum is often observed in heavy-fermion paramagnetic systems along the easy magnetization axis, see e.g.~studies on the CeRu$_2$Si$_2$ family \cite{HAoki2014, Flouquet2005}, as a signature of the development of antiferromagnetic correlations and the formation of the coherent heavy-fermion state. As we will see below, the energy scale of $T_{\rm max}$ corresponds to that of the field induced metamagnetic transition for a field applied along the $b$ axis \cite{Knafo2019, Miyake2019}. At 300~K the $c$ axis susceptibility $\chi_c$ is smallest. On cooling, only a smooth increase of susceptibility $\chi_c$ is observed which is slightly stronger than the Curie-Weiss behavior below 150~K, and only for $T <25$~K the  $b$ axis is the magnetically hard axis with  $\chi_b < \chi_c$. At 2~K, for $H = 7$~T, the magnetization $M_a$ along the $a$ axis reaches 0.5~$\mu_{\rm B}$/U, while $M_b$ and $M_c$ get 0.09 and 0.15~$\mu_{\rm B}$/U (see inset in Fig.~\ref{fig2}). The $a$ axis, which corresponds to the ladder direction, is the initially easy magnetization axis,  and the $b$ axis, which is perpendicular to the ladder rungs, is the hard magnetic axis. The anisotropy $M_a/M_b \sim 5.5$, is not very strongly Ising-type in UTe$_2$. 
It can result from crystal field effects and also by the interaction themselves governed by local properties and the Fermi surface topology. 
High field magnetization measurements along the $a$ axis indicate that the magnetization increases up to 20~T and tends to saturate for $H > 40$~T to a value of $M = 1.1 \mu_{\rm B}/{\rm f.u.}$ in a fully polarized state (see inset of Fig.~\ref{A_coef_a}). It is worthwhile to note that just above $T_{sc}$ the susceptibility $\chi_a = \partial M_a/\partial H$ becomes lower than that of  $\chi_b = \partial M_b/\partial H$ at $H_{a-b} \approx 16$~T, i.e.~roughly the field when the upper critical field $H_{\rm c2}$ along $b$ shows a clear field reinforcement. 

\begin{figure}
\begin{center}
\includegraphics[width=1\columnwidth]{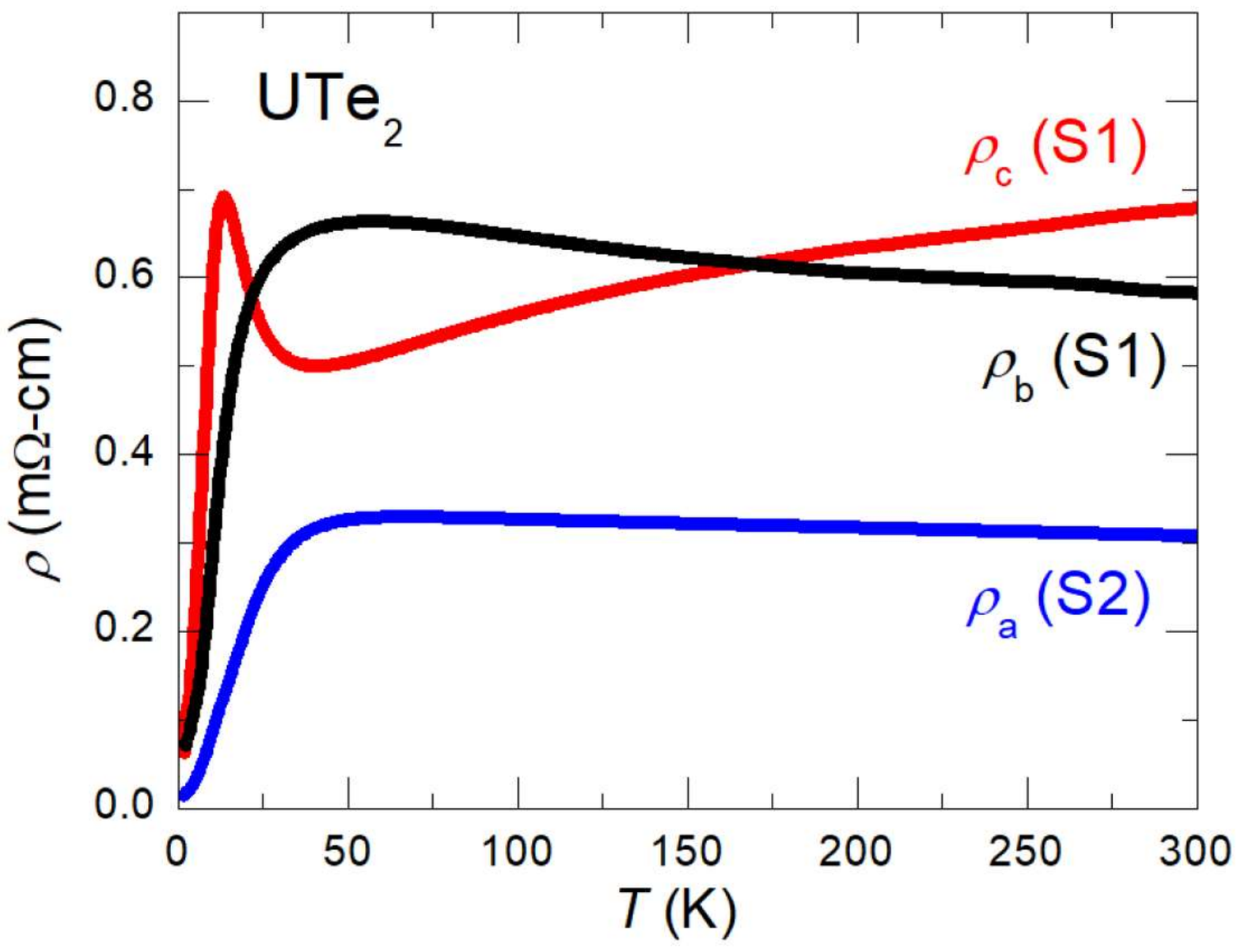}
\caption{Temperature dependence of the electrical resistivity of of UTe$_2$ for a current applied along different crystallographic axes. (taken from Ref. \cite{Eo2021})   }
\label{fig3}
\end{center}
\end{figure}

The electrical resistivity in UTe$_2$ is anisotropic and it depends strongly on the current direction \cite{Eo2021} as shown in Fig.~\ref{fig3}. While for a current injected along $a$ or $b$ the resistivity shows a shallow maximum around 60~K and starts to decrease strongly below 50~K due to the formation of the coherent Kondo-lattice, the resistivity with current direction along the $c$ axis decreases from 300~K with a distinct minimum near 40 K and shows a pronounced maximum near the characteristic temperature $T^\ast \approx 14$~K.  
The derivative of the resistivity $d\rho/dT$ shows a maximum for the three current directions near $T_\rho^* \approx 15$~K for $a$ and $b$ axis current and near 7.5~K for $c$ axis current. Thus, the different location of $T^\ast_\rho$ from the derivative of the resistivity marks clearly the crossover behavior. 
The microscopic origin of this distinct different behavior on current directions is not determined right now, but it may be related to the characteristic Fermi surface topology \cite{Eo2021}. The residual resistivity $\rho_0$ varies at least by a factor five between the current direction $j \parallel a$ axis and $j \parallel c$ axis. In addition the $A$ coefficient of the Fermi liquid resistivity $\rho (T) = \rho_0 + AT^2$, observed below 5~K, depends also from the current direction. For current along $a$, $A < 1$~$\mu\Omega$\,cm\,K$^{-2}$ in all reports, for current along $b$  $A \approx 1.5$~$\mu\Omega$\,cm\,K$^{-2}$, while it is strongly enhanced for current along $c$ axis and  $A \approx 7$~$\mu\Omega$\,cm\,K$^{-2}$ is observed. As roughly $A/\rho_0$ appears current invariant, it suggests that the $(a,b)$ plane is the easy conducting plane. 

\begin{figure}
\begin{center}
\includegraphics[width=1\columnwidth]{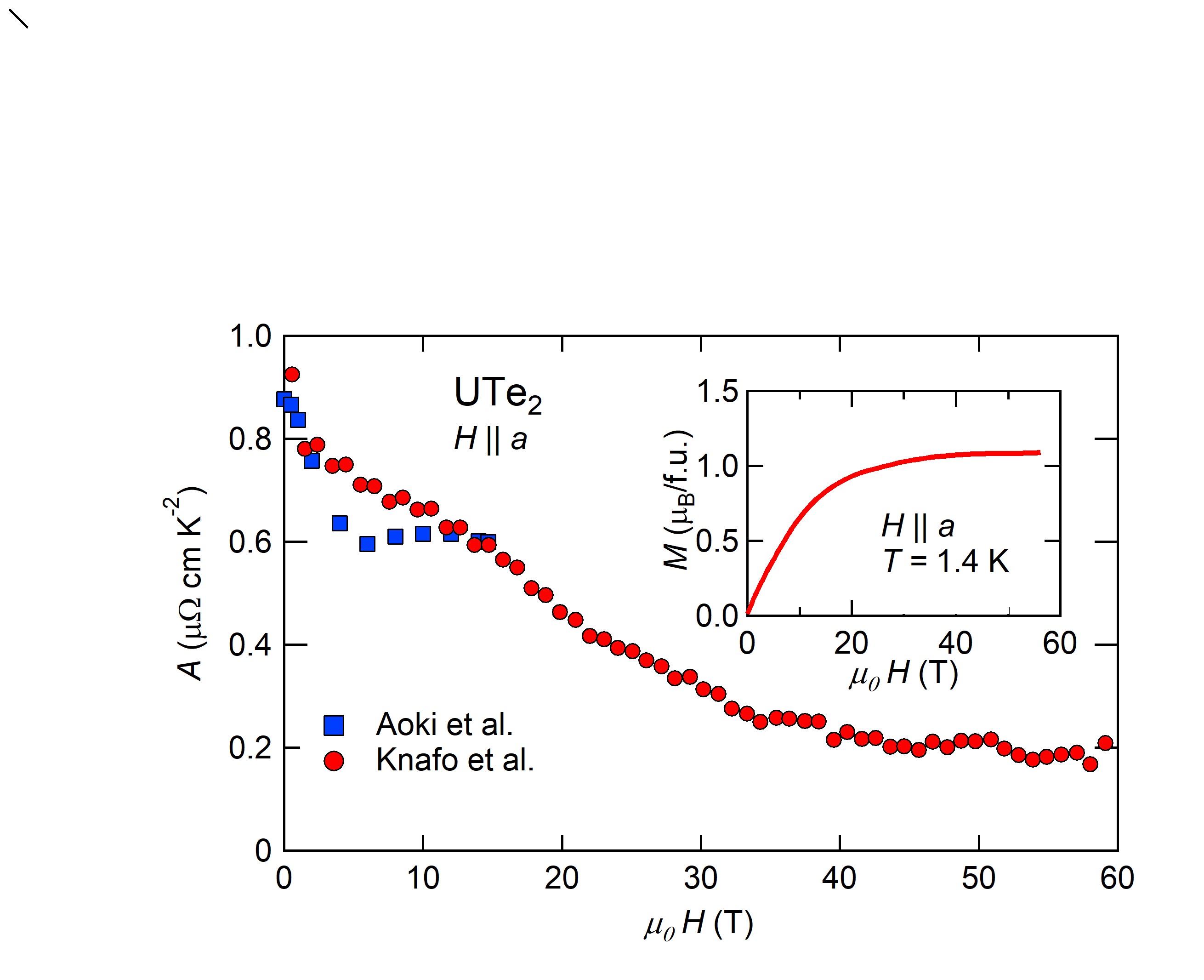}
\caption{Field dependence of the $A$ coefficient of the resistivity for $H \parallel a$ axis, taken from Refs.~\cite{Aoki2019, Knafo2019}. The inset shows the field dependence of the magnetization at $T = 1.4$~K.   }
\label{A_coef_a}
\end{center}
\end{figure}

\begin{figure}
\begin{center}
\includegraphics[angle=-90,width=1\columnwidth]{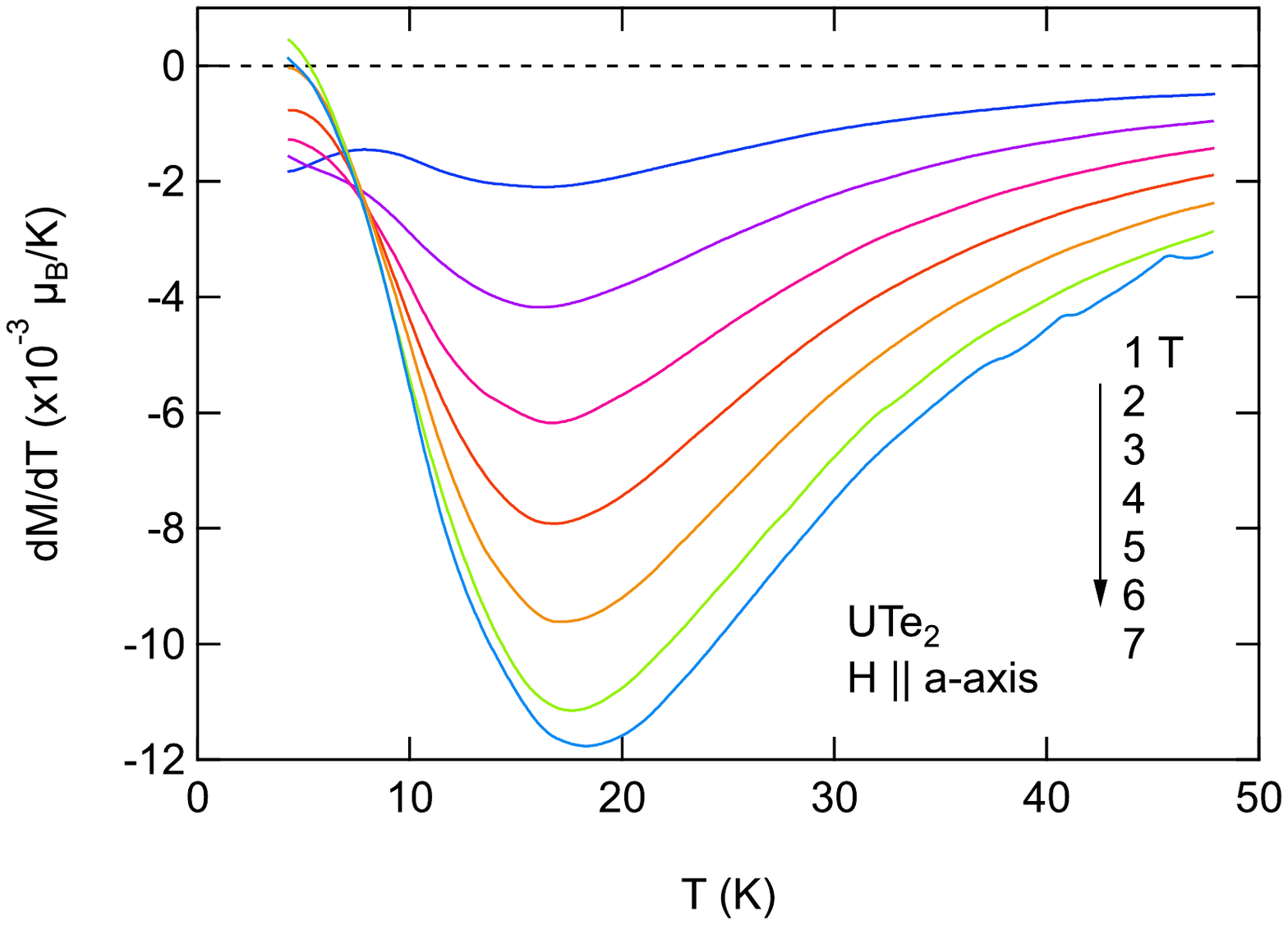}
\caption{Temperature dependence of $dM/dT$ for different fields along the $a$ axis up to 7~T. }
\label{dM_dT}
\end{center}
\end{figure}

In a crude single band scheme, the field dependence of the $A$ coefficient give some hints to the field dependence of the effective mass $m^\ast$. In a first approach the Kadowaki-Woods ratio is fulfilled in UTe$_2$ (The ratio $A/\gamma^2 \approx 10^{-5}\,\mu\Omega\,{\rm cm} ({\rm mJ} \,{\rm mol^{-1}K}^{-2})^{-2}$ has an universal value in many different heavy fermion materials) \cite{Kadowaki1986, Jacko2009}.  Of course, more detailed thermodynamic experiments are needed in future to obtain more profound knowledge of the field dependence of the effective mass $m^\ast$ (see Fig.~\ref{A_coef_a}). In first experiments, the $A$ coefficient with the current along the $a$ axis and a magnetic field $H \parallel a$  show a rapid decrease of $A$ for $H$ up to 6~T followed by an almost constant value for $H >6$~T \cite{Aoki2019}. Measurements using pulsed magnetic fields show that $A$ is continuously decreasing up to the highest field of 60~T \cite{Knafo2019, Knafo2020}. At a field of $H_a \sim 6$~T a field-induced Lifshitz transition has been clearly observed in thermoelectric power experiments  at low temperature \cite{Niu2020}. At the field of the Lifshitz transition  a small anomaly occurs in the field-derivative $dM(T)/dH$ \cite{Miyake2019}. An important  point is that at this field the field-induced magnetization reaches a critical value of 0.4~$\mu_{\rm B}$ \cite{Niu2020}, which is very close to the value of the magnetization where a first order metamagnetic transition along $b$  with a large jump of the magnetization occurs, as discussed below in section \ref{section_Hparallelb}. The thermoelectric power experiments show further anomalies near 10.5 and 21~T, which may hint to further Fermi surface changes, but up to now the details of the Fermi surface topology are  not determined. 

\begin{figure}
\begin{center}
\includegraphics[width=1\columnwidth]{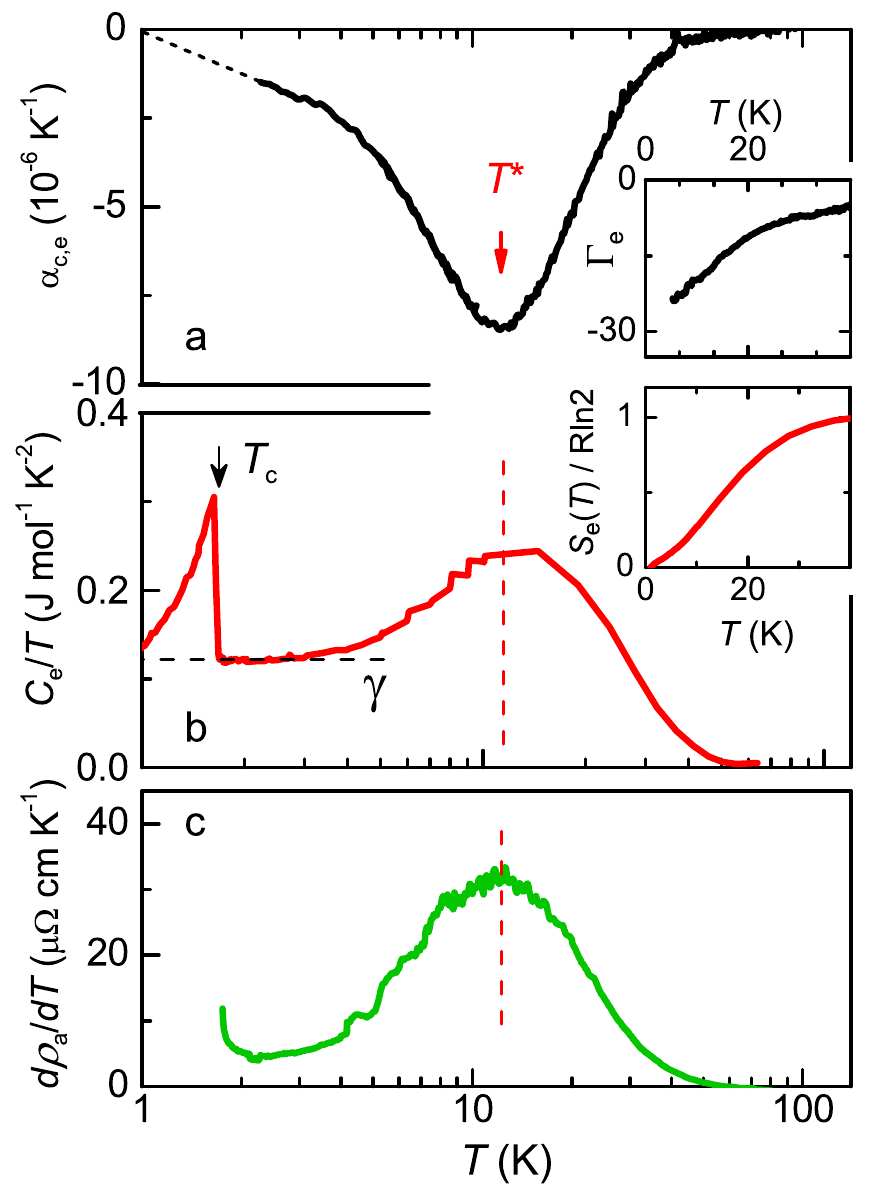}
\caption{Electronic part of the thermal expansion along the $c$ axis (upper panel), the electronic contribution of the specific heat $C_{\rm el}/T$ (middle panel), and the derivatived $d\rho_a/dT$ as function of temperature. The upper inset in (a) shows the temperature dependence of the electronic Gr\"uneisen parameter $\Gamma_{\rm el}$. The lower inset indicates the electronic entropy as function of temperature (after Ref.~\cite{Willa2021}). }
\label{Willa_thermal_expansion}
\end{center}
\end{figure}

We note that for $H \parallel a$, a clear change of regime occurs near $T^\ast \approx 12 - 15$~K, below which the susceptibility starts increasing strongly to lower temperatures at low magnetic fields. Similar to the resistivity, the characteristic temperature $T^\ast$ shows up very clearly as a pronounced minimum in $\frac{d M_a}{d T}$ for $H \to 0$ near 14~K; increasing the magnetic field leads to an increase of $T^*$ which can be followed by a clear minimum of $\frac{\partial M_a}{\partial T}$ as a function of $H$, $T ^\ast$ reaches 20~K at $H = 14$~T (see Fig.~\ref{dM_dT} and Ref.~\cite{Willa2021}). 
In difference to the $a$ axis susceptibility, the temperature dependence of the electronic specific heat  $(C_{\rm el} \propto \gamma T)$ as well as the that of the electrical resistivity follows a Fermi-liquid behavior ($\rho \propto T^2$)  above the superconducting transition. Figure~\ref{Willa_thermal_expansion}(b) shows the electronic specific heat divided by temperature $C_{\rm el}/T$ as a function of temperature. The superconducting anomaly at $T_{\rm sc} = 1.6$~K is very sharp. The Sommerfeld coefficient $\gamma \approx 0.12$~J$\cdot$mol$^{-1}$K$^{-2}$  just above $T_{\rm sc}$ indicates the moderate heavy fermion behavior. In the electronic contribution of the specific heat  $C_{\rm el}/T$ a broad maximum occurs at $T^\ast \approx T^\ast_a \approx 12$~K. Under magnetic field along the $a$ axis this maximum increases with field and can be located at 15 K at 9~T \cite{Willa2021}. 

\begin{figure}
\begin{center}
\includegraphics[width=1\columnwidth]{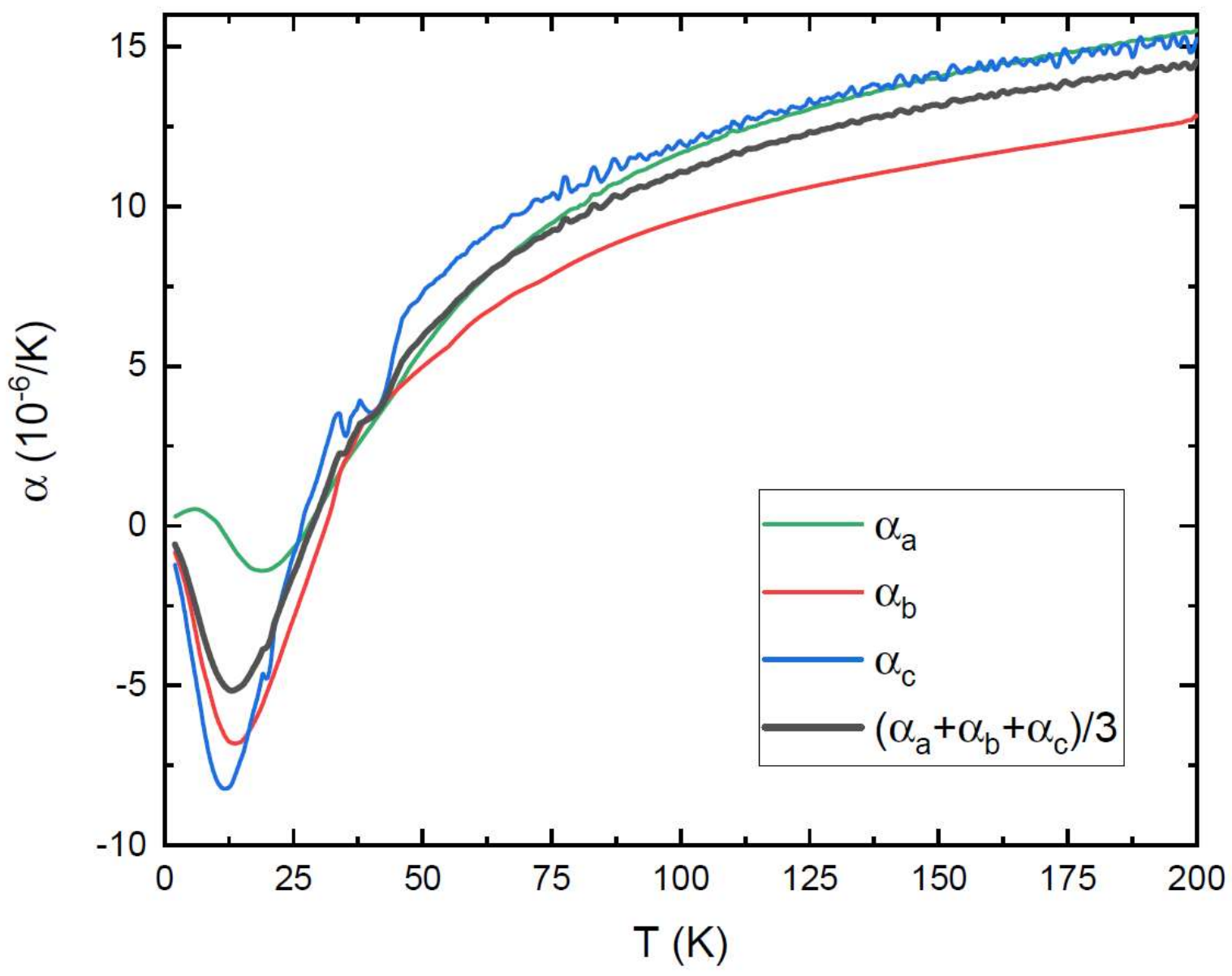}
\caption{ Linear thermal expansion coefficient as function of temperature for the three orthorhombic axes and the volume. (taken from Ref. \cite{Thomas2021})}
\label{thermal_expansion}
\end{center}
\end{figure}

\begin{figure}
\begin{center}
\includegraphics[width=0.95\columnwidth]{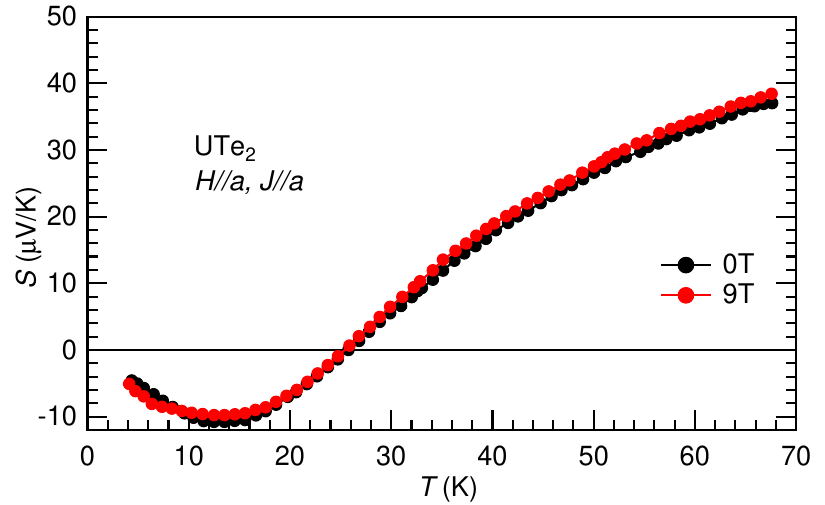}
\caption{Temperature dependence of the longitudinal thermoelectric power $S$ at $H = 0$ and 9~T along the $a$ axis. (taken from Reference \cite{Niu2020})}
\label{thermopower}
\end{center}
\end{figure}

A clear signature of $T^\ast$ is also found in the temperature dependence of the thermal expansion $\alpha_i= \frac{1}{\ell_i}\frac{\partial (\Delta \ell_i /\ell_i)}{\partial T}$ by a minimum for $\alpha_b$ and $\alpha_c$ and a change of sign for $\alpha_a$, as shown in Fig.~\ref{thermal_expansion} \cite{Thomas2021, Willa2021}.  
As the values for  $\alpha_b$ and  $\alpha_c$ are much larger than that for  $\alpha_a$, the volume thermal expansion  $\alpha_{\rm V}$ is negative below 30 K and shows a minimum at $T^\ast$. Knowing the value of the bulk modulus $B = 57$~GPa \cite{Honda2021}, the corresponding Gr\"{u}neisen parameter $\Gamma_{el} = B \frac{\partial T^\ast}{\partial p} = B \frac{\alpha_{\rm el}}{C_{\rm e}l}$ extrapolates to $\Gamma_{\rm el} (T=0) = -30$ while at $T^\ast \approx 12$~K, $\Gamma_{\rm e} \approx -17$ has only half of the extrapolation to $T \to 0$. Just below $T^\ast$, the free energy cannot be reduced to an unique variation of $T/T^\ast$ which would have lead to a constant Gr\"{u}neisen parameter: correlations play a major role.  
Taking $\Gamma_{\rm el} (T =0)$, the decrease of $T^\ast$ with pressure is estimated to be $-6\,{\rm K/GPa}$, which is close to the value detected by pressure experiments, $-7.8\,{\rm K/GPa}$ \cite{Ran2020}, $-5.8\,{\rm K/GPa}$ \cite{Li2021}. Other signatures of $T^\ast$ are also a maximum of the electrical resistivity with current along the $c$ axis \cite{Eo2021} and in thermoelectric power measurements with the thermal gradient along the $a$ axis  which shows a minimum (see Fig.~\ref{thermopower}) \cite{Niu2020}. Other indications of the characteristic temperature $T^\ast$  from microscopic measurements as discussed later (see section \ref{NMR_normal}) are the strong enhancement of the spin-spin relaxation rate $1/T_2$ in the NMR experiment \cite{Tokunaga2019} and the saturation of the antiferromagnetic fluctuations detected by the inelastic neutron scattering experiment \cite{Knafo2021}. In agreement with the field-increase with $H\parallel a$ of $T^\ast$ in $\frac{\partial M_a}{\partial T}$, a similar field dependence is observed for the extreme in $C_{\rm e}/T$ and in $\alpha_c$. Interestingly, the integrated entropy reaches $R\ln 2$ at $T \sim  50$~K indicating a doublet ground state; with the hypothesis of a major U$^{3+}$ configuration (i.e.~$5f^3$), a doublet can be expected in agreement with the measured anisotropy of the magnetization $M_a/M_b$ measured at 2~K \cite{Shick2021}.

The increase of the susceptibility $\chi_a$ on cooling below 50~K and the increase of the crossover temperature $T^\ast$ for field applied along the $a$ axis favor the picture that in addition to the antiferromagnetic fluctuations detected by inelastic neutron scattering  \cite{Duan2020, Knafo2021} low energy ferromagnetic fluctuations develop at low temperatures (see discussion on NMR and $\mu$SR data in section \ref{NMR_normal}) \cite{Tokunaga2019, Sundar2019}. At first glance an attempt may be to try to derive a  Fermi-liquid parameter  $F_0^a$ for $H \parallel a$, as it was done for liquid $^3$He \cite{Leggett1975}. However, a first difficulty is that $\chi_a$ at lowest temperatures does not follow a Fermi-liquid behavior above $T_{\rm sc}$ as discussed in section \ref{subsectionHpara_a}; a second particularity is that, contrary to $^3$He, where the moment of the quasiparticle is known, the corresponding quasiparticle may carry a moment very different than the one Bohr magneton of the free electronic carrier. Thus the value of the Wilson Ratio $R_{\rm W} \propto \chi /\gamma$ as well as the Korringa ratio ${\cal K} \propto 1/(K^2 T_1T)$, where $K$ is the Knight shift of the NMR resonance and $1/T_1$ the spin-lattice relaxation rate, which would indicate a nearly ferromagnetic state in UTe$_2$ (see Ref.~\cite{Willa2021}), can be misleading in the derivation of $F_0^a$ taking into account the dual localized and itinerant character of the $5f$ electrons and the complexity of the multi band structure \cite{Zou1986, Aeppli1987}. 

An open question is  the origin of the strong increase of the susceptibility $\chi_a$ below $T^\ast$. Using simple arguments on an extra contribution of $\chi_a$ from a Fermi liquid estimate (which may be reached above 6~T, or on the necessity to respect the entropy balance at $T_{\rm sc}$, a simple amount of free paramagnetic Ising centers near 1\% is sufficient to give a bath leading  on cooling to an apparent residual $\gamma_{\rm res}$ value in the superconducting state below $T_{\rm sc}$. This extra contribution has been neglected for the discussion of the superconducting properties \cite{Metz2019}. It is regarded as totally separated from the superconducting condensate. The proximity to valence, magnetic and electronic instabilities will favor the creation of defects strongly coupled to the lattice properties. 

\begin{figure}
\begin{center}
\includegraphics[width=0.95\columnwidth]{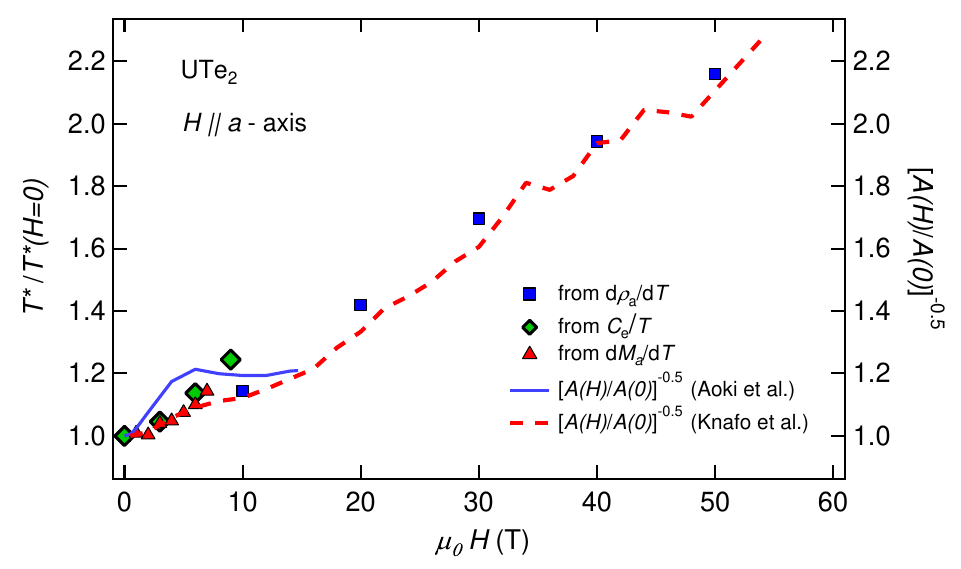}
\caption{Field dependence  of the characteristic temperature $T^\ast$ normalized to its zero field value for $H || a$. $T^\ast$ is obtained from the maximum of $d\rho_a/dT$, the maximum of the electronic specific heat $C_{\rm el}/T$, from the minimum of $dM_a/dT$. We also show the field dependence of $[A(H)/A(0)]^{-0.5}$, which is roughly proportional to the field dependence of $1/m^\ast$ (see also Ref.~\cite{Willa2021}).  }
\label{Tstar_a}
\end{center}
\end{figure}

Figure~\ref{Tstar_a} shows the relative field dependence of the characteristic temperature $T^\ast$ for a field along the easy magnetization axis $a$ defined either by the temperature derivative of the $a$ axis resistivity (data from Ref.~\cite{Knafo2019}), the maximum of the electronic specific heat $C_{\rm e}/T$,  the minimum of the  temperature derivative of the magnetic susceptibility $dM_a/dT$, or the inverse of the square root of the normalized $A$ coefficient of the resistivity. 
At least $T^\ast$ derived from $dM_a/dT$ indicates a change of the regime for $H\sim 6$~T; it is less clear from $T^\ast \propto 1/\sqrt{A}$ scaling. We note that within the accuracy of the determination of the $A$ coefficient from the pulsed field experiments, no clear field saturation of $A$ occurs, despite the saturation of the magnetization above 30~T.  The driving force which may control $T^\ast$ may be the result of the entrance in a coherent motion between the itinerant quasiparticles and the magnetic fluctuations originating from the local $5f$ uranium centers and not a characteristic energy of a given magnetic fluctuation. $T^\ast$ emerges clearly in the temperature dependence of the detected incommensurate antiferromagnetic fluctuations \cite{Knafo2021}. 

\begin{figure}
\begin{center}
\includegraphics[width=0.95\columnwidth]{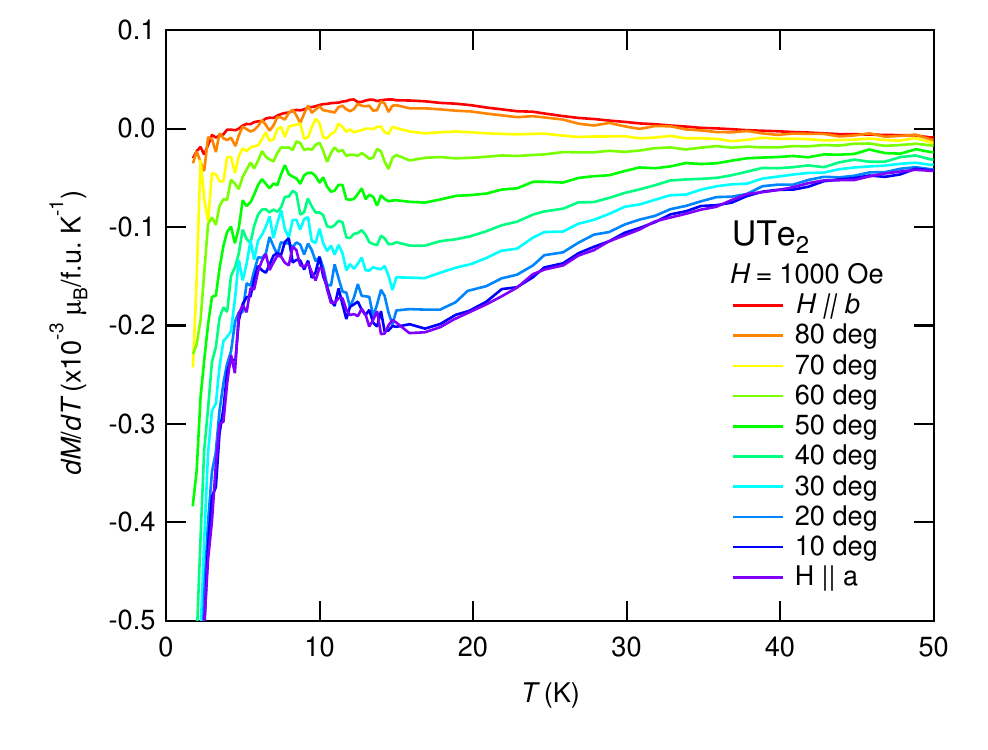}
\caption{Temperature derivative of the magnetization versus temperature measured at 1000 Oe for different angles from the $a$ axis to the $b$ axis in steps of 10 deg. }
\label{dM_dT_ang}
\end{center}
\end{figure}

As we will see from the concomitant analysis of the upper critical field $\Hc$ and the lower critical field $H_{c1}$ one may suspect that low energy ferromagnetic fluctuations play a major role in the superconducting pairing at low pressure with an initial strong field decrease for $H \parallel a$. In a simple classical case, a good way to detect the field dependence of the pairing is to go back to the temperature derivative of the magnetization, as it is discussed in URhGe \cite{Hardy2011a}.
Following the Maxwell relations,  $(\partial M/\partial T)$ at constant field is linked to the field derivative of the entropy as $(\frac{\partial M}{\partial T})_H = (\frac{\partial S}{\partial H})_T$. If the magnetization $M$ has a Fermi liquid $T^2$ dependence, we can derive from the temperature dependence of the magnetization the field derivative of the $\gamma$ value of the specific heat, as  $\frac{\partial M}{\partial T} =  T\frac{\partial \gamma}{\partial H}$. Figure~\ref{dM_dT_ang} shows the temperature dependence of $\partial M/\partial T$ measured at 1000 Oe for different angles from the $a$ axis to the $b$ axis.  For $H \parallel a$ $\partial M/\partial T$ is negative and it shows a pronounced minimum near $T^\ast \approx 15$~K. 
The position of the minimum is unchanged with increasing angle towards the $b$ axis. It gets lower in intensity and near 70 degree from $a$, it has vanished. For lower temperatures a negative maximum occurs on cooling for $H \parallel a$ and the absolute value of $\partial M/\partial T$ is strongly increasing and is a signature of the complexity of the magnetic response. For $H \parallel b$, $\partial M/\partial T$ changes sign from negative to positive near $T_\chi^{\rm max} \approx 35$~K and a positive maximum occurs near 15~K. The sign change at lowest temperature may indicate some paramagnetic impurities. 

Thus, the magnetization measurements give already a first hint on the field dependence of the superconducting pairing interaction, as in a simple picture $\gamma (H) \propto m^\ast (H)\propto \lambda (H)$. 
It is obvious that the field-decrease of $\partial S/\partial H$ occurs for field along the $a$ axis and a field-increase of  $\partial S/\partial H$ for $H \parallel b$ for $H \parallel a$ the magnetic field will lower the pairing while for $H \parallel b$ an increase of the pairing is expected. 
Despite the complexity of the $\partial M/ \partial T$ response for $H\parallel a$, a rough estimation of $\partial \gamma/ H$ is in agreement with the possible  strong collapse of the pairing, as will be discussed later.

\subsection{$H\parallel b$: Metamagnetism and $T_{\rm max}$}
\label{section_Hparallelb}

The remarkable feature in the normal state for a magnetic field applied along the $b$ axis is the occurrence of a metamagnetic transition at $\Hm \approx 35$~T. Its link
with a ferromagnetic instability is not obvious and it remains an intriguing phenomena.  Usually, one would expect that metamagnetism in a nearly ferromagnetic system will occur along the easy magnetic axis, as e.g.~observed in UCoAl \cite{Aoki2011a} or in paramagnetic state above the critical pressure in UGe$_2$ \cite{Pfleiderer2002, Taufour2010}. 
%
%
In a \replaced[id=GK]{simplistic}{ crude} model scheme of \added[id=GK]{the} density of states the pronounced maximum in $C_{\rm el}/T$ at $T^\ast$ \replaced[id=GK]{indicates that the maximum of the density of states is located not exactly at the Fermi energy $E_{\rm F}$ }{is a mark of  the minima of the density of states at $E_{\rm F}$}. \added[id=GK]{This} seems in agreement with electronic structure calculations, where a narrow peak near $E_{\rm F}$ occurs only at low temperature \cite{Miao2020}.  \replaced[id=GK]{It is also supported by the observation}{ also with the observation} of a maximum in the temperature dependence of the susceptibility $\chi_b$ along the $b$ axis at $T_{\rm max}$.  

\begin{figure}
\begin{center}
\includegraphics[angle=-90, width=0.95\columnwidth]{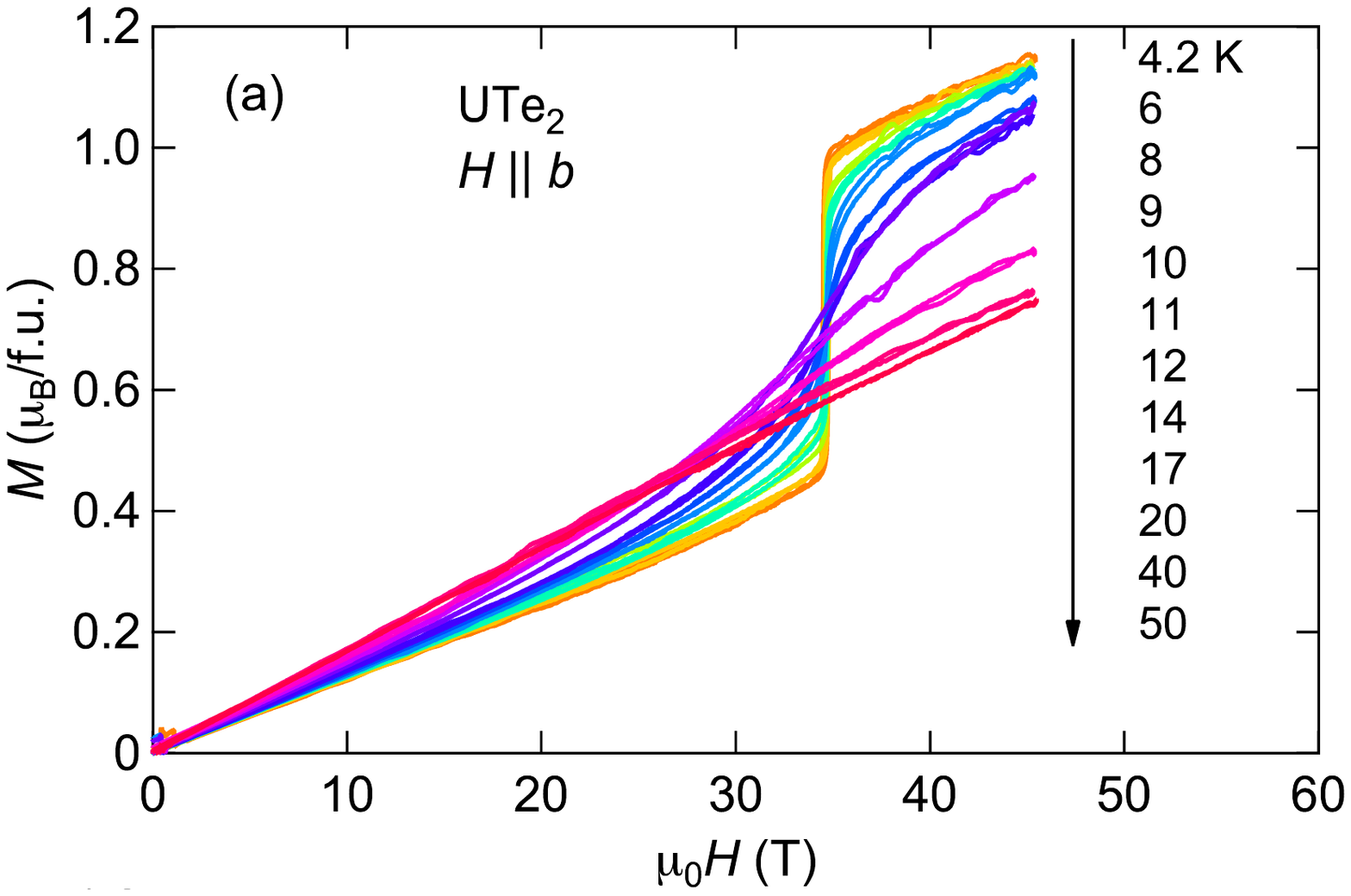}
\caption{Magnetization versus magnetic field applied along the $b$ direction at different temperatures. (taken from Reference \cite{Miyake2019})}
\label{Miyake_magnetization}
\end{center}
\end{figure}

\begin{figure}
\begin{center}
\includegraphics[angle=-90, width=0.95\columnwidth]{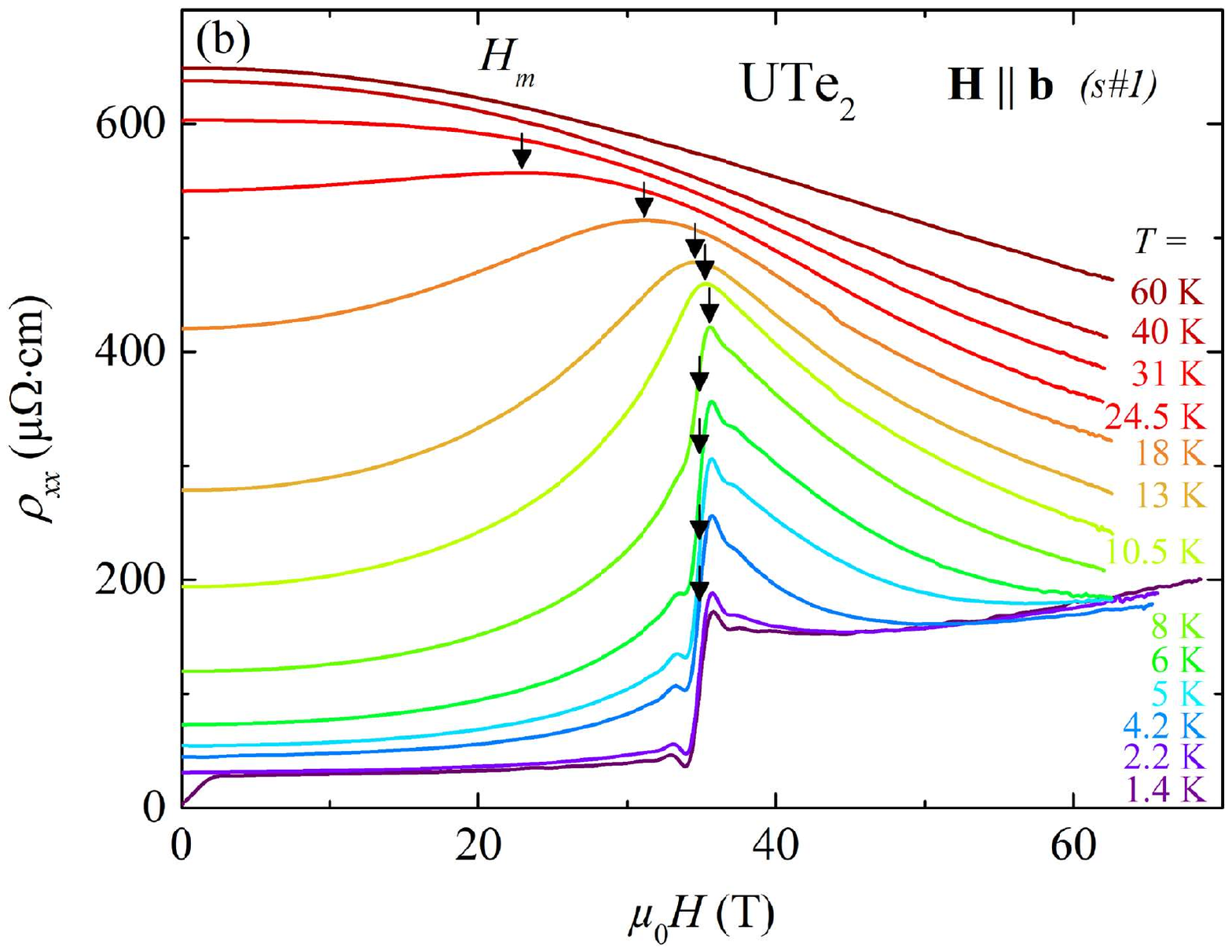}
\caption{Magnetoresistivity of UTe$_2$ for magnetic field applied along the $b$ axis. (taken from Reference \cite{Knafo2019})}
\label{Knafo_resist}
\end{center}
\end{figure}

Below 8~K field dependent magnetization measurements clearly show a first order metamagnetic transition at $\Hm \approx 35$~T at $M_b^\ast \approx 0.4 \mu_{\rm B}$ with a huge jump of $\Delta M_b^\ast \approx 0.5 \mu_{\rm B}$ and a hysteresis between field up and down sweeps, which increases to low temperatures \cite{Miyake2019, Ran2019a} (see Fig.~\ref{Miyake_magnetization}). Interestingly,  the slope $\frac{\partial M}{\partial H}$ is almost the same below and above $\Hm$ (see Fig.~\ref{Miyake_magnetization}) as if the metamagnetic instability corresponds mainly to a jump of the local $5f$ magnetization. Above a critical end point $T_{\rm CEP} \approx 6-8$~K, where the hysteresis of the transition vanishes, the first order transition changes into a crossover and the magnetization $M_b$ shows a marked inflection point as function of field along the $b$ axis. Magnetoresistivity measurements for $H\parallel b$  show that this jump of the magnetization is accompanied by a huge jump of the residual resistivity $\rho_0$ at $\Hm$ by a factor of 4, as shown in Fig.~\ref{Knafo_resist}. Increasing temperature above 7~K the first order nature gets lost and the maximum in the magnetoresistivity and the Hall effect indicates a crossover which seems to be connected to the maximum in the magnetic susceptibility $\chi_b$ \cite{Knafo2019, Niu2020b}.

\begin{figure}
\begin{center}
\includegraphics[angle=-90, width=0.95\columnwidth]{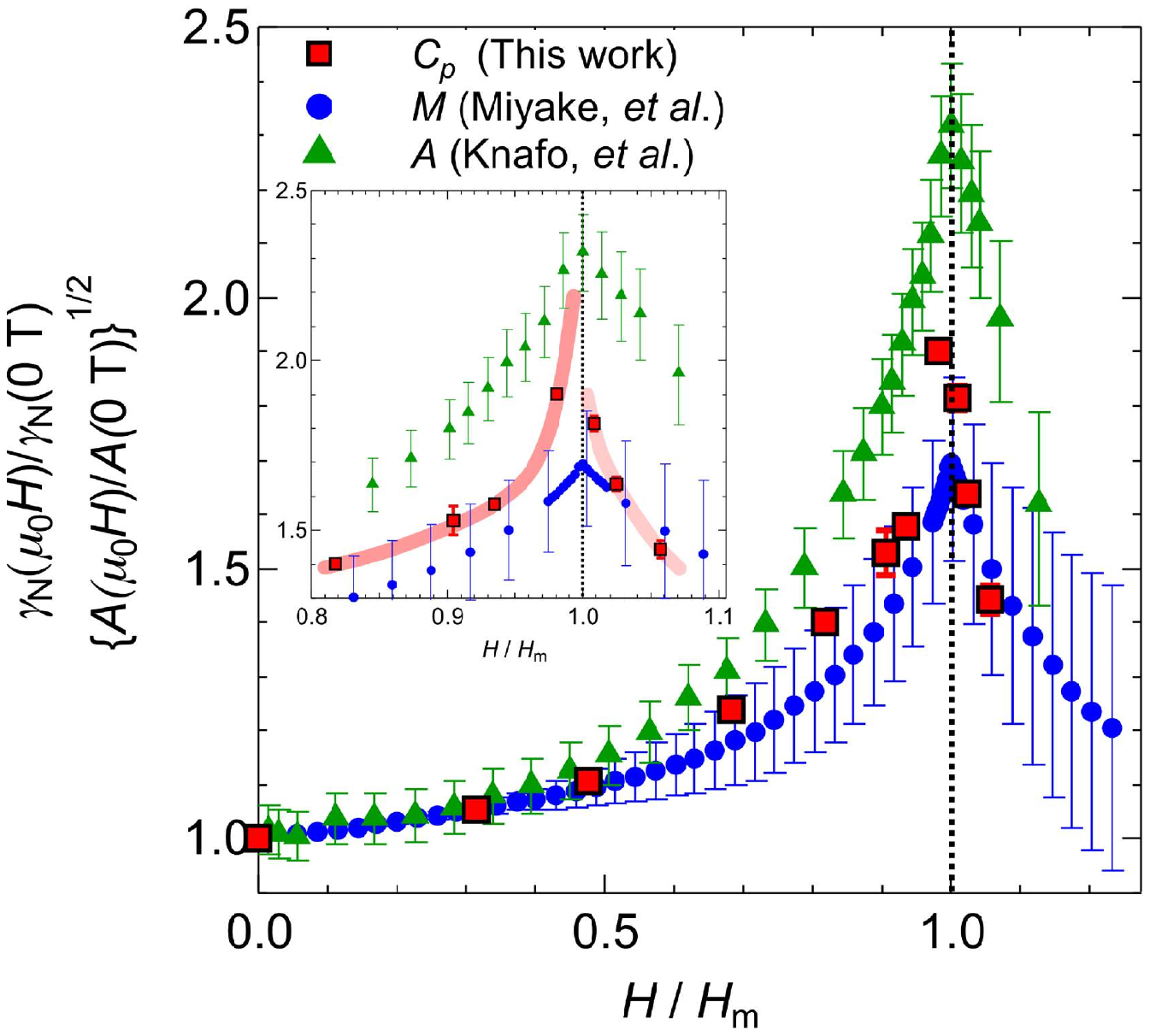}
\caption{Magnetic field dependence of $\gamma_{\rm N}$ normalized to  $\gamma_{\rm N} (H=0)$ as function of field normalized to $\Hm$. The red squares are from specific heat measurements in pulsed magnetic field \cite{Imajo2019}, blue circles from magnetization measurements using the Maxwell relation \cite{Miyake2019}, and green triangles from the $A$ coefficient of the resistivity \cite{Knafo2019}. The inset gives a zoom on the field range near $\Hm$. (Figure taken from Reference \cite{Imajo2019})}
\label{Imajo_gamma}
\end{center}
\end{figure}

In addition, the $A$ coefficient of the resistivity increases strongly on approaching $\Hm$ and has a strong maximum at $\Hm$ \cite{Knafo2019}. A rather similar enhancement is observed in $\gamma (H)$ derived from the temperature dependence of the magnetization $M_b (T)$ by using the Maxwell relation \cite{Miyake2019}, and also by direct measurements of the specific heat at high magnetic fields \cite{Imajo2019} as shown in Fig.~\ref{Imajo_gamma}. From this last measurements, just below $\Hm$ we find $\frac{\gamma (\Hm  -\varepsilon)}{\gamma_0} \sim 1.8$ while just above $\Hm$ the ratio $\frac{\gamma (\Hm  + \varepsilon)}{\gamma_0} \sim 1.6$. Thus there may be a discontinuity of $\gamma_{\rm n}(H)$ at the first order metamagnetic transition. 
Recently, taking into account the magnetocaloric effect and the Clausius-Clapeyron equation, a discontinuity of the Sommerfeld coefficient $\gamma$ through $\Hm$ has been established \cite{Miyake2021b} (see also Fig.~\ref{Comp_mag} in section \ref{Comp_FM-SC}). It is interesting to note that an almost similar field dependence of the $A$ coefficient has been reported for an angle of 30 deg from the $b$ to the $c$ axis, where the field enhanced superconductivity occurs only above the metamagnetic transition \cite{Ran2019a, Knafo2020}. Again, the recent magnetocaloric measurements find a jump of $\gamma$ on crossing $\Hm$ \cite{Miyake2021}. A remaining open question is to analyse the weight of nonuniform and time dependent magnetic fluctuations during the pulse which can be induced on crossing $\Hm$ in pulse magnetic fields \cite{Miyake2021c}.


\begin{figure}
\begin{center}
\includegraphics[width=0.95\columnwidth]{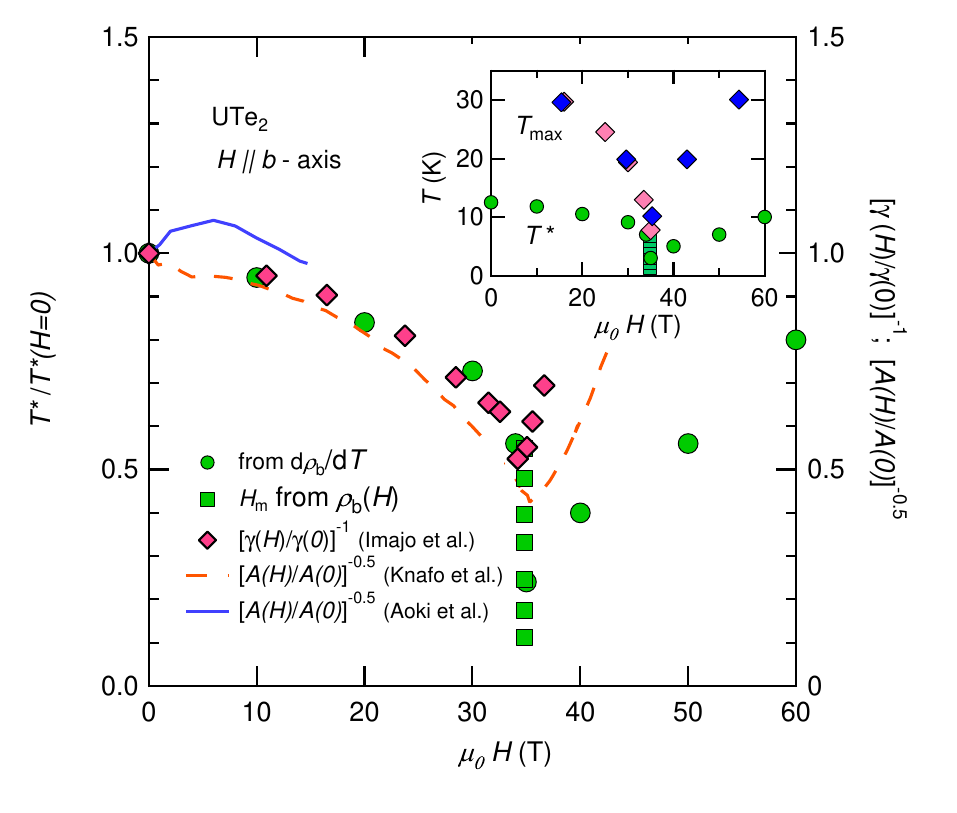}
\caption{Field variation for $H || b$ of the characteristic temperature $T^\ast$ normalized to its zero field value. $T^\ast$ is obtained from the maximum of $d\rho_a/dT$. We also show the field dependence of $[A(H)/A(0)]^{-0.5}$ \cite{Knafo2019}, and the normalized $\gamma$ value $[\gamma (h)/\gamma (0)]^{-1}$ \cite{Imajo2019} which is roughly proportional to the field dependence of $1/m^\ast$. The inset compares the field dependence of $T^\ast$ and $T_{\rm max}$ taken from Ref.~\cite{Niu2020b}.}
\label{Tast_b}
\end{center}
\end{figure}

Following the previous considerations for $H\parallel a$, as the derivative of the resistivity resembles the specific heat anomaly around a magnetic phase transition \cite{Fisher1968} the relative field dependence of $T^\ast_\rho$ for $H \parallel b$ is reported on Fig.~\ref{Tast_b} to track the field variation of the crossover regime. 
We also added the relative field dependence of $T^\ast_\gamma \sim 1/\gamma$, $T^\ast_A \sim 1/\sqrt{A}$ normalized to their zero field values. The inset in Fig.~\ref{Tast_b} compares the field dependence of the crossover temperatures $T^\ast$ and $T_{\max}$. Clearly, $T^\ast_\rho$ and $T_{\max}$ end at the critical end point of the first order transition $T_{\rm CEP} \approx 7$~K at 34.5~T, while $T^\ast_\gamma$ and $T^\ast_A$ show deep minima. A basic question is why such an enhancement of the effective mass (reflected in the increase of $\gamma \propto  \sqrt{A}$) on approaching the metamagnetic field can occur at a first order phase transition. In the framework of ferromagnetic fluctuations an enhancement of $\gamma$ on both sides of $\Hm$ \cite{Miyake2021} has been proposed. However, the microscopic origin of the metamagnetism remains an open question. 


Interestingly, we calculate the Gr\"{u}neisen parameter of $\Hm$, $\Gamma_{\Hm} = - \frac{\partial \ln \Hm}{\partial \ln V} = -21$ taking the observed pressure decrease of $\Hm$, which is about - 15~T/GPa \cite{Knebel2020}. It is quite near the values of $\Gamma_{\rm el}(T^\ast)$ and $\Gamma_{\rm el} (T=0)$. Thus the collapse of $T^\ast$ and $\Hm$ will occur in the same pressure range, as shown experimentally \cite{Ran2020, Knebel2020} (see section \ref{section_Pressure} below). Furthermore, the comparable values of the different Gr\"{u}neisen parameters $\Gamma$ stress that the driving phenomena must be the same up to $\Hm$ despite possible changes in the magnetic coupling. Knowing the pressure variation of $\Hm$ and the jump of the magnetization at $\Hm$, a relative volume shrinking near $10^{-3}$ is associated to the first order transition at $\Hm$. This shrinking is 30 times lower than the volume contraction going from $p=0$ to the critical pressure $p_{\rm c}$ \cite{Honda2021}.        

It is obvious that under a magnetic field the Zeeman splitting of the heavy bands drives an eletronic instability for a critical value of the magnetization $M^\ast_b  \sim 0.4 \mu_{\rm B}$ at $\Hm$.   At the opposite to the behavior of $\gamma$ and the $A$ coefficient on crossing the metamagnetic transition, the thermoelectric power changes steplike at $\Hm$ and a sign change from negative to positive occurs at low temperature \cite{Niu2020b}. This is a clear indication that the majority heat carriers change at the metamagnetic transition and gives a strong hint to a Fermi surface instability. Additional support for this is given by Hall effect measurements which also indicate a decrease of the charge carrier number by a factor of nine, which presents a clear hint to a Fermi surface instability occurring at $\Hm$ \cite{Niu2020b}.  


\subsection{Spin dynamics probed by nuclear magnetic resonance}
\label{NMR_normal}

\begin{figure}[htbp]
\begin{center}
\includegraphics[width=8.0cm]{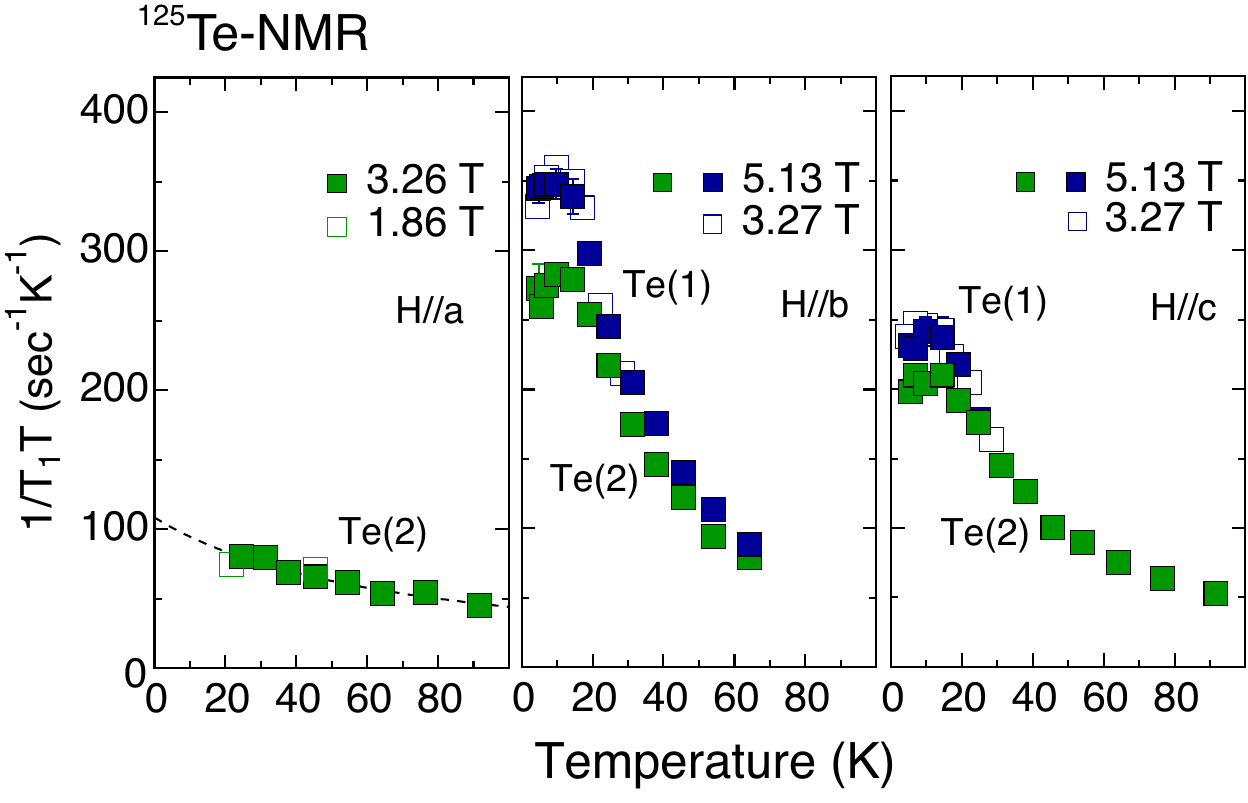}
\end{center}
 \vspace*{-10pt}
\caption{The temperature dependence of $(1/T_1T)_\alpha$ for fields applied along the three crystalline axes \cite{Tokunaga2019}.}
  \label{NMRF1}
\end{figure}

 
The nature of the spin fluctuations in UTe$_2$ has been investigated by means of $^{125}$Te-NMR \cite{Tokunaga2019}. $^{125}$Te nuclei ($I=1/2$) carries no quadrupole moment and has a gyromagnetic ratio $\gamma_{\rm N}/2\pi = 13.45$ MHz/T, so that
the nuclear relaxations occur purely in magnetic origin. 
Figure~\ref{NMRF1} shows the temperature dependence of $(1/T_1T)_\alpha$ measured in a single crystal for fields applied along all three crystal axis directions ($\alpha$= $a$, $b$ and $c$). 
$(1/T_1T)_{b,c}$ exhibit a strong temperature dependence in contrast with flat behavior for $(1/T_1T)_{a}$.
With decreasing temperature, $(1/T_1T)_{b,c}$ increase rapidly and tend to saturate below 10 K.
There is no obvious field dependence below $5$ T in low fields.

\begin{figure}[htbp]
\begin{center}
\includegraphics[width=7.5cm]{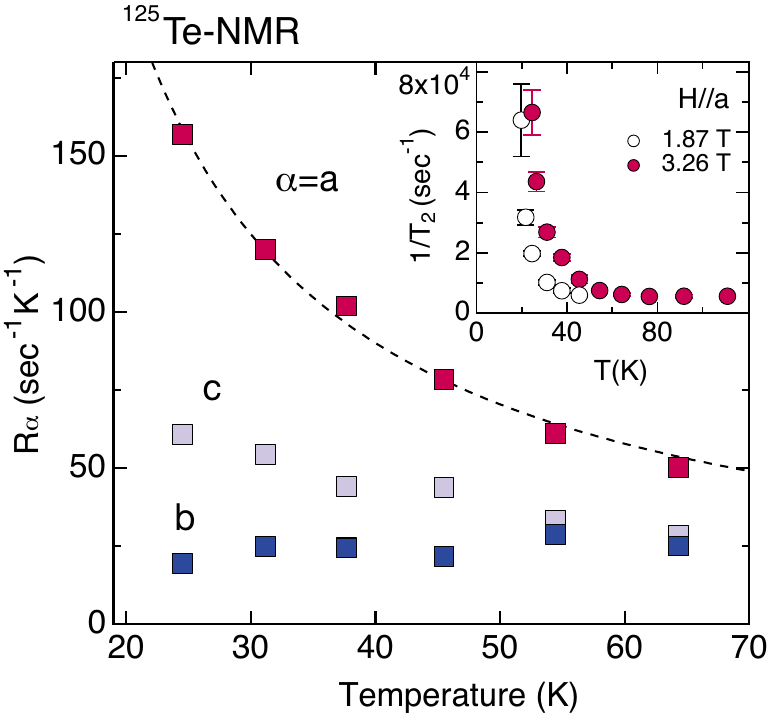}
\end{center}
 \vspace*{-10pt}
\caption{Orientation-resolved dynamic susceptibility $R_\alpha$ (see text) along the three crystalline axis directions in UTe$_2$.
The inset shows the temperature dependence of $1/T_2$ for fields applied along the the $a$- axis \cite{Tokunaga2019}.}
  \label{NMRF2}
\end{figure}

In general, $1/T_1T$ measured in a field along the $\alpha$ direction is associated with the imaginary part of the
dynamic susceptibility $\chi^{''}_{\beta,\gamma}({\bm{q}}, \omega_{\rm n})$ along the $\beta$ and $\gamma$ directions perpendicular
to $\alpha$ \cite{Moriya1963,Ihara2010}.
Hence  the directional dynamic susceptibility components for each orthorhombic crystal axis,
\begin{equation}
 R_ \alpha=\frac{\gamma_{\rm n}^{2}k_{\rm B}}{2}\sum_{{\bm q}}|A^{\alpha}_{\rm hf}|^2\frac{\chi^{''}_{\alpha}({\bm q},\omega_{\rm n})}{\omega_{\rm n}}
 \end{equation} 
can be evaluated using the relations $(1/T_1T)_a=R_b+R_c$, $(1/T_1T)_b=R_a+R_c$ and $(1/T_1T)_c=R_a+R_b$, 
where $\omega_n$ is the NMR resonance frequency  and $A_{\rm hf}^\alpha$ the hyperfine coupling constant. The results are shown in Fig.~\ref{NMRF2}.
Since there is no $(1/T_1T)_a$ data below 20 K (due to the divergence of $1/T_2$ as discussed later), $R_{\alpha}$ was not estimated in that temperature region.
Figure~\ref{NMRF2} demonstrates that UTe$_2$ exhibits a strong  anisotropy for the dynamical spin susceptibilities, i.e., $R_{a}\gg R_{c}> R_{b}$ above 20 K,
which is, however, moderate, compared to the case for UCoGe  \cite{Ihara2010}.
Note that the static  ($\chi_{\alpha}$) and dynamical spin susceptibilities ($R_{\alpha}$) often possess contrasting anisotropies in $f$-electron systems \cite{Kambe2007,Sakai2010,Baek2010,Sakai2012}.  In the present case, however, nearly the same anisotropy has been found for them.


The inset of Fig.~\ref{NMRF2} shows the temperature dependence of the spin-spin relaxation rate $1/T_2$ at 1.87 and 3.26~T. 
$1/T_2$  starts to increase upon cooling below about 40 K, and 
diverges near 20 K.  
The $1/T_2$ observed here reflects the magnitude of the longitudinal component (parallel to the applied field) of slow spin fluctuations near zero frequency ($\omega\sim0$).
That is, $1/T_2\propto G_{\|}(0)$, where $G_{\alpha}(\omega)=\int_{-\infty}^{\infty}\left<h_{\alpha}(t) h_{\alpha}(0) \right>\exp(i\omega t)dt$
is the spectral density of the fluctuating hyperfine field \cite{Tokunaga2015,Tokunaga2016}.
In general, the development of such a low frequency mode of fluctuations implies the onset of static order along the $a$-axis.
However, neither specific heat nor other bulk measurements have detected any signature of a phase transition around 10 - 20~K \cite{Ran2019,Aoki2019,Ikeda2006}, but a broad anomaly in specific heat at $T^\ast \approx 12$~K.
Clearly, there is a crossover temperature for magnetic fluctuations near 20~K for $H \parallel a$, below which the time scale of the
longitudinal fluctuations along the $a$ axis become extremely slow, of the order of NMR frequencies ($\sim$MHz), i.e. very slow by comparison to the antiferromagnetic correlations detected in neutron scattering \cite{Knafo2021} (see section \ref{section_neutron}).

Slow spin fluctuations has been also detected by $\mu$SR at zero external field.  
However, These spin fluctuations develop only below 5 K \cite{Sundar2019}.
The discrepancy in temperature between NMR (performed in fields $H>1.8$ T) and $\mu$SR ($H=0$ T) suggests that there exists strong field-dependence of the slow fluctuations at low temperatures. 
The $\mu$SR experiment also demonstrates the coexistence of the slow fluctuations with superconductivity below $T_{\rm SC}$.

 
So far it has not been concluded from the NMR data whether the observed slow fluctuations are ferromagnetic in character.
The $\mu$SR experiment suggests that the temperature dependence of the muon relaxations rate follows  to the power-law equation $\propto T^{-4/3}$  in an intermediate temperature range of $0.4 <T<4.9$~K \cite{Sundar2019}; the behavior is expected from the self-consistent renormalization (SCR) theory near a ferromagnetic quantum critical point in a three-dimensional metal \cite{Moriya1991}. However, the deviation from this power law at the lowest temperatures has still to be clarified;  the occurrence of ferromagnetic fluctuations in UTe$_2$ is still puzzling.

\subsection{Spin dynamics probed by neutron scattering} 
\label{section_neutron}

\begin{figure}[htbp]
\begin{center}
\includegraphics[width=0.9\columnwidth]{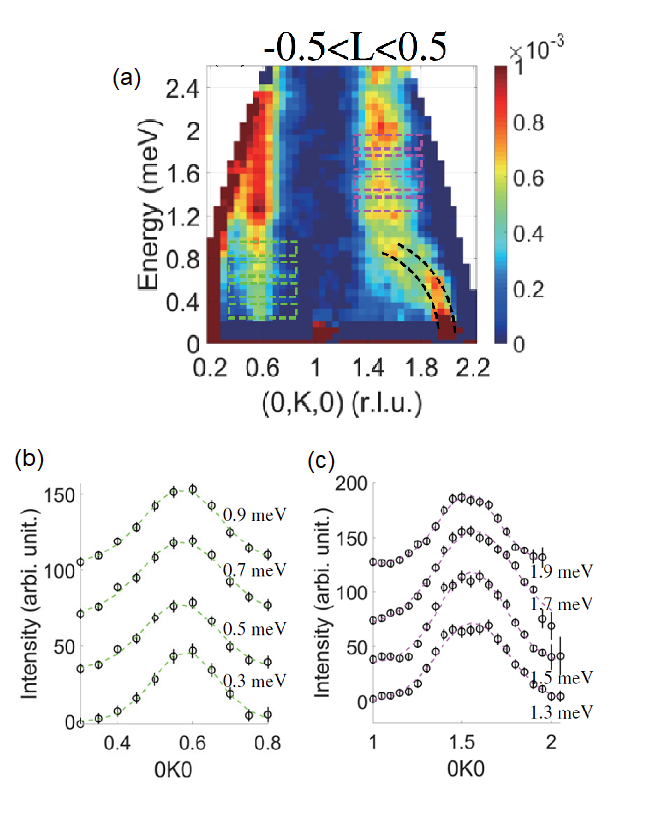}
\end{center}
\caption{(a) Energy dependence of the magnetic excitations measured  along the $[0, K, 0]$ direction revealing two excitations at  $\mathbf{k_1} = (0, 0.57, 0)$ and  $\mathbf{k_2} = (0, 1.57, 0)$ integrated between $-0.5<L<0.5$. Panels (b) and (c) show 1D cuts along the $K$ direction at different energies $E$ through the vectors $\mathbf{k_1}$ and $\mathbf{k_2}$. (Figure taken from Ref.~\cite{Duan2020}.) }
  \label{Duan}
\end{figure}

As inelastic neutron scattering is wave vector resolved it allows to characterize fully the magnetic excitations. A first inelastic neutron scattering experiment has been performed at the Cold Neutron Chopper Spectrometer at Oak Ridge National Laboratory on a large assembly of 61 oriented single crystals to obtain a total mass of 700~mg, but with the constraint of a mosaicity of about 15 degrees of the full assembly \cite{Duan2020}. Surprisingly, it led to the discovery of antiferromagnetic fluctuations at the incommensurate wavevectors $\mathbf{k_1} = (0, 0.57, 0)$ and  $\mathbf{k_2} = (0, 0.43, 0)$ (in reciprocal lattice units) which extends at least up to 2.6~meV as shown in Fig.~\ref{Duan}. The obtained incommensurate wave vectors are very close to the commensurate antiferromagnetic vector $(0,0.5,0)$. The emergence of these antiferromagnetic fluctuations is explained by electronic structure calculations and it has been concluded that the RKKY interaction between the 5$f$  moments drive the antiferromagnetic fluctuations \cite{Duan2020}.  

\begin{figure}[htbp]
\begin{center}
\includegraphics[width=0.8\columnwidth]{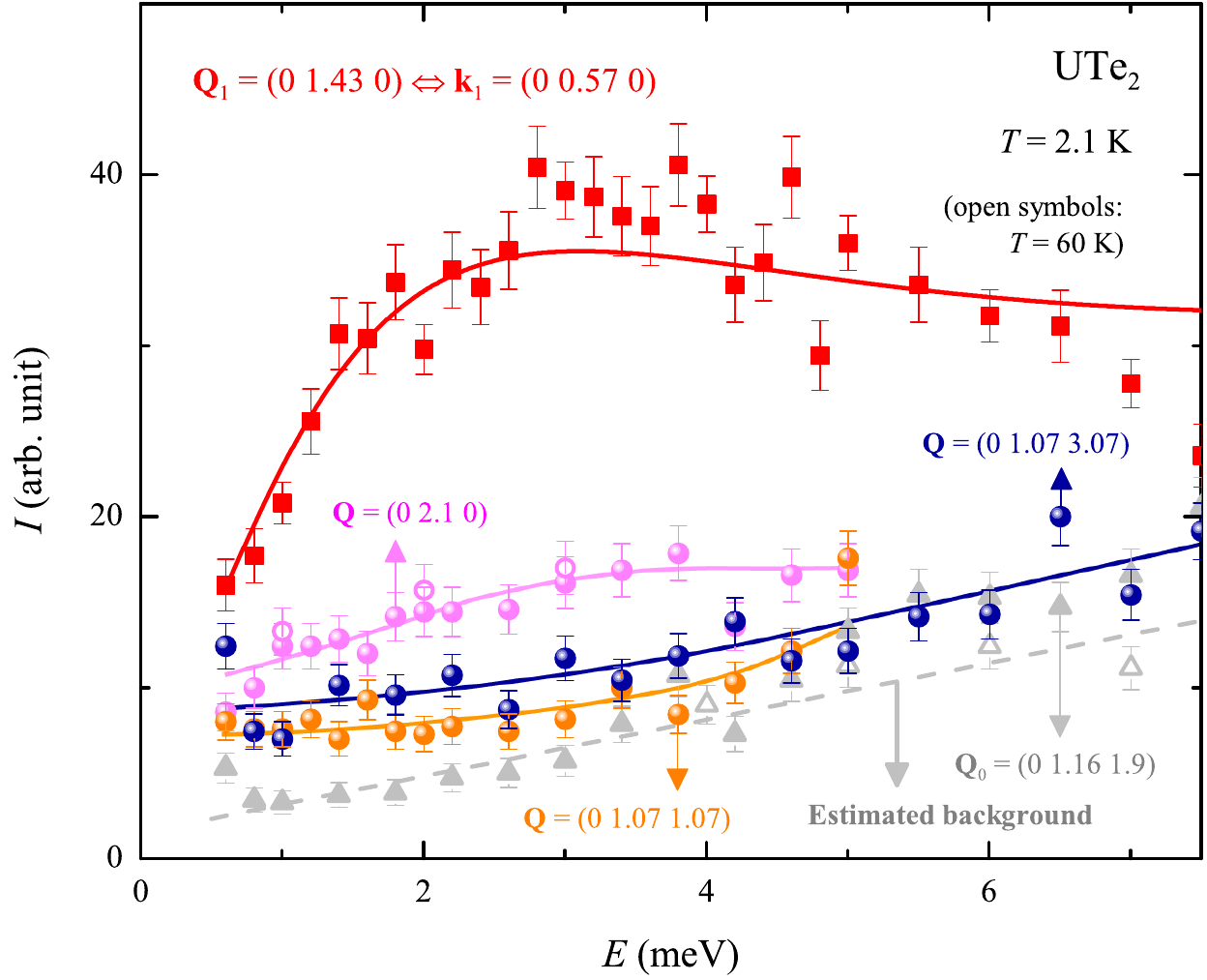}
\end{center}
\caption{Energy scans measured at the temperature $T = 2.1$~K and the momentum transfers $\mathbf{Q_1} = (0, 1.43, 0)$ characteristic of antiferromagnetic fluctuations (red squares). The other $\mathbf{Q}$ vectors are characteristic for ferromagnetic fluctuations, but no remarkable signal compared to the background (taken at $\mathbf{Q_0} = (0, 0.16, 0.19$) or the signal taken at $T=60$~K has been observed. For $\mathbf{Q_1}$ the line is a fit to the data by a quasi-elastic Lorentzian shape. (Figure taken from Ref.~\cite{Knafo2021}.)  }
  \label{Knafo_energy_scan}
\end{figure}

Recently, neutron scattering experiments using a one single crystal confirmed the incommensurate magnetic fluctuations. This experiment has been performed on the triple axis spectrometer Thales at the Institute Laue Langevin with energy transfers in the range $0.6<E<7.5$~meV using a  crystal with a total mass of 241~mg with a mosaicity lower than 2 degrees \cite{Knafo2021}. Figure \ref{Knafo_energy_scan} represents energy scans at $T=2.1$~K at momentum transfers $\mathbf{Q_1} = (0, 1.43, 0)$ which corresponds to the incommensurate wave vector $\mathbf{k_1}$ ($\mathbf{Q_1} = \tau - \mathbf{k_1}$, where $\tau =(0, 2, 0)$ is a nuclear Bragg peak) and $\mathbf{Q} = (0, 2.1, 0)$, $\mathbf{Q} = (0, 1.07, 1.07)$, and $\mathbf{Q} = (0, 1.07, 3.07)$ which  are characteristic for ferromagnetic fluctuations. Obviously, within the experimental resolution, the intensities at the ferromagnetic wave vectors are very close to the background  level in this energy window. 

\begin{figure}[htbp]
\begin{center}
\includegraphics[width=0.95\columnwidth]{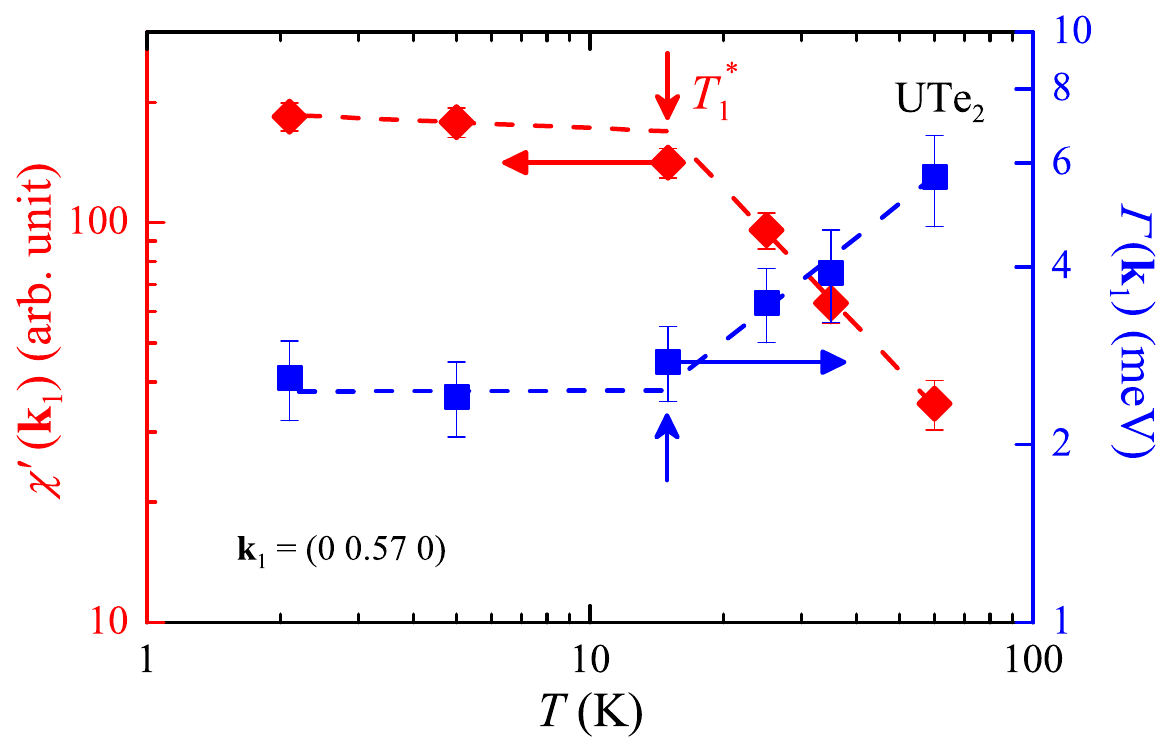}
\end{center}
\caption{Temperature dependence, in a double logarithmic presentation, of the relaxation rate $\Gamma (\mathbf{k_1})$ (blue) and of the real part of the static susceptibility at $\mathbf{k_1}$ (red).  (Figure taken from Ref.~\cite{Knafo2021}.)  }
 \label{Knafo_linewidth}
\end{figure}

The energy scans at $\mathbf{Q_1} = (0, 1.43, 0)$ have been performed up to 60~K. Figure \ref{Knafo_linewidth} shows the temperature dependence of the relaxation rate of the incommensurate fluctuations at $\mathbf{k_1}$ and of the static susceptibility $\chi'$ at $\mathbf{k_1}$. Both quantities point out again a characteristic temperature of 15~K, close to the value of $T^\ast$ detected in thermodynamic, electric transport and NMR measurements. With increasing temperature $\Gamma (\mathbf{k_1})$  increases strongly; an open question is whether antiferromagnetic correlations persists above 100~K as suggested by the Curie-Weiss temperature.  

\begin{figure}[htbp]
\begin{center}
\includegraphics[width=0.95\columnwidth]{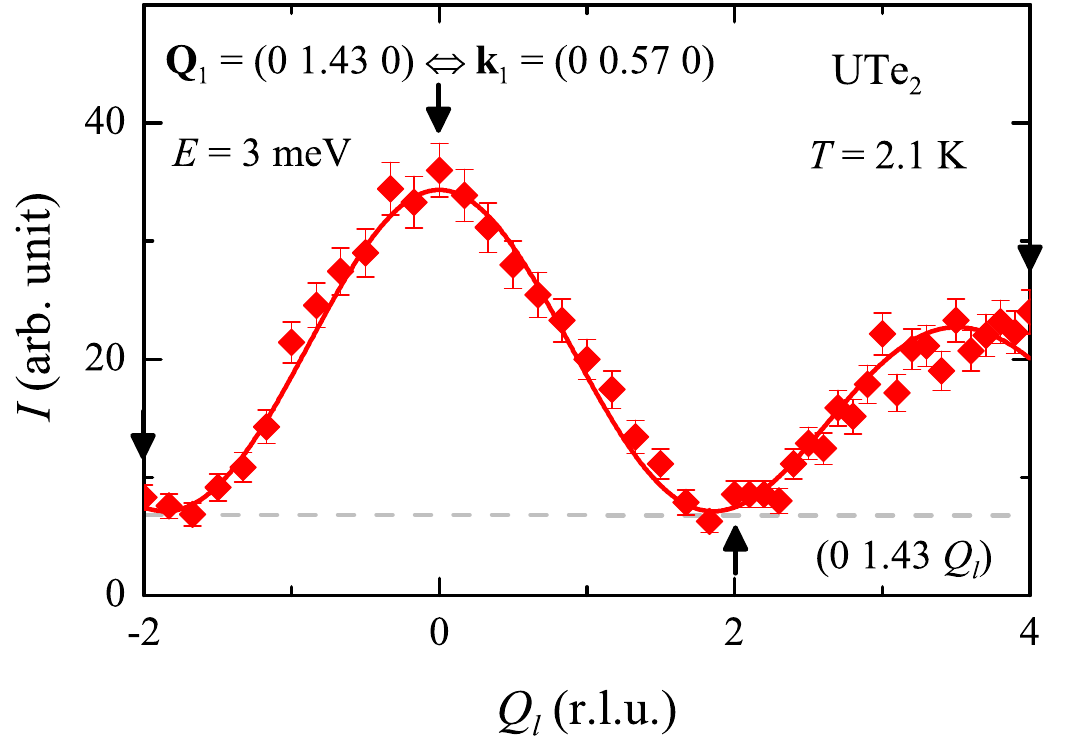}
\end{center}
\caption{$(0, 1.43, Q_L)$ scan at the energy transfer $E = 3$~meV and the temperature $T = 2.1$~K.  (Figure taken from Ref.~\cite{Knafo2021}.)  }
 \label{Knafo_Ql_scan}
\end{figure}

Fig.~\ref{Knafo_Ql_scan} displays a $Q_L$ scan at the momentum transfer $\mathbf{Q} = (0, 1.43, Q_L)$ at the energy transfer 3~meV. 
It shows a sine-wave evolution of the signal as discussed in Ref.~\cite{Knafo2021}, best fit to the data is made by assuming fluctuating magnetic moments $\mu \parallel a$ and that the magnetic moments on the ladder rung are ferromagnetically coupled \cite{Knafo2021}. The sine-wave modulation indicates that the magnetic correlations along the $c$ axis, i.e. the inter-ladder coupling in this directions can be neglected.  

In difference to the initially expected ferromagnetic correlations, neutron scattering experiments detect only incommensurate magnetic fluctuations close to an antiferromagnetic wave vector. These antiferromagnetic correlations involve ferromagnetically ordered moments $\mu \parallel a$ on the ladder rung, with antiferromagnetic inter-ladder coupling along the $b$ axis, and a quasi magnetic decoupling along the $c$. Future neutron experiments in the normal state above $\Tsc$  have to address the field dependence of these fluctuations in a magnetic field applied along the $a$ and $b$ axes.


\subsection{Comparison to the normal state properties of URhGe and UCoGe}

\label{Comp_FM-SC}

After the discovery of superconductivity, the paramagnetic UTe$_2$ has been considered to show very similar properties to the ferromagnetic superconductors URhGe and UCoGe \cite{Ran2019}. However, there are very basic differences between them. Firstly, the  URhGe and UCoGe crystallize in the TiNiSi-type orthorhombic structure with the space group $Pnma$ (\#62, $D^{16}_{2h}$), which belongs to the non-symmorphic space group. A particularity of this structure is that the U atoms form a zig-zag chain along the $a$ direction and thus,  the local inversion symmetry at the U site is broken. The shortest distances between U atoms are 3.497 and 3.481\AA\ for URhGe and UCoGe, respectively. It is much smaller than in UTe$_2$ and thus much closer to the Hill limit. 
Secondly, both, URhGe and UCoGe order ferromagnetic with respective  Curie temperatures of $T_{\rm Curie} = 9.5$ and 2.5~K. Bulk superconductivity occurs below $T_{\rm sc} \approx 0.25$~K in URhGe \cite{Aoki2001} and below $T_{\rm sc}\approx 0.6$~K in UCoGe \cite{Huy2007, Aoki2011, Taupin2014, Avers2021}. Most spectacularly, re-entrance of superconductivity in URhGe and a strong enhancement of the superconducting upper critical field in UCoGe along the $b$ axis near the field where the ferromagnetic order collapses in the transverse field \cite{Levy2005, Aoki2009}. 

\begin{figure}[htbp]
\begin{center}
\includegraphics[width=0.95\columnwidth]{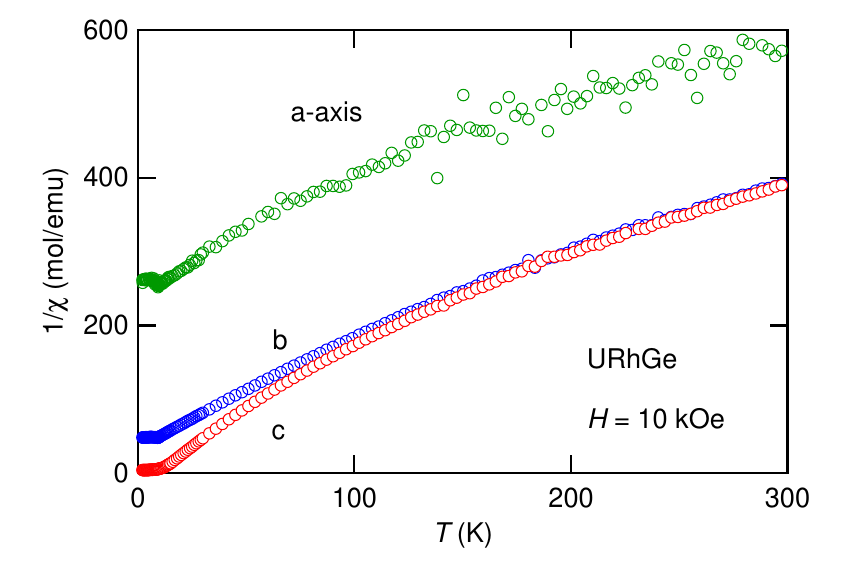}
\end{center}
\caption{Temperature dependence of the inverse susceptibility $1/\chi (T) $ of URhGe measured at a magnetic field of $H=10$~kOe for the $a$, $b$ and $c$ axis.  }
 \label{URhGe_inv_sus}
\end{figure}

In URhGe the re-entrance of superconductivity is accompanied by a reorientation of the magnetic moment from the easy $c$ axis under magnetic field to the $b$ axis, which results to a strong increase of the magnetization for $H \parallel b$ at $H_{\rm m}\sim 12\,{\rm T}$  \cite{Levy2005}. The origin of the metamagnetic transition along the hard magnetization $b$ axis is well understood (see also \cite{Mineev2021}). The transverse magnetic field along the $b$ axis leads to a field decrease of its Curie temperature \cite{Levy2005, Hardy2011a}. At zero field, the ferromagnetic moment is 0.4~$\mu_B$/U atom. In the temperature dependence of the inverse of the static susceptibility $\chi$ (see Fig.~\ref{URhGe_inv_sus}), it is obvious that $1/\chi$ along the three crystallographic axes has never a linear $T$ dependence over a significant temperature range from which definitely a Curie Weiss term can be extracted. Near $T \sim 300$~K the respective Curie Weiss temperatures would be $\Theta = -496$~K, -137~K and $-122$~K for the $a$,$b$ and $c$ axes, respectively. On cooling, for both $c$ and $b$ axes a strong decrease of $|\Theta |$  can be observed and finally for the $ c$ axis a dominant  ferromagnetic interactions towards $T_{\rm Curie} \approx 9.5$~K. A striking point is that in low fields due to the formation of a ferromagnetic order with a magnetic moment of $0.4 \mu_{\rm B}$ aligned along the $c$ axis, the susceptibility $\chi_c$ is small, but the susceptibility $\chi_b$ along the $b$ axis is three times larger than $\chi_c$. Thus, the comparison of the susceptibilities suggests that in high enough field the $b$ axis will become the easy magnetization axis and even a favorable axis for the development of a ferromagnetic component along $b$ \cite{Levy2005}.

\begin{figure}[htbp]
\begin{center}
\includegraphics[width=0.95\columnwidth]{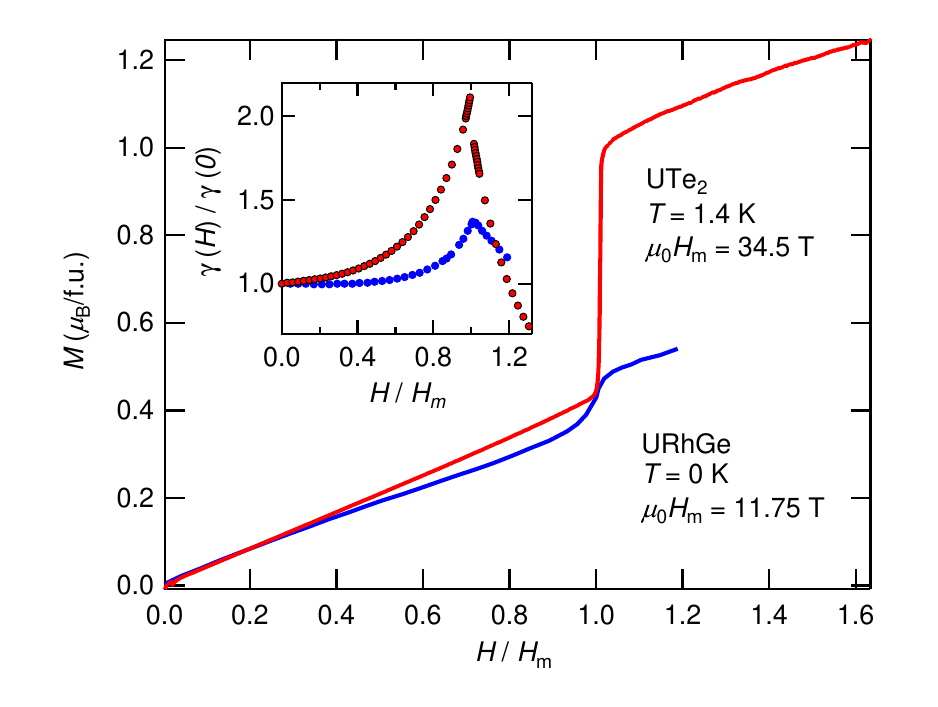}
\end{center}
\caption{Comparison of the magnetization along the $b$ axis of UTe$_2$ and URhGe normalized to the respective metamagnetic field. The inset shows the field dependence of the normalized specific heat coefficient $\gamma (H)/\gamma (0)$ as function of the normalized field $H/\Hm$ along the $b$ axis. (Data for URhGe are taken from Ref.~\cite{Hardy2011a}, data for UTe$_2$ from Ref.~\cite{Miyake2021b}.) }
 \label{Comp_mag}
\end{figure}

As URhGe orders ferromagnetically, on cooling no characteristic  crossover temperature will occur above the ferromagnetic Curie temperature $T_{\rm Curie}$ in thermodynamic properties such as $C/T$ or the thermal expansion $\alpha_{\rm V}$ at least from 30~K to $T_{\rm Curie}$. At 30~K the electronic Gr\"{u}neisen parameter $\Gamma_{\rm el} (30 {\rm K}) \approx +2$ and it rises up to $+15$ below $T_{\rm Curie}$ \cite{Sakarya2003, Gasparini2010}. As pointed out,  in URhGe $\Gamma_{\rm e}$ and $\Gamma_{\rm T_{\rm sc}}$ are quite opposite in excellent agreement that there is only one dominant mechanism in the effective mass enhancement and in the pairing mechanism (see below) \cite{Wu2017}. 

Similarly to UTe$_2$, metamagnetism is observed in URhGe for a critical value of the magnetization $M_b^\ast \approx 0.4 \mu_{\rm B}$, however the jump of the $\Delta M_b^\ast$ is five to six times smaller than in UTe$_2$ (see Fig.~\ref{Comp_mag}. This weak $\Delta M_b^\ast$ value may explain the rather tiny Fermi surface change \cite{Yelland2011} and thus the persistence of field reentrant superconductivity on a narrow field range above $H_m$. Qualitatively the relative enhancement of the Sommerfeld coefficient $\frac{\gamma (H)}{\gamma (H_m)}$ as a function of $H/H_m$ are rather similar, however with a larger ratio for UTe$_2$ \cite{Hardy09}. 

In URhGe, spin fluctuations are strongly enhanced around $\Hm$ along both the $b$-axis (longitudinal mode, increase of $1/T_2$ ) and the $c$-axis (transverse mode, increase of $1/T_1$). As a consequence of the strong enhancement of the magnetic fluctuations,  the reentrant superconductivity has a higher $T_{\rm sc} \approx 0.45$~K than at zero field ($\Tsc \approx 0.25$~K).

In UCoGe, the spin fluctuations develop at low temperatures as a feature of a system close to the ferromagnetic instability \cite{Ihara2010, Ohta2010,Hattori2012}. As in URhGe only ferromagnetic interactions are involved.
 In the normal state the temperature dependence of $1/T_1$ measured by $^{59}$Co-NQR is nearly constant above a characteristic temperature of 40~K, which is regarded cross-over temperature to a heavy fermion state at low temperatures \cite{Ohta2010}. Below 10~K $1/T_1$ increases strongly and shows a large peak at the ferromagnetic transition at $T_{\rm Curie} \approx 2.5$~K. From the temperature dependence of the $^{59}$Co nuclear quadrupole resonance spectrum however, it has been concluded that the ferromagnetic transition in UCoGe is already at ambient pressure of weakly first order. $1/T_1$ and Knight shift measured by NMR confirm the strong Ising type character of the magnetic fluctuations.  The observed fluctuations are in longitudinal mode ($H\parallel c$) and they are drastically suppressed by external fields applied along the same direction (i.e., parallel to the easy-magnetization axis) \cite{Hattori2012}. 
 The anisotropy of the superconducting upper critical field can be explained by the observed magnetic anisotropy of the fluctuations \cite{Hattori2012, Hattori2014}. 
 Such a strong characteristic feature has not been observed for URhGe. 

A key difference of UTe$_2$ compared to the ferromagnetic superconductors URhGe and UCoGe is the absence of long-range magnetic order in the ground state. As a consequence  the characteristic energy of spin fluctuations is expected to be much higher at low temperatures around $T_{\rm sc}=1.6$ K, and  indeed, in UTe$_2$ there is no signature of continuous softening of fluctuations in $1/T_1T$.
 Instead, $1/T_1T$ shows a nearly $T$-independent behavior below 10 K in UTe$_2$ \cite{Tokunaga2019}; the behavior expected for an ordinary heavy electron metal. Interestingly, however,  $1/T_2$ for $H \parallel a$ detects slow fluctuations in longitudinal mode along the easy-magnetization axis in the temperatures region where superconductivity appears. 
 To understand the field-enhancement of superconductivity it will be crucial to explore the spin dynamics in UTe$_2$ for a magnetic field above 16~T for $H\parallel b$. 
 
 At least UCoGe is at ambient pressure ($p=0$) at the verge of a ferromagnetic--paramagnetic instability which will occur at $p_{\rm c} \approx 1$~GPa \cite{Hassinger2008, Slooten2009, Bastien2016}. Already at  $p=0$ the ordered moment of 0.06 $\mu_{\rm B}$ is weak. The susceptibility $\chi_c$ along the $c$ axis is under high magnetic field up to 50~T higher than for the other directions \cite{Knafo2012}. In difference to URhGe, 
 for a transverse field applied along the  $b$ axis, which suppresses $T_{\rm Curie}$ near 13~T no metamagnetism has been observed and thus the enhancement of superconductivity in UCoGe near 15~T is not connected to a metamagnetic transition. Under pressure the ferromagnetic order strongly increases in URhGe and a magnetic quantum critical regime can not been attained by pressure at leat up to 14~GPa \cite{Hardy2005}. 
It is worthwhile to notice that around $p_c$ the resistivity in UCoGe does not show a Fermi liquid $T^2$ term, but a linear $T$ dependence on approaching $\Tsc$; the recovery of a $T^2$ temperature dependence  is observed only above 4~GPa far above $\pc$ \cite{Bastien2016}.


\section{Superconductivity}
\label{section_supra}

After the initial excitement triggered by the remarkable properties of \UTe under high magnetic field, strongly suggesting a spin triplet superconducting state, experimental and theoretical works have focused on three main issues:
\begin {itemize}
\item evidences for a spin triplet superconducting state, covering not only the remarkable high-field behaviour of $\Hc$ for $H \parallel b$, but also the detection of the spin state by NMR, as well as that of gap nodes and the search for ferromagnetic fluctuations,
\item topological properties, which quickly focused on the possibility of a chiral (time reversal symmetry breaking) state,
\item mechanisms for the high field-reinforced or induced superconducting phases.
\end{itemize}

\subsection{Basics of unconventional superconductivity}
\label{basicSuperconductivity}

Before discussing the experimental and theoretical investigations on these issues in UTe$_2$, let us introduce some basics of unconventional superconductivity.  
Superconducting phases are characterized by the pair potential, which may also be called gap function or superconducting order parameter. 
The pair potential appears in the off-diagonal part of the Bogoliubov-de Gennes Hamiltonian. 
For spin $1/2$ fermions \cite{note1,Khim2021}, it is described by the $2 \times 2$ matrix~\cite{Leggett1975}, \begin{align}
\hat{\Delta}({\bm k}) = \left(\psi({\bm k}) + {\bm d}({\bm k}) \cdot {\bm \sigma}\right) i \sigma_y, 
\end{align}
with the Pauli matrix, ${\bm \sigma} =(\sigma_x,\sigma_y,\sigma_z)$.
The vector component ${\bm d}({\bm k})$, called ${\bm d}$-vector, represents the spin-triplet Cooper pairs, while the scalar component $\psi({\bm k})$ represents the spin-singlet Cooper pairs. 
To fulfill the fermions anti-commutation relation, the ${\bm d}$-vector must be an odd function with respect to the momentum ${\bm k}$. Furthermore, the ${\bm d}$-vector is an axial vector as the angular momentum is. Therefore, the ${\bm d}$-vector, namely, the order parameter of spin-triplet superconductivity has odd parity under spatial inversion. By contrast, the scalar order parameter $\psi({\bm k})$ for spin-singlet superconductivity has even parity. Thus, there is a close link between the parity and spin of Cooper pairs; spin-singlet superconductivity is even-parity superconductivity, while spin-triplet is odd-parity. This correspondence ensures that the spin-triplet superconductivity can be distinguished from the spin-singlet superconductivity when space inversion symmetry is preserved. Thus, the spin-triplet superconductivity is well-defined even when the spin is not a good quantum number due to spin-orbit coupling or magnetic order.  

In superconductors, various crystalline symmetries other than space inversion may be preserved. Indeed, the superconducting phases are classified by the crystalline point group symmetry~\cite{Sigrist-Ueda1991}; according to the Landau theory for phase transitions, the order parameter at the second order phase transition must belong to an irreducible representation (IR) of the total symmetry group. In other words, transition temperatures of the superconducting order in different IRs are nonequivalent. 
For example, the IRs of $D_{4h}$ and $D_{2h}$ point groups are listed in Table~\ref{tab:classification}. Basis functions reveal typical order parameters of the corresponding IRs. 
The $D_{4h}$ group corresponds to tetragonal crystals such as CeCoIn$_5$~\cite{115_review} and CeRh$_2$As$_2$~\cite{Khim2021}, while the $D_{2h}$ group is the symmetry of the spin-triplet superconductor candidates UGe$_2$, URhGe, UCoGe~\cite{Aoki2019a}, and (the main topic of this article) UTe$_2$. 

\begin{table}[bht]
\begin{center}
\caption{\label{tab:classification} Classification of superconducting phases in the (a) $D_{4h}$ and (b) $D_{2h}$ point group symmetry. Spin-orbit coupling is taken into account. Irreducible representations (IR), basis functions, and parity for spatial inversion are listed. A superconducting order parameter of the corresponding IR is a linear combination of the basis functions as Eqs.~\eqref{d-vector_Au} and \eqref{d-vector_B3u}, with coefficients determined by microscopic properties.}
\vspace{5mm}
\scalebox{0.9}[0.9]{ 
  \begin{tabular}{ccc}\hline \hline 
 \multicolumn{3}{c}{(a) $D_{4h}$} \\ \hline 
IR &  Basis functions & Parity\\ \hline
$A_{1u}$ &  $k_a \hat{a} + k_b \hat{b}$, $k_c \hat{c}$ & -\\
$A_{2u}$ & $k_b\hat{a} - k_a\hat{b}$ & -\\
$B_{1u}$ & $k_a \hat{a} - k_b \hat{b}$ & -\\
$B_{2u}$ & $k_b \hat{a} + k_a \hat{b}$ & -\\
$E_{u}$ & $\{k_a \hat{c}, k_b \hat{c}\}$, $\{k_c \hat{a}, k_c \hat{b}\}$ & -\\
\hline
$A_{1g}$ &  $1$ & +\\
$A_{2g}$ &  $k_a k_b (k_a^2-k_b^2)$ & +\\
$B_{1g}$ &  $k_a^2-k_b^2$ & +\\
$B_{2g}$ &  $k_a k_b$ & +\\
$E_{g}$ &  $\{k_a k_c, k_b k_c\}$ & +\\
\hline \hline
\\ \\
  \end{tabular}
}
\hspace{10mm}
\scalebox{0.9}[0.9]{ 
 \begin{tabular}{ccc}\hline \hline 
 \multicolumn{3}{c}{(b) $D_{2h}$} \\ \hline 
IR &  Basis functions & Parity \\ \hline
$A_u$ &  $k_a \hat{a}$, $k_b \hat{b}$, $k_c \hat{c}$ & -\\
$B_{1u}$ & $k_b\hat{a}$, $k_a\hat{b}$, $k_a k_b k_c \hat{c}$ & -\\
$B_{2u}$ & $k_a \hat{c}$, $k_c \hat{a}$, $k_a k_b k_c \hat{b}$ & -\\
$B_{3u}$ & $k_c \hat{b}$, $k_b \hat{c}$, $k_a k_b k_c \hat{a}$ & -\\ 
\hline
$A_{g}$ & 1 & +\\ 
$B_{1g}$ & $k_x k_y$ & +\\ 
$B_{2g}$ & $k_x k_z$ & +\\ 
$B_{3g}$ & $k_y k_z$ & +\\ 
\hline \hline
\\ \\
\end{tabular}
  }
\end{center}
\end{table}

\subsection{Spontaneous symmetry breaking in unconventional superconductors}
\label{SectionSCandSymmetries}

In superconductors, the relationship between the symmetry of order parameter and the symmetry breaking is different from other phase transitions. 
This is because the hidden $U(1)$ gauge degree of freedom becomes apparent in superconductors. 
For example, odd-parity superconductivity preserves space inversion symmetry in combination with a $U(1)$ gauge operation, although the space inversion symmetry in the naive sense is broken. 
Since the $U(1)$ gauge degree of freedom is inactive in most properties, it is regarded that the space inversion symmetry is preserved in odd-parity superconductors. 
This should be contrasted to the diagonal long-range order, such as parity-violating antiferromagnets and odd-parity multipole order~\cite{Watanabe2018,Hayami2018}, where the odd-parity order parameter breaks the space inversion symmetry. 
In the same sense, all the superconducting transitions of one-dimensional IR break neither the crystalline symmetry nor the time-reversal symmetry. 
Spontaneous symmetry breaking occurs (1), when the order parameter belongs to a multi-dimensional IR, or (2), when transition temperatures of nonequivalent IRs are accidentally coinciding (degenerate IRs) and order parameters distinguished by symmetry coexist.

An example of case (1) is seen in the $D_{4h}$ group; it contains the two-dimensional $E_{g}$ and $E_{u}$ representations. In the case of the $E_{u}$ representation, the time-reversal symmetry is broken when the ${\bm d}$-vector is ${\bm d}({\bm k}) = (k_a \pm i k_b) \hat{c}$ (chiral superconductivity). On the other hand, the fourfold rotation symmetry is broken when ${\bm d}({\bm k}) = k_a \hat{c}$ (nematic superconductivity). The spontaneous symmetry breaking of this kind has been investigated for a long time~\cite{Sigrist-Ueda1991}. 
On the other hand, in the $D_{2h}$ group, all the IRs are one-dimensional. 
Thus, only the case (2) can occur, and the two-stage (double) superconducting transition or first-order one is required for the symmetry breaking. 
Positive and negative evidence for a two-stage transition in UTe$_2$ will be discussed later (section \ref{DoubleTransition}), and the reliability of the positive evidence is one of the actual central issues. 
The double superconducting transitions at least under the pressure reached a consensus, although those at ambient pressure are under debate. 
If the double transition indeed occurs, the combination of the IRs determines which symmetries are broken. 
For example, the space inversion parity is different between the odd-parity representation (spin-triplet superconductivity) and the even-parity (spin-singlet superconductivity). Therefore, the coexistent mixed-parity state does not have a definite parity, indicating a broken inversion symmetry. 
Such a possibility will be discussed below for UTe$_2$. 
Similarly, a $A_u + B_{3u}$ state breaks the mirror and rotation symmetries, while maintaining the inversion symmetry. 
The time-reversal symmetry is broken when the relative phase of the order parameters is neither $0$ nor $\pi$. 
All the odd-parity superconducting states with broken time-reversal symmetry are chiral superconducting states, allowing a non-zero orbital magnetic moment. 
The chiral axis depends on the combination: For example, the $B_{3u} + i B_{2u}$ or $B_{1u} + i A_{u}$ state has the chiral axis along the $c$-axis. The polar Kerr rotation measurement would imply such an orientation of the chiral axis \cite{Wei2021}. On the other hand, in the $B_{3u} + i A_{u}$ state supported by the London penetration depth measurements~\cite{Ishihara2021}, the chiral axis is along the $a$-axis. 
The chiral axis corresponds to the direction of spontaneous magnetic (orbital) moment. 

The above discussion also applies to the mixing of order parameters by external perturbation. For example, the space inversion symmetry may be broken by the crystalline structure, and an odd-parity superconducting order parameter can be admixed with an even-parity one, as has been studied in the literature~\cite{Bauer2012NCS,Smidman2017}. 
In the same way, the magnetic field may break the crystalline symmetry. In the case of UTe$_2$, the $D_{2h}$ group is reduced to $C_{2h}$ when the magnetic field is applied along a symmetry axis. Therefore, the superconducting state may be described by the combination of IRs, such as $A_{u} + i B_{3u}$ for ${\bm H} \parallel \hat{a}$. The relative phase should be $\pm \frac{\pi}{2}$ because the time-reversal symmetry is broken in the magnetic field. Mixed superconducting order parameters for magnetic fields along the symmetry axes are listed in Table~\ref{tab:mixing}. The field direction is equivalent to the chiral axis of the superconducting state. Intuitively speaking, the spontaneous magnetic moment along the chiral axis is induced by the external magnetic field. 

\begin{table}[htb]
\begin{center}
\caption{\label{tab:mixing} Classification of odd-parity superconducting phases in UTe$_2$ under the magnetic field. Typical order parameters belonging to each IR are listed in Table~\ref{tab:classification}.}
\vspace{5mm}
\begingroup
\renewcommand{\arraystretch}{1.2}
  \begin{tabular}{ccc}\hline \hline 
IR of $C_{2h}$ &  $A_u$ & $B_u$ \\ \hline
${\bm H} \parallel \hat{a}$ & $A_{u} + i B_{3u}$ & $B_{1u} + i B_{2u}$ \\ 
${\bm H} \parallel \hat{b}$ & $A_{u} + i B_{2u}$ & $B_{1u} + i B_{3u}$ \\ 
${\bm H} \parallel \hat{c}$ & $A_{u} + i B_{1u}$ & $B_{2u} + i B_{3u}$ \\ 
\hline \hline
  \end{tabular}
\endgroup
\end{center}
\end{table}

\subsection{Spin susceptibility of spin triplet superconductors}

As opposed to spin singlet superconductors, spin-triplet superconductors possess spin degrees of freedom, well represented by their superconducting order parameter. 
This order parameter can have different representations, the most natural one being to write it as:
\begin{equation*}
\begin{aligned}
\ket{\Psi(\mathbf{\hat{k}})} & = \Delta^{\uparrow}(\mathbf{\hat{k}}) \ket{ \uparrow\uparrow } + \Delta^{\downarrow}(\mathbf{\hat{k}}) \ket{\downarrow\downarrow} + \Delta^{0}(\mathbf{\hat{k}}) (\ket{ \uparrow\downarrow} + \ket{ \downarrow\uparrow})
\end{aligned}
\end{equation*} 
where $\Delta^{\uparrow}$,  $\Delta^{\downarrow}$, and $\Delta^{0}$ are the spin-up, spin-down and spin-zero amplitudes of the superconducting order parameter for the chosen quantization axis.  
However, the ${\bm d}$-vector representations used here is most commonly used for spin triplet superconductors, as it is "independent" from the chosen quantization axis.  
(see Ref.\cite{Leggett1975} for its definition and use in the seminal example of superfluid $^3$He,  or \cite{Brison2021} for a detailed connection between both representations). 
Indeed, for a quantization axis in a direction characterized by a unit vector $\bm{u}$, the amplitude of the $S_z=0$ component of the spin triplet state for this quantization axis is simply proportional to ${\bm d} \cdot \bm{u}$.

Hence, the spin susceptibility in the superconducting state is closely related to the direction of the ${\bm d}$-vector; the spin susceptibility decreases when the ${\bm d}$-vector is not perpendicular to the magnetic field  $\bm{B}$  ($\vert {\bm d} \cdot \bm{B} \vert$ finite).
Experimentally, the NMR Knight shift measurement reflects the electronic spin susceptibility, giving access to the ${\bm d}$-vector direction in the spin-triplet superconducting state. 
According to the group-theoretical classification in Table~\ref{tab:classification}, all the odd-parity superconducting states possess ${\bm d}$-vector components along the $a$, $b$, and $c$ axis. For instance, the ${\bm d}$-vectors in the $A_u$ and $B_{3u}$ states are written as 
\begin{align}
{\bm d}^{A_{u}}({\bm k})=&
\alpha \, k_a \, \hat{a} +
\beta \, k_b \, \hat{b} +
\gamma\, k_c \, \hat{c} \nonumber \\
&+
{\rm (higher \,\, order \,\, terms)},
\label{d-vector_Au}
\\
{\bm d}^{B_{3u}}({\bm k})=&
\alpha \, k_a k_b k_c \, \hat{a} +
\beta \, k_c \, \hat{b} +
\gamma\, k_b \, \hat{c} \nonumber \\
&+
{\rm (higher \,\, order \,\, terms)}.
\label{d-vector_B3u}
\end{align}
Thus, the spin susceptibility for all the symmetry axes should decrease, when we 
take into account not only the $p$-wave (first order in $k$) and also the $f$-wave (third order in $k$) components. This is also true for the other $B_{1u}$ and $B_{2u}$ states.


Generally speaking, the ${\bm d}$-vector component parallel to the magnetic field is suppressed as
\begin{align}
    \frac{d |\alpha|}{d H_a^2} <0, 
    \hspace{5mm}
    \frac{d |\beta|}{d H_b^2} <0, 
    \hspace{5mm}
    \frac{d |\gamma|}{d H_c^2} <0,
\end{align}
with the coefficients in eqs.~\eqref{d-vector_Au} and \eqref{d-vector_B3u}.
When the above ${\bm d}$-vector component vanishes, the decrease in the Knight shift should disappear. 
The robustness against the magnetic field is related to the mechanism of superconductivity. For instance, the $\beta$-term is robust when the superconductivity is mainly stabilized by the pairing interaction leading to the $\beta$-term, and vice versa. 





%
%

\subsection{Evidences for spin triplet superconductivity}
\label{subsection_triplet}

Experimentally, the two most direct probes of the Cooper pairs spin state are the measurement of the Knight shift by NMR discussed theoretically above, and the determination of the upper critical field, $\Hc$ revealing (or not) paramagnetic limitation. Let us start with the upper critical field, which is one of the most salient and puzzling property of \UTe.

%
%
\subsubsection{Upper critical field}
\label{subsection_Hc2}

\begin{figure}[h]
\begin{minipage}[t]{.48\textwidth}
  \centering
\includegraphics[width=0.89\columnwidth]{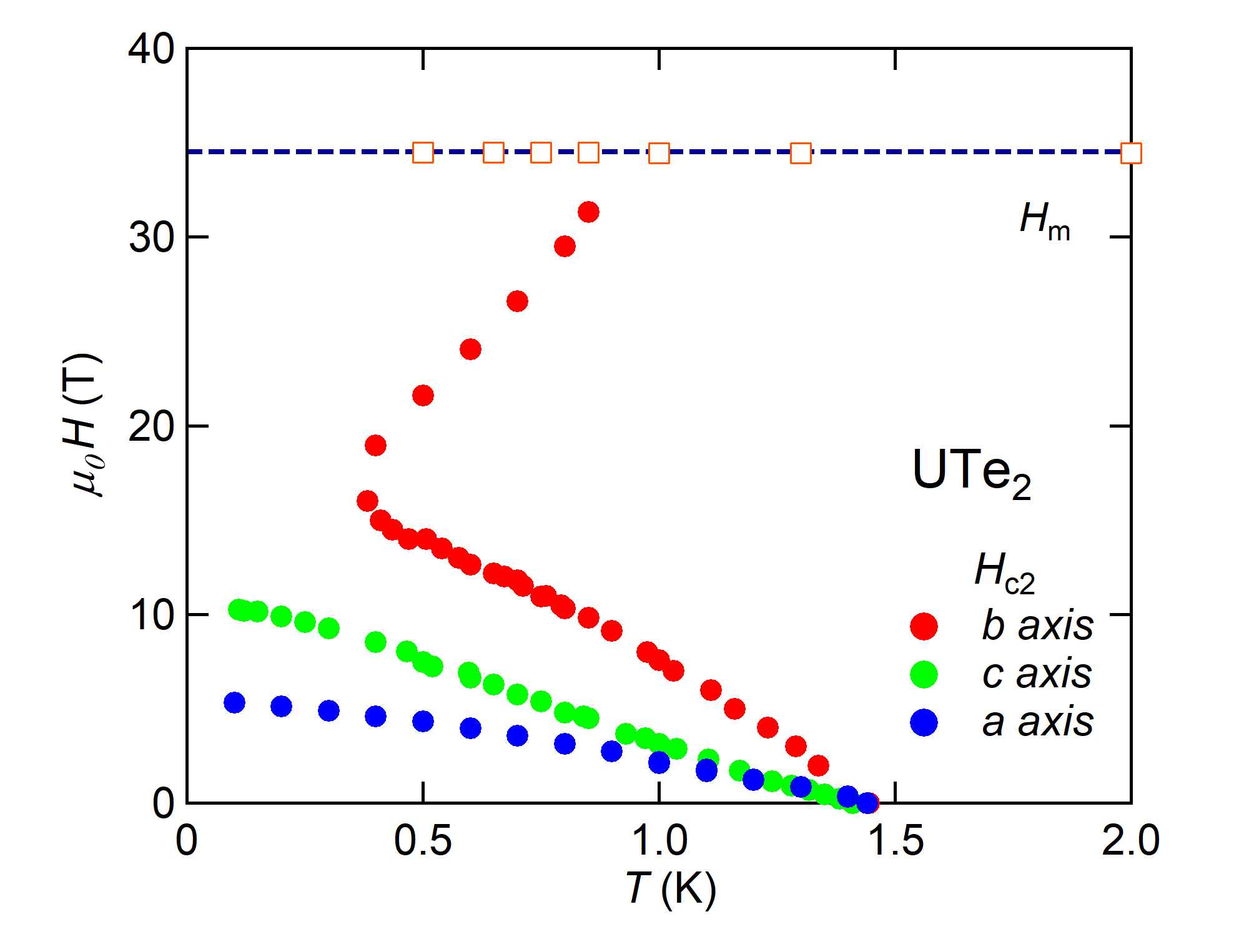}
(a)
\end{minipage}\quad
\begin{minipage}[t]{.48\textwidth}
  \centering
\includegraphics[width=0.8\columnwidth]{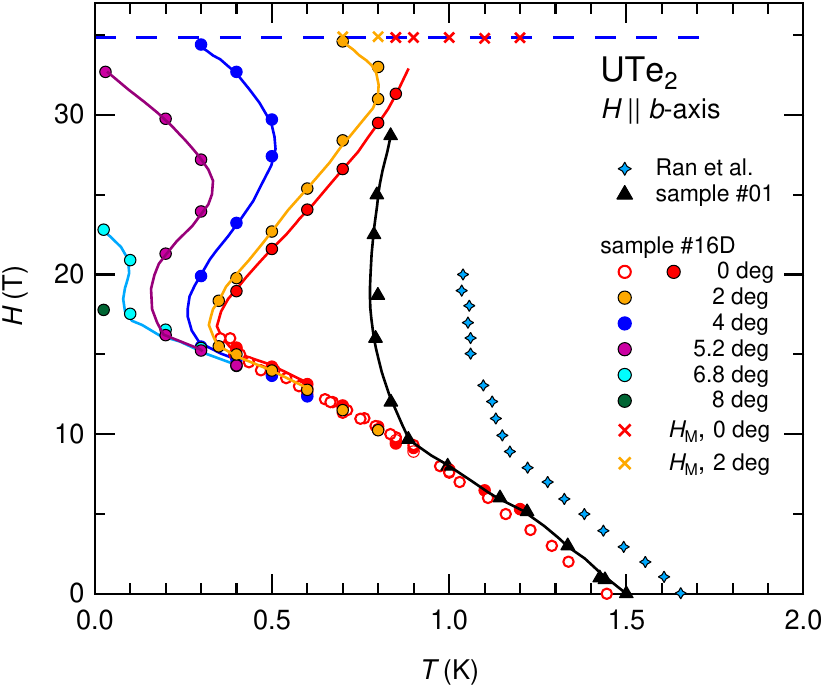}
(b)
\end{minipage}\quad
\caption{(a)-$\Hc$ measured in the three crystallographic directions by electrical transport measurement. Note the strong reinforcement of $\Hc$ for $H \parallel b$ up to the metamagnetic transition (open square symbols) at $\Hm \approx 34.5$~T. Data from \protect{\cite{Knebel2019, Knebel2020}}.
(b)-Angular dependence of $\Hc$ from the $b$ up to 8 deg turned to the $a$ axis. In addition we added the $\Hc$ from another sample, and those taken from Ref.~\protect{\cite{Ran2019}}. Figure taken from \protect{\cite{Knebel2019}}
}
\label{figHc2Georg}
\end{figure}

A striking feature of the upper critical field in \UTe [see Fig.~\ref{figHc2Georg}(a)], mentioned immediately in the first papers reporting its superconducting phase \cite{Ran2019,Aoki2019}, is that the paramagnetic limit is violated in the three principal crystallographic directions, which is taken as a strong support for spin triplet superconductivity. 
Indeed, for a free electron value of the gyromagnetic factor ($g=2$), and a the weak-coupling (BCS) limit for the superconducting state, the Pauli limit is related to the critical temperature by the relation $H_{\rm P} [T]= 1.84 \Tsc [K]$. 
Hence, this limit should be of order 3~T in \UTe, whereas experimental values of $\Hc$ are above, 6, 11 and 15~T along the $a$, $c$ and $b$-axis respectively. The violation is strongest in the hard axis $b$ direction, and even for field above 15~T a strong reinforcement of superconductivity occurs up to the metamagnetic transition at $\Hm$.

\begin{figure}[h]
\begin{center}
\includegraphics[width=0.95\columnwidth]{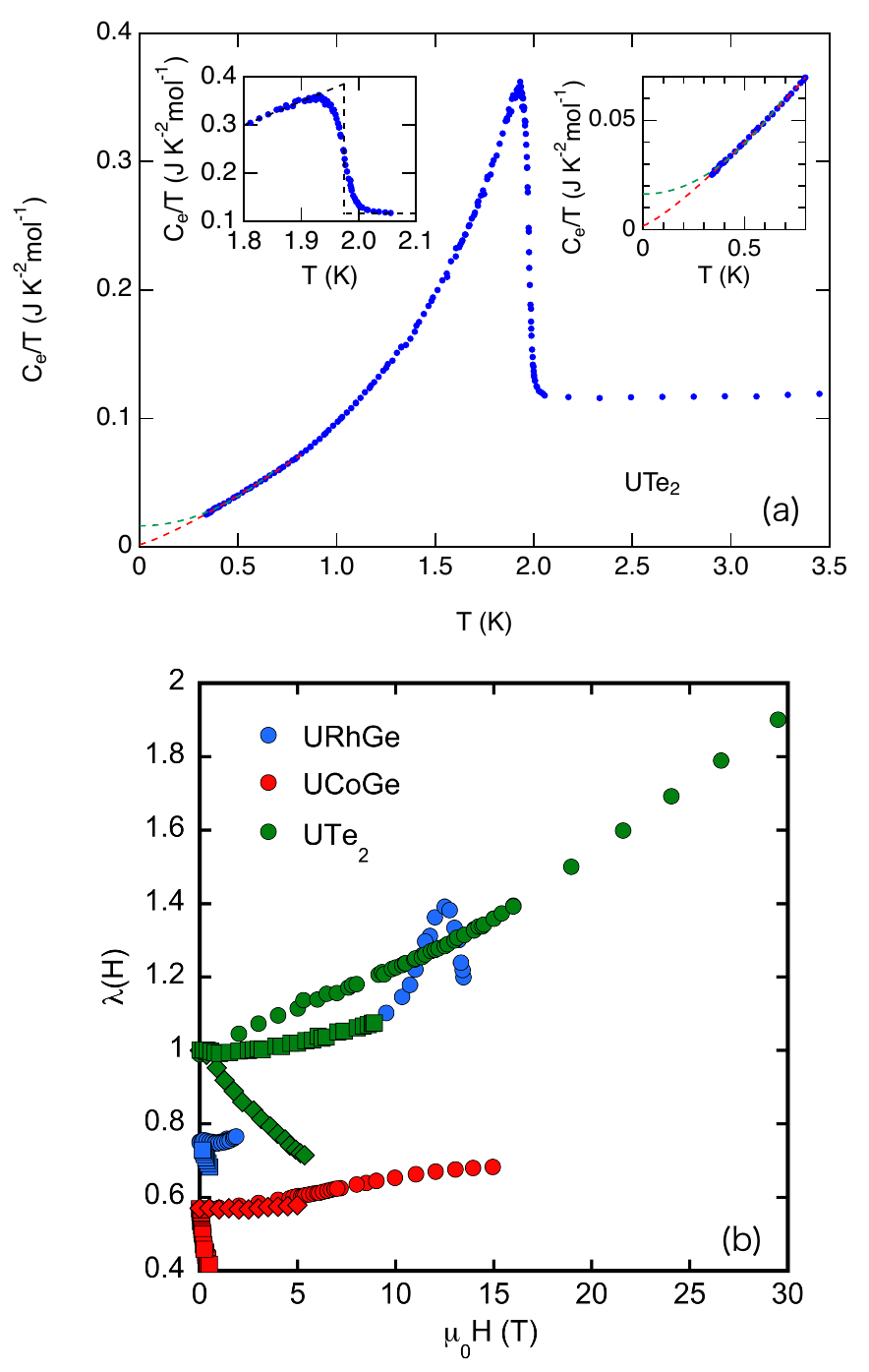}
\caption{\added[id=DA]{(a)Electronic specific heat of high quality \UTe single crystal, showing a large and sharp specific heat jump ($\Delta C_{\rm e}/\gamma T_{\rm sc}=2.29$) at $T_{\rm sc}=1.98\,{\rm K}$, classifying \UTe as a strong coupling superconductor.
The residual $\gamma$-value is about $14\,{\%}$ ($2\,{\%}$) compared to the $\gamma_{\rm N}=0.117\,{\rm J\,K^{-2}mol^{-1}}$ in the normal state, depending on the fitting function with $C_{\rm e}/T \sim T^2$ ($C_{\rm e}/T\sim T^n,\, n\sim1.2$) below $0.8\,{\rm K}$.} 
(b) Field dependence of the pairing strength in \UTe in the three directions (green), as deduced from strong coupling fits of $\Hc$. Note that $\lambda$ is almost doubled between zero field and $\Hm$. 
Also reported is our calculations for the ferromagnetic superconductors UCoGe (red) and URhGe (blue). Diamonds: $a$-axis, circles: $b$-axis, squares: $c$-axis.
In \UTe, for $H \parallel a$, the field dependence is deduced from an estimation of the slope of $\Hc$ compatible with the measurements of $H_{\rm c1}$ \cite{Paulsen2021}. 
A very similar behaviour would be obtained from the fit of $\Hc \parallel a$ as measured by specific heat (see Fig.~\ref{Hc2Kittaka}).
}
\label{figCpEtCompLambda}
\end{center}
\end{figure}

Despite this large possible influence of the paramagnetic limitation on the upper critical field of \UTe, deriving the spin state of the Cooper pairs from the measured upper critical field is not straightforward.
First of all, owing to the large specific heat jump (Fig. \ref{figCpEtCompLambda}(a)), it is clear that \UTe is a strong coupling superconductor, which pushes the paramagnetic limit to larger field values. 
For example, a strong coupling constant $\lambda =1$ already doubles the weak-coupling Pauli limit \cite{Bulaevskii1988}. Hence, the data reported in \cite{Aoki2019} for $H_{\rm c2}^a$ (we use the notation $\Hc^i$ for the upper critical field along the $i$-direction)  can also be fitted for a moderate strong coupling constant $\lambda = 0.75$ with a finite paramagnetic limit, if the $g$ factor is kept below $g <0.5$. 
This emphasises that the interpretation of the paramagnetic limit of the upper critical field requires a knowledge of the effective $g$-factor of the charge carriers, as well as of its anisotropy. 
A well-known difficulty with uranium-based superconductors is the estimation of the spin-orbit interaction. 
For spin-triplet superconductors, it acts potentially to lock (strong spin-orbit limit) the ${\bm d}$-vector direction to the crystallographic axes, and to yield anisotropic $g$-factors.
For example, in the singlet superconductor of URu$_2$Si$_2$, $g$, measured by quantum oscillations, is of order 2 for field along c and vanishes for fields in the basal plane \cite{Altarawneh2012,Bastien2019}. 
Experimentally, not much can be said yet on the value and on the effects of this spin-orbit interaction in \UTe.
Theoretically both the weak and strong spin-orbit limit are considered regarding the orientation of the ${\bm d}$-vector, and some models \cite{Machida2020, Hiranuma2021} explicitly predict a  strongly reduced paramagnetic limit in the transverse ($b$ and $c$) directions due to a very anisotropic spin-orbit interaction.
Indeed, it is quite puzzling that $H_{\rm c2}$ so strongly violates the paramagnetic limit for field along $c$ or $b$, as most models of the order parameter assume, at least in zero field, that the ${\bm d}$-vector has no component along the $a$ axis but non-zero components along both the $c$ and $b$ axes. 
To reconcile these order parameters with the apparent absence of paramagnetic limitation of $H_{\rm c2}$, a very anisotropic $g$-factor \cite{Hiranuma2021}, and/or  a rotation of the ${\bm d}$-vector under field is proposed \cite{Ishizuka2019, Shishidou2021, Kittaka2020}.
However, another factor could play a key role in \UTe: the field dependence of the pairing strength. 

This effect, and notably its reinforcement under fields perpendicular to the easy axis have been documented in ferromagnetic superconductors \cite{Wu2017, Aoki2019a}, and proposed again for \UTe due to the strong reinforcement observed for fields along the $b$-axis \cite{Ran2019,Knebel2019}.
In the first report \cite{Ran2019} a strong increase of $H_{\rm c2}^b$ is found around 1.2~K. 
Since then, it has been shown that the increase of $H_{\rm c2}^b$, is sample dependent (see Fig.~\ref{figHc2Georg}(b) and Refs.~\cite{Knebel2019, Niu2020b}).  
To the best of our knowledge, no bulk measurement has confirmed this field reinforcement of superconductivity for $H\parallel b$-axis.
In Ref.~\cite{Knebel2019}, at $T = 0.4$~K for $H > 16$~T $\Hc^b$ along the $b$ axis reaches a turning point and for a perfect alignment along the $b$ axis $\Hc^b$ increases up to $\Hm$. 
Tilting the magnetic field by 8 deg from the $b$ to the $a$ axis leads to a rapid suppression of the field reinforced superconductivity (see Fig.~\ref{figHc2Georg}(b)). 
The field- reinforcement of superconductivity for $H \parallel b$ is clearly correlated with the strong increase of $\gamma$ on approaching $\Hm$ \cite{Miyake2019, Imajo2019, Knebel2019}. 
The strong coupling parameter $\lambda (H)$ deduced from calculations of the upper critical field is shown in Fig.~\ref{figCpEtCompLambda}(b) and compared to that observed in the ferromagnetic superconductors URhGe and UCoGe. 
For a perfect alignment along the $b$ axis $\lambda$ varies from 1 at $H=0$ to 2 at $\Hm$. 

Note also that the reinforcement of $\Hc$ along the $b$-axis remains to be confirmed by bulk thermodynamic measurements; up to now, it has only been detected by electrical transport measurements \cite{Ran2019, Knebel2019, Ran2020, Knafo2021} or by the Seebeck effect \cite{Niu2020}.

\begin{figure}[h]
\begin{center}
\includegraphics[width=0.95\columnwidth]{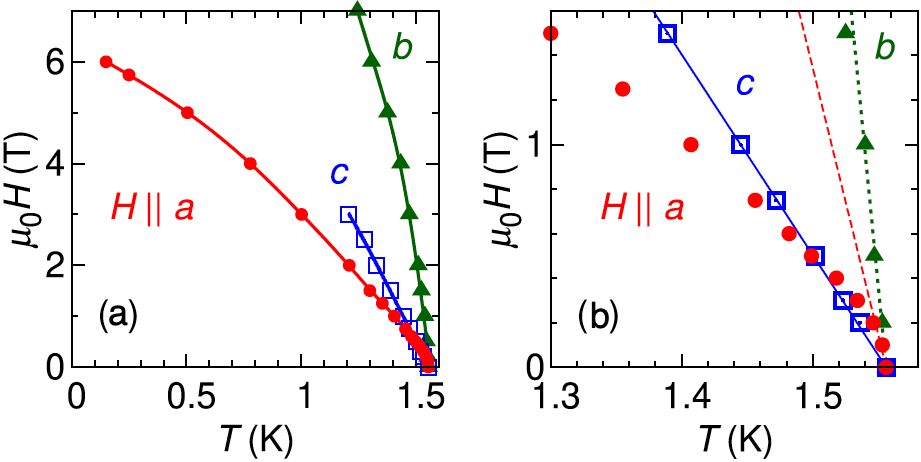}
\caption{ (a) $\Hc$ determined by specific heat measurements along the three crystallographic directions.
(b) On the zoom close to $\Tsc$, note the very strong  slope of $\Hc \parallel a$-axis, comparable to that along the $b$-axis, as well as the anomalous very strong curvature of $\Hc \parallel a$ at very low fields \cite{Kittaka2020}.
}
\label{Hc2Kittaka}
\end{center}
\end{figure}

This is an important issue, as thermodynamic measurements seem also to yield a quantitatively different behaviour of $\Hc$ from that of transport measurements at low fields. 
For example, the  determination of  $\Hc$ from specific heat measurements in \cite{Kittaka2020} suggests a slope of $\Hc $ at $\Tsc$ along the $a$-axis much larger than along the $c$-axis, and rather close to the value along the $b$-axis (see Fig.~\ref{Hc2Kittaka}). 
So, contrary to the determination of $\Hc$ from transport measurements, which are easily polluted close to $\Tsc$ by inhomogeneities and the presence of a small amount of higher-$\Tsc$ phases, bulk specific heat measurements of the initial slope of $\Hc$ are in better agreement with band calculations predicting a ``2D-like'' anisotropy, with lower Fermi velocity along the $c$ axis than along $a$ or $b$. 
This anisotropy of the Fermi velocities is also supported by the anisotropy of the resistivity (see section \ref{subsectionHpara_a}).
This remains true as long as the effects of a potential variation of the pairing strength with field are ignored.

\begin{figure}[h]
\begin{center}
\includegraphics[width=0.9\columnwidth]{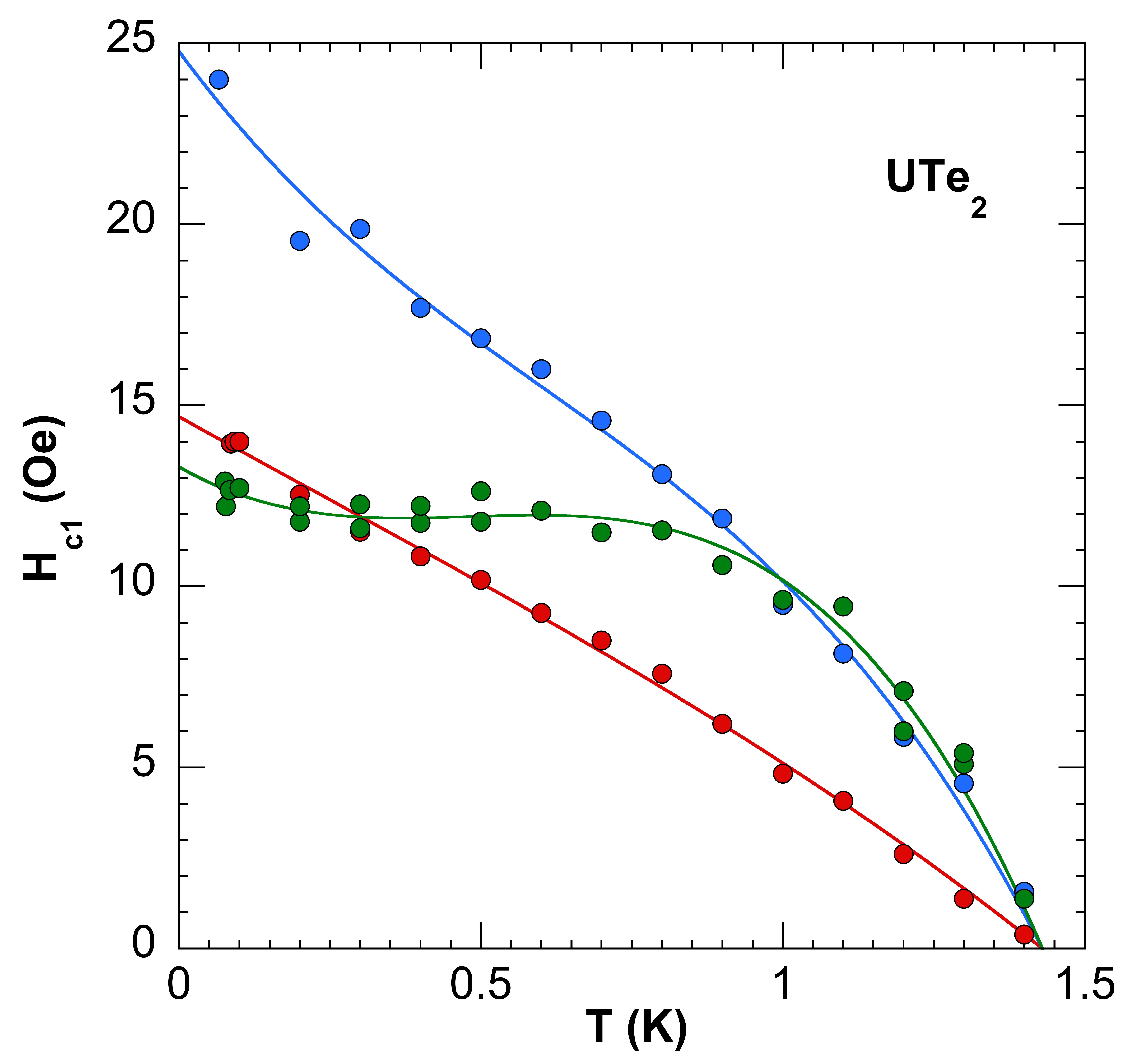}
\caption{ $H_{\rm c1}$ determined by analysis of magnetization measurements: red, $H \parallel a$, blue, $H \parallel b$, green, $H \parallel c$, lines are guides to the eye. Note that $H_{\rm c1} \parallel a$  and $H_{\rm c1} \parallel b$ have the same anisotropy than $\Hc$, whereas it should be opposite. And $H_{\rm c1} \parallel c$ is larger than in the two other directions, which can be reconciled by thermodynamic measurements of $\Hc$  (see Fig. \ref{Hc2Kittaka}) but not by transport  (see Fig.~\ref{figHc2Georg}-(a)).
}
\label{Paulsen_Hc1}
\end{center}
\end{figure}

Nevertheless, $\Hc$ for $H \parallel a$ as determined in \cite{Kittaka2020} cannot be fitted in the whole temperature range by any combination of orbital and paramagnetic limit, owing to the strong curvature close to $\Tsc$ which does not match the ``large value'' of $H_{\rm c2}(0)$.
So it is likely that the very strong curvature observed for $H \parallel a$ results from a strong suppression of the pairing strength along the easy axis, as also suggested by lower critical field measurements (see Fig.~\ref{Paulsen_Hc1} and their comparison to $\Hc$ data \cite{Paulsen2021}). 
This does not help to identify the spin state of the Cooper pairs, however, a strong suppression of the pairing strength along the easy axis is a support to ferromagnetic fluctuations as playing a key role in the pairing mechanism, and hence to a spin-triplet state with vanishing component of the ${\bm d}$-vector along the $a$-axis.

%
%
\subsection{Mechanisms for the high field-reinforced or induced superconducting phases}

\subsubsection{Field-reinforced superconductivity.}

\label{field_reinforcement}

Regarding the possible mechanism behind the field reinforcement of $\Hc$, we want first to stress that this question is possibly related, however different from that of a possible change of symmetry of the superconducting order parameter under field.
Indeed several theoretical works have proposed a field-induced transition between different irreducible representations for field applied along the $b$-axis \cite{Ishizuka2019,Shishidou2021}, notably because most proposed ${\bm d}$-vector for the ground state of \UTe have a non-zero component along the $b$-axis (and no component along the easy $a$-axis), which imply a finite paramagnetic limitation for $H$ along $b$.
As experiment shows that the paramagnetic limit is strongly violated for this axis (by more than a factor 10), it strongly suggests that the reinforced phase above 15T requires a change of symmetry of the superconducting order parameter.
For example, as already mentioned, the saturation of $\Hc$ imposed by the paramagnetic limit could be lifted by a rotation of the ${\bm d}$-vectors, such that the upper phase has no component of ${\bm d}$ along the $b$-axis (e.g. switching from $B_{3u}$ to $B_{2u}$ in high fields \cite{Shishidou2021,Ishizuka2019}).
Nevertheless, such a transition between different symmetry states alone will not explain the upturn of $\Hc$:
the enhancement of $\Tsc$ for fields above $15$~T requires in addition a field-reinforcement of the dominant pairing mechanism above that field.

 This issue did not trigger as many theoretical works as the putative topological properties of the superconducting ground state of \UTe. 
 Yet, these striking experimental behaviours call for explanations. 
 An original proposal for the field reinforcement of $\Hc$ along the $b$-axis, different from that of a field-induced increase of the paring strength, is the Lebed mechanism \cite{Mineev2020, Lebed2020} .
 This mechanism could lead to a complete suppression of the orbital limitation in quasi 1D \cite{Lebed2014} or quasi 2D \cite{Lebed2020} superconductors. 
 Hence, in quasi 1D \cite{Lebed2014} or quasi 2D triplet superconductors with no paramagnetic limitation, $\Tsc$ in high fields could be the same as in zero field.
As well explained by Lebed himself \cite{Lebed2020}, the physical origin of this suppression of the orbital limitation is that in the quasi-classical picture, electron trajectories oscillate in the directions perpendicular to the applied field, with an amplitude inversely proportional to the field. 
So in a quasi-2D superconductor with field in the conducting plane, under high field, electrons become confined in the plane, a geometry for which the orbital effect is suppressed.
This physical picture shows that this mechanism can apply if well defined conducting planes separated by ``insulating regions'' exist, like in the cuprates or in some organic materials.
This is required for the confinement of the charge carriers within these planes at high fields to make sense.
However, \UTe is a 3D system, notably as regards the flat f-bands which are dominant for superconductivity, even though some Fermi sheets are predicted to have a 2D or even 1D character \cite{Xu2019, Ishizuka2019}.
The ``failure'' of the Lebed mechanism applied to \UTe is best seen through an estimation of the characteristic field $B^{\*}$ where field induced reinforcement should appear.
The quasi 2D behaviour of \UTe, if any, would appear within the ($a,b$) planes with limited coupling along the $c$-axis.
For fields along $b$, if $d$ is the distance between  ($a,b$) planes, $\xi_0^b$ the coherence length controlling $\Hc \parallel b$, and $H^{'}_{c2}$ the slope of $\Hc$ at $\Tsc$,   $B^{*}$ should be of order:
\begin{equation}
\begin{aligned}
	B^{*} & =\frac{\hbar v_c}{v_a q d^2} = \frac{\Phi_0}{d^2}\frac{v_c}{\pi v_a} \\
		& = 2 H_{c2}(\parallel b) \left( \frac{\xi_0^b}{d}\right)^2  \frac{H^{'}_{c2}(\parallel c)}{H^{'}_{c2}(\parallel a)}
\end{aligned}
\end{equation}
Putting numbers, with the largest distance between uranium ions along c of order $0.38$~nm, $B^{*}$ is above $1000$~T.
\UTe is not 2D enough for this mechanism to help understanding the re-entrant behaviour of $\Hc \parallel b$.

Microscopically, the only work we are aware of trying to understand how the metamagnetic transition could trigger the reinforcement of $\Hc$ is that of Ref.~\cite{Miyake2021}, where it is predicted that ferromagnetic longitudinal fluctuations might appear along the $b$-axis close to $\Hm$, leading also to the observed increase of the specific heat Sommerfeld coefficient on approaching $\Hm$ \cite{Miyake2019, Imajo2019, Knafo2019, Miyake2021b}.
It gives some justifications for the estimates of the evolution of the strong coupling parameter made with crude models from the behaviour of $\Hc$ (\cite{Ran2019, Knebel2019, Knebel2020}) as discussed in section \ref{subsection_Hc2} (and see Fig.~\ref{figCpEtCompLambda}(b)). 
Nevertheless, many arbitrary choices made for these estimations, still require to be checked or invalidated by future experiments.
Notably: 
\begin{itemize}
\item $\frac{\partial \lambda}{\partial H}$ is assumed finite and positive for $H\parallel b$, in the initial paper \cite{Ran2019}, and to reconcile $\Hc$ and $H_{\rm c1}$ anisotropies \cite{Paulsen2021}. 
This seems also coherent with the evolution of the Sommerfeld $\gamma$ coefficient \cite{Miyake2019, Imajo2019}, however, precise low field measurements of  $\gamma(H)$ in all three principal crystallographic directions are still missing.
If $\gamma(H)$ in \UTe is also governed by the field dependence of the interaction responsible for the pairing, which might be less obvious than in the ferromagnetic superconductors due to the presence of competing interactions  (see section \ref{normal_state_properties}), then this measurement is a good test of the hypothesis on $\frac{\partial \lambda}{\partial H}$ (at least qualitatively).
\item when considering the relationship between $\lambda$ and $\gamma$, an important parameter is the zero field value of $\lambda$. 
Indeed, in the weak coupling limit, $\lambda$ could be field dependent and have a strong influence on the field dependence of $\Hc$, with almost no noticeable effects on $\gamma$: 
in the weak coupling limit ($\lambda \rightarrow 0$), $\Tsc$ depends exponentially on $\lambda$ through a BCS-like relation $T_{\rm sc} \propto \exp \left(-\frac{1}{\lambda - \mu^*}\right)$, whereas $\gamma \propto (1+\lambda)$.
Moreover, for paring mediated by magnetic fluctuations, the relation between $\gamma$ and  $\lambda$ is not as simple as for the phonon case notably in the strong coupling regime \cite{MineevAnnPhys2020}.
\item With $\lambda$ increasing strongly with field along the $b$-axis, it is also possible to add a paramagnetic limitation, which will be enhanced by the strong coupling regime being further reinforced under field.
For example, the reported field dependence of $\lambda$ for \UTe in Fig. \ref{figCpEtCompLambda}(b) assumes a paramagnetic limitation corresponding, for s-wave superconductivity, to a value of the gyromagnetic factor $g\approx 0.9$ for $H \parallel b$ (and $g=0$ for $H \parallel a$ and 
 $H \parallel c$). 
 The resulting smooth behaviour of $\lambda(H)$ is even qualitatively more consistent with the reported field dependence of the Sommerfeld coefficient \cite{Miyake2019, Imajo2019} than  $\lambda(H)$ deduced without paramagnetic limitation (see \cite{Knebel2019}).
 \end{itemize}
Hence, the combined strong coupling regime with increased coupling strength along the $b$-axis, as well as the spin-orbit effects strongly weakens the argument that the clear violation of the paramagnetic limit along the three axis is a smoking gun for $p$-wave superconductivity in \UTe.

%
%
\subsubsection{Superconductivity above $\Hm$}

\begin{figure}
\begin{center}
\includegraphics[width=0.9\columnwidth]{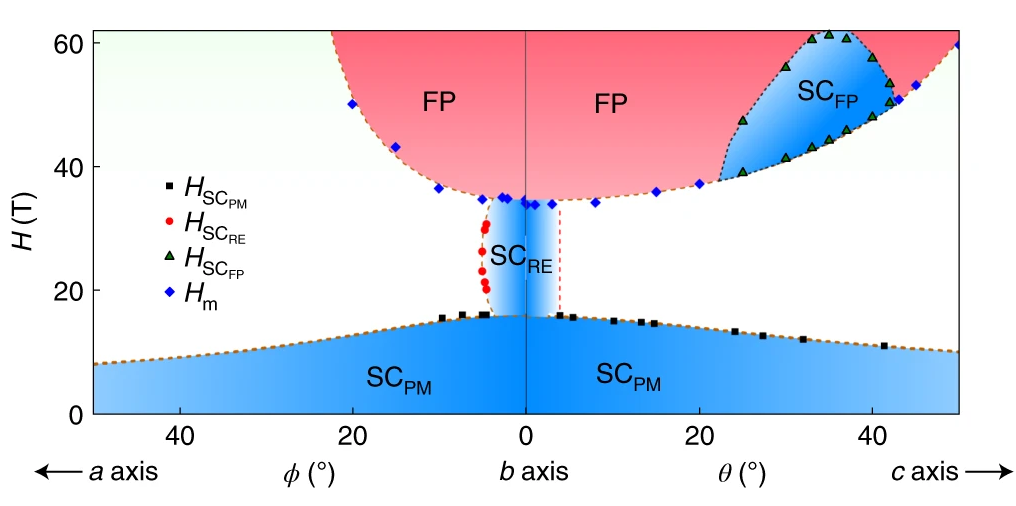}
\caption{Magnetic field - angle phase diagram of the upper critical field $\Hc$ indicating different superconducting phases. Remarkably, in the ($b-c$) plane superconductivity occurs above $\Hm$ in the angular range from 23 to 45 degrees. Figure taken from Ref.~ \cite{Ran2019a}).} 
\label{Ran_Hc2_angle}
\end{center}
\end{figure}

\begin{figure}
\begin{center}
\includegraphics[width=0.9\columnwidth]{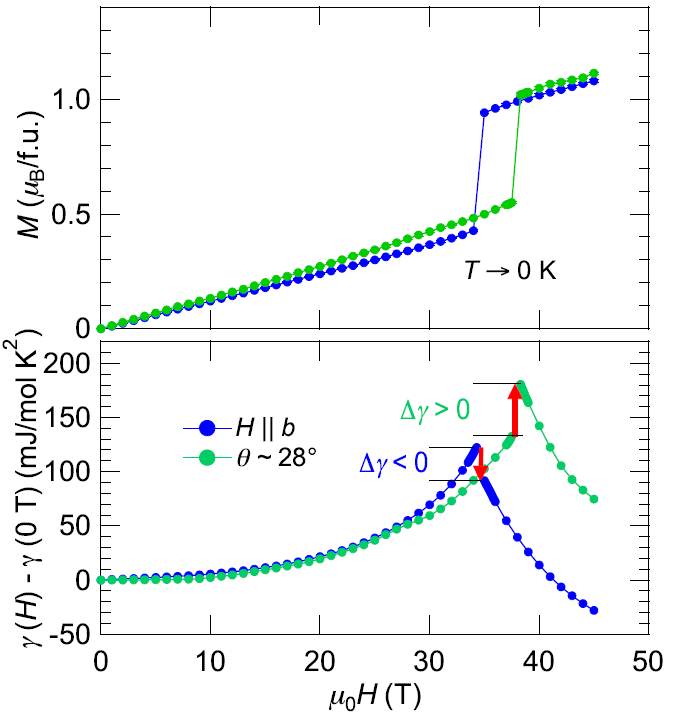}
\caption{(upper panel) Magnetization curves extrapolated to 0~K of \UTe for $H \parallel b$ and for $\theta \approx 28$~deg from the $b$ to the $c$ axis. (lower panel) Magnetic field dependence of $\Delta \gamma$ derived from the magnetization data. (Figure taken from Ref.
~\cite{Miyake2021b}}  
\label{Magnetization_30deg}
\end{center}
\end{figure}

The disappearance of superconductivity above $\Hm$ for $b$-axis seems associated with a strong increase of the residual resistivity linked to the Fermi surface reconstruction at $\Hm$ (see section \ref{section_Hparallelb}).
It may also point out a change of the pairing mechanism on crossing $\Hm$ associated to a change in the nature of the interactions.
Surprisingly, when tilting the magnetic field from the $b$-axis toward the $c$-axis, pulsed field experiment at the National High Magnetic Field Laboratory in Los Alamos  revealed a new superconducting pocket above $\Hm$ \cite{Ran2019a}, for angles $\theta$ between 25$^{\circ}$ and 45$^{\circ}$ hence around the  to $[011]$ direction, with a maximum $\Hc$ above 60~T close to 35$^{\circ}$, see Fig.~\ref{Ran_Hc2_angle}.
This result was confirmed by further experiments made in the High Magnetic Field Laboratory in Toulouse \cite{Knafo2021}. 
Magnetization  measurements, realized in the different groups show little dependence of the magnetization jump with angle: experiments in the US report $\Delta M \sim 0.3 \mu_{\rm B}$ for $H\parallel b$ or $H$ at $\theta \approx 35^{\circ}$ \cite{Ran2019a}, whereas experiments in Japan  report a larger jump $\Delta M \sim 0.6 \mu_{\rm B}$  for $H\parallel b$ \cite{Miyake2019} or at $\theta \approx 28^{\circ}$ \cite{Miyake2021b} (see Fig.~\ref{Magnetization_30deg}).

As discussed in section \ref{section_Hparallelb}, some normal state properties like the $A$ coefficient of resistivity for current along the $a$-axis, or the specific heat Sommerfeld coefficient deduced thanks to Maxwell relations from $M(T,H)$ seem, at first glance,  rather ``symmetric'' with respect to $\Hm$, and change little with angle; not only $\Delta M$ \cite{Ran2019a, Miyake2021b}, but also the $A$ coefficient of the resistivity \cite{Knafo2021}.
This would suggest that no drastic changes occur as regards the density of states either across $\Hm$ or when rotating the field from $b$ toward $[011]$, which contrasts with Hall effect or Seebeck measurement detecting a Fermi surface instability and carrier number change at $\Hm$.
More detailed direct specific heat measurement however seem to indicate the presence of a sizeable jump at $\Hm$, of order 20\% \cite{Imajo2019}, and new magnetization measurements confirm this negative jump of the $\gamma$ term above $\Hm$ for $H \parallel b$ and predict a positive jump of the $\gamma$ term of the same order for $H$ applied at $\theta \approx 28^{\circ}$ (Fig.~\ref{Magnetization_30deg}).
Clearly, more work is required to clarify the normal state behaviour at $\Hm$.

Up to now, the origin of this new superconducting phase was considered as a real puzzle, notably when contrasted to the disappearance of superconductivity above $\Hm$ for $H \parallel b$. 
The Lebed mechanism has been proposed as a possible route \cite{Ran2020}, however, the same arguments as given above for the field reinforced phase ($H \parallel b$) apply as well for H at an angle $\theta \approx 35^{\circ}$ from the $b$-axis.
Another possibility, evoked but refuted in Ref. \cite{Ran2019a}, is a Jaccarino-Peter mechanism.
This mechanism, very efficient and well documented in some Chevrel phases (see e.g. \cite{Meul1984,Konoike2004}), is a compensation of the Zeeman effect of the conduction electrons in the Cooper pairs by an internal exchange field arising from the field-induced polarization of local magnetic moments.
It could fit with the expected local nature of the uranium moments in \UTe (see section \ref{sectionCrystalStructure}), and the negative exchange field expected to have led to the heavy (Kondo) bands.
It is possible to set the requirements for such a mechanism to be effective at a finite angle but not for $H \parallel b$ \cite{Brison2021b}.
If this mechanism is at work, it would imply that $\Hc$ along the $b$-axis is also influenced by the paramagnetic effect (see discussion above) either due to a non-zero component of the ${\bm d}$-vector along $b$ or to spin-singlet pairing.

Naturally, if it is confirmed that contrary to the initial indications from transport measurements, the density of states has an opposite jump at $\Hm$ for $H \parallel b$ or $H$ at 30$^{\circ}$, it would be easier to understand this new high-field induced phase, without a Jaccarino-Peter mechanism.
A new pairing mechanism would also be expected, either dominated by antiferromagnetic fluctuations if they are responsible for the metamagnetic transition, or by ferromagnetic fluctuations along the $b$-axis if a mechanism such as proposed in \cite{Miyake2021} is at work. 
In both cases, it calls for a microscopic investigations on the magnetic properties above $\Hm$, either to explore the validity of the Jaccarino-Peter hypothesis or to understand the interplay between Fermi surface anomalies, metamagnetic transition and the pairing mechanism.

%
%
\subsubsection{Knight-shift measurements in the superconducting state} 

Knight shift measures the effective field produced by the electrons, and is proportional to the local spin susceptibility at the nucleus. 
As opposed to the upper critical field, which is sensitive to the spin state only for large-enough fields, when the paramagnetic effect can be dominant compared to the orbital effect, Knight shift measurements by NMR have larger accuracy in the low or intermediate field range, and moreover are performed at constant field, which avoids the caveats connected to the possible field-dependence of the pairing strength.

In an even parity spin-singlet superconductor the spin susceptibility is isotropically suppressed in the superconducting state. 
For the spin-triplet case, the spin part of the Knight shift decreases in the superconducting state when the magnetic field  $\mbox{\boldmath $H$}$ is parallel to the ${\bm d}$ vector ($\mbox{\boldmath $H$} \parallel {\bm d}$); however, it remains unchanged in the case of  $\mbox{\boldmath $H$} \perp {\bm d}$.
Hence, the ${\bm d}$-vector component can be derived from the measurements of the temperature dependence of the Knight shift along each crystalline axis.

Already in the first report on superconductivity in UTe$_2$, the NMR Knight shift measured on a powdered sample has been presented and no change in the NMR frequency through $T_{\rm sc}$ has been observed \cite{Ran2019}. 
However, the very tiny Knight-shift value in the normal state ($K_n \sim 0.01$ \%) is unexpected for a U-based compound near the magnetic instability.
Moreover, to identify the spin state thoroughly, NMR measurements on a single crystalline sample in field applied along each axis are required.
A very good sensitivity on high quality single crystals was obtained in the next studies \cite{Nakamine2019, Nakamine2021, Nakamine2021b}, using in addition $^{125}$Te-enriched samples (99.9\% of $^{125}$Te): the signal intensity of the enriched sample was roughly one order of magnitude larger than that of the Te-natural sample. 

Up to now, the NMR spectrum when $H \parallel a$ could not be observed below 20 K due to the quasi-divergence of the nuclear spin-spin relaxation rate $1/T_2$ \cite{Tokunaga2019}, as discussed in section \ref{NMR_normal}. 
Along the $b$ and $c$ axes, the NMR measurements support the spin-triplet pairing and reveal that the ${\bm d}$-vector has $\hat{b}$ and $\hat{c}$ components that show an anisotropic response against $H$ in the $bc$ plane.  

\begin{figure}[h]
\begin{center}
\includegraphics[width=0.9\columnwidth]{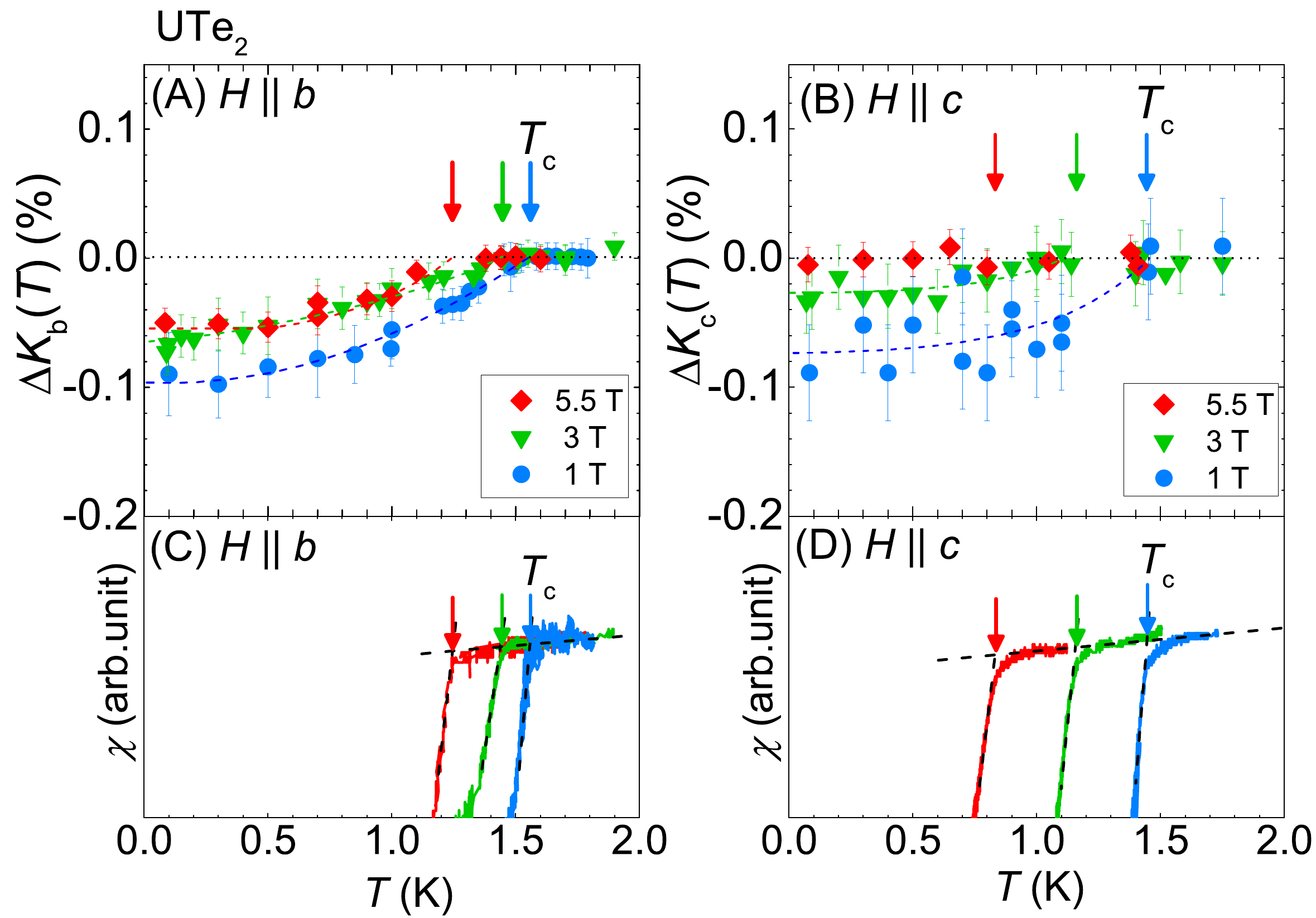}
\caption{\label{fig:fig3} Temperature dependence of $\Delta K_b$ (A) and $\Delta K_c$ (B) measured at 1, 3, and 5.5 T (see in the text)\cite{Nakamine2021}.
Dashed lines are added as a guide to the eye. 
Temperature dependence of the AC susceptibility $\chi$ at the same magnetic fields along the $b$ (C) and $c$ (D) axes. (Figure from Ref.~\cite{Nakamine2021})}
\end{center}
\end{figure}
Figures \ref{fig:fig3} (A) and (B) present the temperature variation of $\Delta K_b(T)$ (A) and $\Delta K_c(T)$ (B) at $\mu_0 H$ = 1, 3, and 5.5 T, which are compared with the temperature dependence of the AC susceptibility.
Here, $\Delta K_i (T)$ ($i$ = $b$ and $c$) is defined as $\Delta K_i (T) \equiv K_i (T) - K_{n, i}$ with the normal-state Knight shift denoted as $K_{n, i}$. 
The decrease in $\Delta K_b(T)$ observed at 1 T is $\sim$ 0.1\% \cite{Nakamine2019}, and almost the same for $\Delta K_c$.
The decrease in $\Delta K_b$ in the SC state is almost field-independent above 3 T.  
By contrast, when the field increases, $\Delta K_c$ vanishes even at the lowest temperatures, as shown in Fig.~\ref{fig:fig3} (B).

\begin{figure}[h]
\begin{center}
\includegraphics[width=0.9\columnwidth]{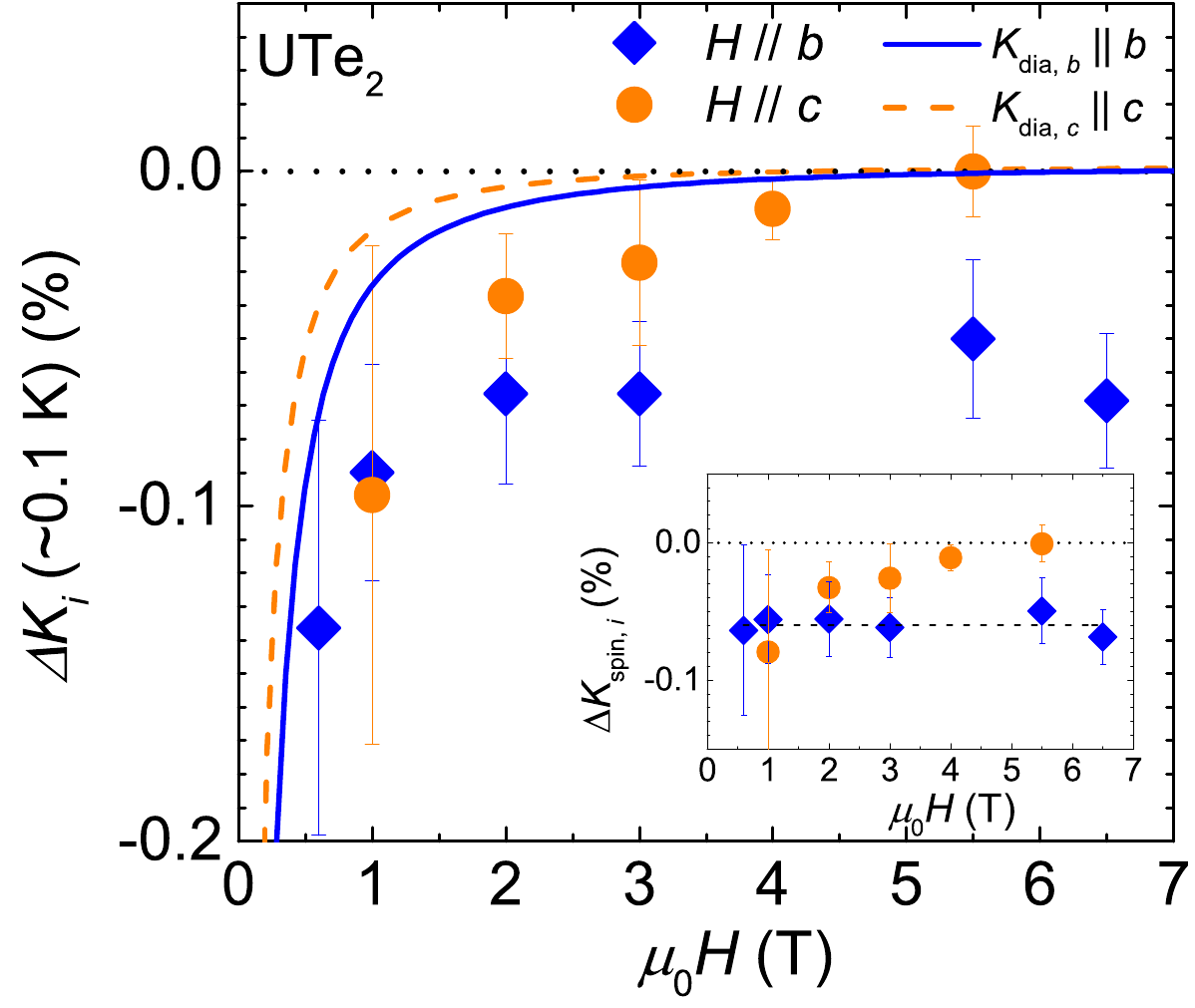}
\caption{\label{fig:fig4}
The magnetic field dependence of the decrease in $\Delta K_b$ (diamonds) and $\Delta K_c$ (circles)\cite{Nakamine2021}.
The solid and dashed lines represent the calculated SC diamagnetic shielding effects $\Delta K _{{\rm dia}, b}$ and $\Delta K _{{\rm dia}, c}$, respectively.
The inset shows the $H$ dependence of the decrease in $K_{{\rm spin}, i}$ ($\Delta K_{{\rm spin}, i}$) estimated by subtracting $\Delta K_{{\rm dia}, i}$ from $\Delta K_i$. }
\end{center}
\end{figure}
The field-dependence of $\Delta  K_i$ at the lowest temperature ($\sim 0.1$~K) is plotted in Fig.~\ref{fig:fig4}.
In general, $\Delta K_i$ is attributed to two contributions: a decrease in the spin part of $K_i$ ($\Delta K _{{\rm spin,} i}$) in the SC state and a change of the internal field due to the SC diamagnetic shielding effect $\Delta K _{{\rm dia}, i}$  \cite{deGennesSC}. 

The field dependence of $\Delta K_{{\rm dia}, i}$ ($i$ = $b$ and $c$) is estimated from the experimental results, and depicted with the solid and dotted curves in Fig.~\ref{fig:fig4}.
The details are given in Ref.~\cite{Nakamine2021}.
As highlighted by Paulsen {\it et al.}, the $H_{\rm c1}$ value along the $b$ axis is unexpectedly large and is two times the estimated value from $H_c$\cite{Paulsen2021}.
Even if such an unexpectedly large $H_{c1}$ is adopted for $H \parallel b$, the observed field dependence of the $K_b$ decrease cannot be accounted for solely by the field dependence of $\Delta K_{{\rm dia}, b}$; hence, it indicates that $K_{\rm spin}$ along the $b$ and $c$ axes decreases in the SC state, and the decrease in $K_{\rm spin}$ shows an anisotropic response against applied field, as shown in the inset of Fig.~\ref{fig:fig4}.    
This suggests that the decrease in the spin susceptibility is maintained at least up to 6.5 T when $H \parallel b$. By contrast, it is suppressed gradually when $H \parallel c$, and the spin susceptibility does not change anymore with temperature across $\Tsc$ at 5.5~T.

A crucial point to interpret the amplitude of the observed decrease of $\Delta K$ is to estimate the spin contribution of the itinerant quasiparticles to the normal-state $K$. 
In NMR studies of transition metals, the spin-part of the Knight shift has been evaluated from the $K-\chi$ plot, where it was assumed that the orbital shift is temperature-independent and $K_{\rm spin}$ is temperature-dependent\cite{Clogston1964}. 
However, it has been understood that such a simple estimation is not valid in heavy-fermion superconductors due to the strong spin-orbit coupling, particularly for the estimation of $K_{\rm spin}$ related to superconductivity.  
In such a case, $\Delta K_{\rm spin}$ can also be roughly estimated from the change of the electronic term $\gamma_{\rm el}$ in the specific heat below $\Tsc$\cite{Tou2005} 
\[
\Delta~K_{\rm \gamma} = \frac{A_{\rm hf}}{N_A \mu_B}\Delta\chi_{\rm \gamma}=\frac{A_{\rm hf}}{N_A \mu_B}\frac{\Delta\gamma_{\rm el}(\mu_{\rm eff})^2}{\pi^2k_B^2}R. 
\]
Here, $N_A$ is Avogadro's number, $\mu_B$ is Bohr magneton, $\Delta\chi_{\gamma}$ is the change of the quasi-particle spin susceptibility below $\Tsc$ evaluated from the decrease of $\gamma_{\rm el}$, $\mu_{\rm eff}$ is the effective moment, ($k_B$ is Boltzmann's constant,) and  $R$ is the Wilson ratio. 
In UTe$_2$, by adopting $\Delta \gamma_{\rm el} \sim$ 60 mJ/mol$\cdot$K$^{2}$\cite{Aoki2019a} and $A_{\rm hf}$ determined from a $K-\chi$ plot\cite{Tokunaga2019}, $\Delta K_{\gamma}$ along the $b$ and $c$ axes are estimated and are shown in Table \ref{tab:T1}, where we used $(\mu_{\rm eff})^2 = g^2\,j(j+1) \mu_B^2 = 3\,\mu_B^2$ for the free-electron value ($g$ = 2, $j$ = 1/2) and $R = 1$.
In the Table \ref{tab:T1}, the experimental reduction of $K$ in the SC state $\Delta K_{\rm spin}$ and these values in UPd$_2$Al$_3$ \cite{Tou1995, Kitagawa2018} and URu$_2$Si$_2$ \cite{Hattori2018} are also shown.
It is noted that $\Delta K_{\gamma}$ determined by this method is in good agreement with $\Delta K_{\rm spin}$ in UPd$_2$Al$_3$ \cite{Tou2005} and URu$_2$Si$_2$ \cite{Hattori2018}, which are identified as a spin-singlet superconductors.
In UTe$_2$, $K_{\rm spin}$ along both axes is approximately one order of magnitude smaller than $\Delta K_{\gamma}$, suggestive of a large residual spin component at $T$ = 0. 
The presence of such a large spin component seems incompatible with a spin-singlet state, and favors the spin-triplet superconductivity scenario. 
We point out that such a tiny decrease of the Knight shift in the superconducting state was also reported in the multi-phase superconductor UPt$_3$, where the decrease in the superconducting state measured in low fields ($\sim 0.2$ T) is only $0.5 \sim 1.4$\% of $\Delta K_{\gamma}$ \cite{Tou1998}.

However in \UTe, even for a spin-triplet scenario, the observed decrease in directions perpendicular to the easy axis is far too small, and it shows the limitations of the estimate from $\Delta \gamma_{\rm el}$ and the Wilson ratio;
taking proper account of the spin orbit effect on the spin susceptibility anisotropy and the band splitting is required.
This is well illustrated by a recent theoretical calculation based on the spin-triplet superconducting scenario\cite{Hiranuma2021}, and anisotropic spin-orbit coupling, which does reproduce such a tiny $\Delta K_{\rm spin}$. 
The main ingredient of the model is precisely the band splitting induced by a strongly anisotropic spin-orbit coupling, which lifts the degeneracy between the different values of $\vert j_z \vert$, with quantization axis along the easy $a$-axis. 
For transverse field directions, the normal state susceptibility is governed by interband terms (indexed by different values of $\vert j_z \vert$), so that the decrease of the spin susceptibility in the superconducting state, only affected by the intra-band term with $\vert j_z \vert = 1/2 $, remains very small. 
Hence in these strongly correlated systems, a precise microscopic modeling of the multi-orbital $f$-electron bands is required to properly interpret the anisotropy and magnitude of the decrease of the Knight-shift in the superconducting state.


\begin{table*}
\begin{center}
\caption{\label{tab:T1}Physical quantities on U-based superconductors. Observed reduction of the electronic term in the specific-heat measurement ($\Delta \gamma_{\rm el}$), spin susceptibility ( $\Delta \chi_{\gamma}$) estimated from the Wilson ration with $\Delta \gamma_{\rm el}$, Hyperfine coupling constant ($A_{\rm hf}$), spin-part Knight shift ($\Delta K_{\gamma}$) estimated from the $\Delta \gamma_{\rm el}$, observed Knight-shift reduction in the SC state ($\Delta K_{\rm spin}$) are shown.}
\scalebox{0.95}[0.95]{ 
\begin{tabular}{c|ccccccc}\hline \hline
 \hspace{1cm} & $\Tsc$      & $\Delta \gamma_{\rm el}$ & $\Delta \chi_{\gamma}$ & $H$   & $A_{\rm hf}$                 & $\Delta K_{\gamma}$  & $\Delta K_{\rm spin}$  \\ 
                     &   (K)       &   (mJ/mol K$^2$)            &  ($10^{-3}$ emu/mol)    &          &   (T / $\mu_{\rm B}$)    &    (\%)                      &       (\%)                      \\ \hline   
UTe$_{2}$        & 1.6         &      60                            &    0.82                 &  $\parallel b$ &   5.18                       &      0.76                   &     0.058                       \\ 
                     &              &                                    &                           &  $\parallel c$ &   3.9                        &       0.57                   &    0.08                         \\
UPd$_2$Al$_3$  &  2          &    150                         &     2.0                   &   $\perp c$     &  0.35                       &      0.13                    &    0.11                         \\                  
                      &             &                                 &                             &   $\parallel c$ &  0.35                        &       0.13                  &    0.08                          \\  
URu$_2$Si$_2$  &  1.2        &    65.5                      &   0.90                      &   $\perp c$     &  0.36                       &      0.058                  &   $\sim 0$ *                  \\   
                     &               &                               &                              &   $\parallel c$ &  0.36                        &      0.058                 &    0.05                           \\ 
UPt$_3$          & 0.55/0.5   &    420                      &    6.0                       &  $\perp c$    & -7.1                          &       -8.8                 &    -0.05                          \\ 
                     &               &                              &                               &   $\parallel c$ &   -8.4                      &       -7.3                   &   -0.10                       \\  \hline \hline  
\end{tabular}}
\\ *Absence of $\Delta K_{\rm spin}$ in URu$_2$Si$_2$ $H \perp c$ is interpreted as the anisotropy of the $g$-factor  \cite{Altarawneh2012,Bastien2019}. 
\end{center}
\end{table*}


\begin{figure}[h]
 \begin{center}
 \includegraphics[width=0.9\columnwidth]{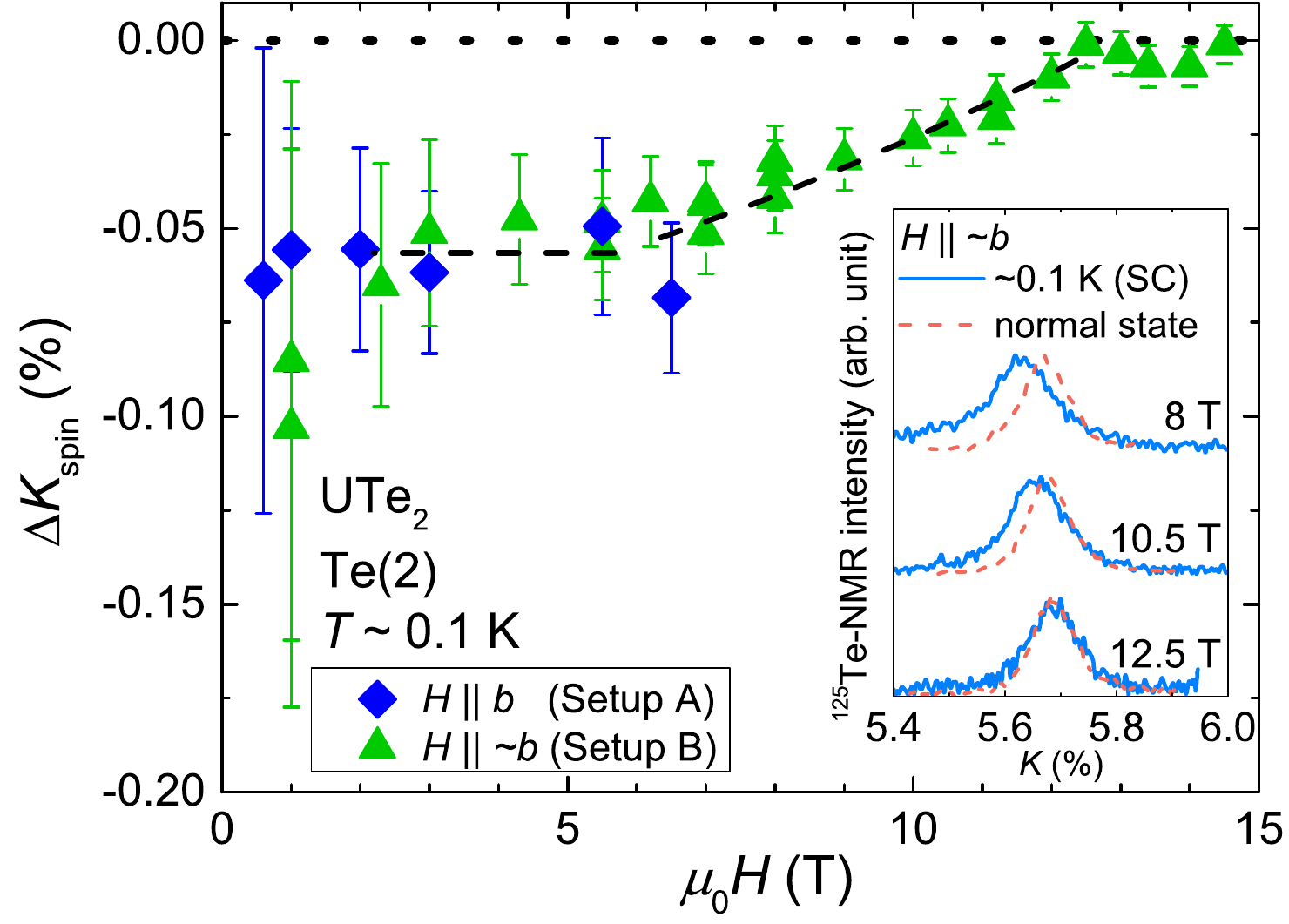}
 \caption{\label{fig6}
Magnetic field dependence of the change in the spin component of the Knight shift in the SC state, $\Delta K_{\rm spin}$\cite{Nakamine2021b}.
The dotted line indicates $\Delta K_{\rm spin}$ = 0.
The dashed lines are a guide for the eye.
(Inset) $^{125}$Te-NMR spectrum at $\sim 0.1$~K (SC state) measured at 8, 10.5, and 12.5~T.
For comparison, the normal-state NMR spectrum is also shown.}
 \end{center}
\end{figure}


Considering that the spin susceptibility decreases with a similar magnitude for $H \parallel b$ and $H \parallel c$ at low fields, a ${\bm d}$-vector containing non-zero $\hat{b}$ and  $\hat{c}$ components at low $H$ is very likely.
In addition, with the strong anisotropy along the $a$ axis in the normal-state spin susceptibility, the ${\bm d}$ vector could well be perpendicular to the $a$ axis.
Hence, considering the possible SC order parameters for $D_{2h}$ point group symmetry, presented in section \ref{basicSuperconductivity}, Table \ref{tab:classification}, the $B_{3u}$ state is a promising candidate: 
see also discussion below (section \ref{sectionGapNodes}) on the possible gap nodes.
However with the present experiments, an $A_u$ state with non-zero $k$-dependent components of ${\bm d}$ along the $a$, $b$ and $c$-axes cannot be excluded.
To distinguish the two possibilities, Knight-shift measurements along the $a$ axis, probing the $a$ component of the ${\bm d}$ vector, would provide valuable information.

Another major open issue is the behaviour of the ${\bm d}$-vector under high field for $H\parallel b$ or $c$-axes. 
It is expected that the ${\bm d}$-vector orientation changes under field, and that the observed suppression of $\Delta K$ in both directions under high field could result from this re-orientation.
How strongly the ${\bm d}$-vector is pinned in its low-field orientation, and whether the re-orientation occurs as a crossover or a phase transitions, are some of the main points to be investigated.
This should lead to additional constraints on the possible symmetries for the order parameter.

The anisotropic response of the superconducting spin susceptibility shown above suggests that the $c$ component in the ${\bm d}$-vector is continuously suppressed when $H \parallel c$ and vanishes at 5.5 T.
For $H \parallel b$, the field dependence of $\Delta K_{\rm spin}$ in Fig. \ref{fig6}, suggests that the $b$ component is constant up to 7~T, and that a ${\bm d}$-vector rotation occurs above 7 T, leading to a vanishing $b$ component of the ${\bm d}$-vector at 12.5 T.
We point out that 12.5 T is 
not so far from the critical value of $H_{\rm {c2}} \parallel b$ at which $\Tsc$ is minimum, until it increases with $H$ (see Fig.~\ref{figHc2Georg}).
Hence as already stated,  the field-reinforced superconductivity might coincide with a different SC state where the ${\bm d}$-vector can now rotate away from the $b$ axis.
A further experimental investigation by other measurements is required to probe the presence of an anomaly between these two SC states.




%
%
\subsubsection{Gap nodes}

\label{sectionGapNodes}
Unconventional superconductivity is very often characterised by the presence of gap nodes imposed by the broken crystal symmetries, and the most common types of nodes are line nodes. However, in the case of triplet superconductors, Blount's theorem \cite{Blount1985} states that the most general order parameter can only have point nodes (or no nodes), not line nodes. It has been shown that line nodes can nevertheless exist  for odd-parity superconductors in non-symmorphic crystals (those having symmetries combining point group elements with non primitive translations), like UPt$_3$, URhGe, UCoGe \cite{Micklitz2009}. \UTe, which is symmorphic, should be constrained by Blount's theorem. Hence in \UTe, identifying the types of nodes of the gap, if any, can be another route to probe spin-triplet superconductivity.

Several measurements have claimed to identify point nodes in the gap of \UTe: thermal conductivity ($\kappa$) \cite{Metz2019}, London penetration depth ($\lambda_L$) \cite{Metz2019, Bae2021, Ishihara2021}, specific heat ($C_p$) \cite{Metz2019, Kittaka2020}, NMR $1/T_1T$ relaxation rate \cite{Nakamine2019}. Table \ref{tab:Nodes} summarises the different order parameter symmetries and node types and positions proposed for the various experiments.

\begin{table*}[htb]
\begin{center}
\caption{\label{tab:Nodes} Proposals for the gap nodes and corresponding spin-triplet order parameters, deduced from various measurements.}
\vspace{5mm}
\scalebox{0.8}[0.8]{ 
  \begin{tabular}{ccccc}\hline \hline 
Experiment & type of nodes& position & order parameter  & reference \\ \hline
$\kappa$ & points & $k_a$ \& $k_b$ & ? & \cite{Metz2019}\\
$\lambda_L$ & points & in (a,b) plane & $B_{2u}$ or $B_{3u}$ & \cite{Metz2019}\\
$\lambda_L$ & points & close to (a,b) plane & chiral & \cite{Bae2021}\\
$\lambda_L$ & points &  close to $k_b$ \& $k_c$ & $B_{3u} + i A_{u}$ & \cite{Ishihara2021}\\
$C_p$ & points & ? & ? & \cite{Metz2019}\\
\makecell{$C_p$ (angular \\H-dependance)}& points & $k_a$ & \makecell{$k_b + i k_c$ or \\ $k_b \bm{c}+ k_c \bm{b}$ ($B_{3u}$)}&  \cite{Kittaka2020}\\
$1/T_1T$ & \makecell{points or \\fully gapped} &  ? & \makecell{multigap with \\$\Delta_2 \approx   \Delta_1/10$} & \cite{Nakamine2019}\\
\hline \hline
  \end{tabular}
 }
  \end{center}
\end{table*}

\begin{figure}[h]
\begin{center}
\includegraphics[width=0.9\columnwidth]{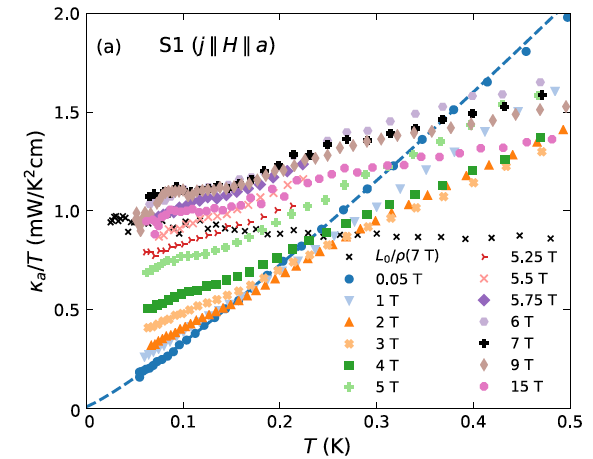}
\caption{Thermal conductivity at low temperatures in the superconducting state of \UTe.
It shows only a very small residual term at zero temperature in zero field. 
Under field, the residual term increases rapidly, which is also taken as supporting point nodes. Figure taken from Ref.~\cite{Metz2019}.
}
\label{figKappaMetz}
\end{center}
\end{figure}

As can be seen from Tab.~\ref{tab:Nodes}, if all measurements essentially converge to the presence of point nodes of gap, the position of these nodes and the proposals for the corresponding superconducting order parameters largely vary. 
It is quite remarkable that so many experiments could already probe the nodal structure of this system so soon after its discovery; this is in strong contrast to the case of URhGe and UCoGe, where the existence of line or point nodes is still today a completely open question. The much higher $\Tsc$ of \UTe certainly helps. But most remarkably, the extrapolation of the thermal conductivity down to the lowest temperatures reveals a negligible residual term (Fig. \ref{figKappaMetz}) \cite{Metz2019}. 

This contrasts with other systems like UCoGe, where the residual terms, even for samples with residual resistivity ratios (RRR) above 150, prevent the observation of the intrinsic power laws and their potential anisotropy \cite{Taupin2014}. 
The reason for such small residual terms despite a RRR barely larger than 40 in \UTe (the samples studied had RRR $\approx 22$) remains unclear, like the sharpness of the specific heat transition (except for some samples displaying double or multiple transitions, see Fig. \ref{figDoubleTransition}). 
A path for progress is to probe gap nodes along the $c$-axis; \UTe crystals have usually a much smaller dimension along the $c$-axis, so that transport measurements are difficult in this direction.

Regarding the London penetration depth $\lambda_L$, the situation is similar because the largest surfaces parallel to the field have the largest contribution:  extracting $\lambda_L$ for the $c$-axis requires high precision to subtract two large quantities \cite{Ishihara2021}.
Today, the  London penetration depth measurements presented in Ref. \cite{Ishihara2021} are the only one probing the nodes in all directions, and surprisingly, the observed power laws suggest nodes close to the three axis directions (see Fig \ref{ShibauchiLambda} in section \ref{section:chirality} below), which is not easily enforced by symmetry or topological arguments .

Beyond the question of the nodes, that of a possible multigap structure is also very acute as it can strongly influence the low temperature density of excitations, independently of the issue of spin triplet or singlet superconductivity.
Indeed, it has been realized that in most heavy-fermion superconductors, multigap superconductivity sets-in, with large gaps on the heavy $f$-bands and smaller gaps on the light-band mass (see e.g.~UNi$_2$Al$_3$ \cite{Jourdan2004}, PrOs$_4$Sb$_{12}$ \cite{Seyfarth2005}, CeCoIn$_5$ \cite{Seyfarth2008}, URu$_2$Si$_2$ \cite{Kasahara2007} and others). 
\UTe should be such a case, with its Fermi sheets having strong $f$ character while $p$ or $d$ bands from Te should yield 1D or 2D like Fermi sheets with light masses (see Fig. \ref{GGA+U-FS}). 
Presently, the only measurement which strongly supports the existence of a very small gap, preventing to distinguish between nodes or fully gapped excitations, is the NMR relaxation rate.
All other measurements claimed the absence of low energy gaps, mainly from the field dependence of $\kappa$ \cite{Metz2019} or $C_p$ \cite{Kittaka2020} at very low temperatures.
Nevertheless, due to the absence of a quantitative comparison to theoretical predictions, it is not clear if the contradiction with $1/T_1T$ is a real deep problem; further studies should certainly clarify this important point, which could otherwise change all conclusions drawn from the observed power laws.

\begin{figure}[h]
\begin{center}
\includegraphics[width=\columnwidth]{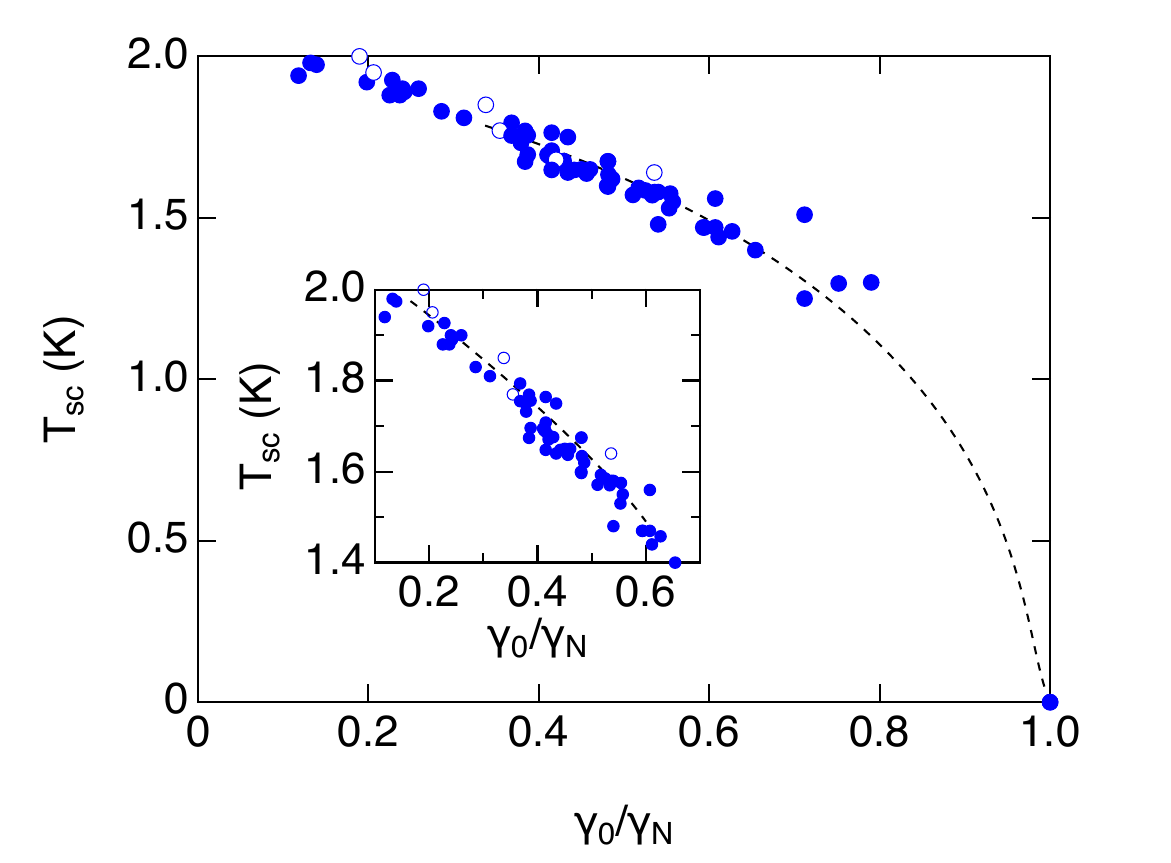}
\caption{
Correlations between transition temperature 
with sample quality as measured by the amplitude of the residual $C/T$ term, $\gamma_0$ scaled by $\gamma_{\rm N}$ in the normal state obtained from \added[id=DA]{50 different batches} in our experiments.
The inset is the zoom at low residual $C/T$.
The data of open circles are taken from Ref.\cite{Rosa2021}
}
\label{CorrelationsCp}
\end{center}
\end{figure}

Another factor which should be taken into account is the question of the strong coupling regime.
The same NMR measurement of $1/T_1T$ showed the absence of a Hebel-Slichter peak, consistent with an unconventional pairing state, and a fast decrease of the relaxation rate close to $\Tsc$, evidencing the presence of a gap at least twice larger than the BCS weak-coupling value \cite{Nakamine2019}.
This strong coupling regime is also supported by the specific heat measurements of the superconducting transition. 
Indeed, for the lowest $\Tsc$ samples, the specific heat jump at $\Tsc$ is $\frac{\Delta C}{\gamma T_{\rm sc}} \sim 1.5$ instead of 1.43 predicted for a weak-coupling BCS singlet state \cite{Ran2019}.
And this value raises up to above 1.8 for samples with $\Tsc$ at 1.77~K \cite{Cairns2020}.
This is all the more remarkable that \UTe displays a low temperature anomaly, apparently independent of the superconducting state \cite{Kittaka2020}, which may contribute to the normal state specific heat at $\Tsc$ \cite{Metz2019}, leading to an over-estimation of the Sommerfeld coefficient $\gamma$ and an under-estimation of  $\frac{\Delta C}{\gamma T_{\rm sc}}$.
Moreover, if gap nodes are indeed present, the predicted weak coupling specific-heat jump should also be lower than 1.43 \cite{Hirschfeld1988}, as can be easily understood from the entropy balance.
So the observed large values of $\frac{\Delta C}{\gamma T_{\rm sc}}$ strongly push \UTe toward the strong coupling regime.

As regards the discussion of gap nodes from the specific heat measurement, they are complicated by the presence of this low temperature anomaly. 
In an intermediate temperature range, between 0.1 and 0.5~K, $C/T$ seems to extrapolate to a residual value $\gamma_{\rm res} \approx$ 60 ~mJ$\cdot$mol$^{-1}$K$^{-2}$, which is roughly half of the value of the normal state value $\gamma_{\rm N}$.
This residual term measured on different crystals and in different laboratories seems at first site quite reproducible, almost independent on sample quality.\cite{Ran2019, Aoki2019, Aokia, Metz2019, Cairns2020, Kittaka2020}. 
However, systematic careful studies have revealed a correlation between sample quality, $\Tsc$ value and amplitude of the residual term (see Fig. \ref{CorrelationsCp}, see also Refs. \cite{Aokia, Cairns2020}). 
As already emphasized, determining the origin of $\gamma_{\rm res}$ is today a fully open problem. 
Empirically, it has been proposed to subtract a ``divergent quantum critical contribution"  to the density of states with a $T^{-1/3}$ temperature dependence \cite{Metz2019}.
This could  reconcile the specific heat temperature dependence with the presence of gap nodes, and with the observed low temperature behaviour of the thermal conductivity.
However, there is presently no theoretical explanation for such a contribution. 

Finally, let us note that with a strong coupling regime, and the possible existence of small gaps as in other heavy-fermion superconductors, the interpretation of power laws in terms of gap nodes requires probably more caution than met in the presently published works.

%
%
\subsubsection{Hunt for ferromagnetic fluctuations}
\label{subsectionMagneticFluctuations}

A last "proof" of triplet superconductivity, is the proximity to a ferromagnetic instability which should favour ferromagnetic fluctuations, and a pairing mechanism through the exchange of ferromagnetic fluctuations. 
As explained in section \ref{section_neutron}, this proximity to a ferromagnetic instability is nowadays not universally accepted, and experimentally, only incommensurate  fluctuations near an antiferromagnetic wave vector  have been observed (see section \ref{section_neutron}).

So let us make a few points clear. 
First, even if ferromagnetic fluctuations would certainly favour spin triplet pairing, spin triplet superconductivity can also arise without ferromagnetic fluctuations; this possibility has even been predicted long ago \cite{Anderson1984}, and continues to stimulate theoretical work (see \cite{Hu2021} for a quite recent example, referring also to \UTe).
Experimentally, a good example is UPt$_3$, where ferromagnetic excitations have never been observed \cite{Aeppli1988}.
Nevertheless, the most popular symmetry for the superconducting order parameter is the odd-parity ($f$-wave) $E_{2u}$ model \cite{Sauls1994}.
Furthermore, in another field, twisted bylayer graphene is strongly suspected to present a triplet superconducting phase, surviving also in high fields much above the paramagnetic limit \cite{CaoArXiv2021}, without any sign of ferromagnetic fluctuations.

As discussed in section \ref{subsection_Hc2}, the reinforcement of superconductivity for fields along the $b$-axis implying a field induced increase of the pairing strength resembles the situation of URhGe or UCoGe. 
Nevertheless, this does not prove that ferromagnetic fluctuations are at the origin of the pairing. 
In \UTe, the microscopic mechanism responsible for the field-induced boost of the pairing strength is unknown, and seems connected to the meta-magnetic transition, which bares more resemblance to that found in antiferromagnetic systems than in ferromagnets (see section \ref{section_Hparallelb}). 

Maybe a stronger case for the role of ferromagnetic fluctuations in the pairing mechanism of \UTe would be the observation of a decrease of the pairing strength along the "easy" a-axis.
Such a strong decrease has been identified in UCoGe as responsible for the very strong anisotropy of the upper critical field and the acute angular dependence of this upper critical field, but also of $1/T_1T$ \cite{Hattori2012,Wu2017}.
It was predicted theoretically for a pairing mechanism arising from ferromagnetic fluctuations, and is much less model-dependent than the behaviour for fields along the intermediate of hard axis \cite{Mineev2011}.
Such a decrease is suspected in \UTe as it could explain the anomalous anisotropy of the lower critical field $H_{\rm c1}$, opposite to expectations from that of $\Hc$ even in the Ginzburg-Landau regime close to $\Tsc$ \cite{Paulsen2021}.
It is also supported by the anomalous curvature of  $\Hc$ for $H \parallel a$-axis as measured by specific heat \cite{Kittaka2020}.
More quantitative work is required to confirm this point, which appears nevertheless as a promising way to demonstrate that the putative ferromagnetic fluctuations would play a strong role in the pairing mechanism of \UTe at ambient pressure.
Besides, several predictions were made, based on realistic theoretical models of the electronic band structure of \UTe and its associated Fermi surface, for the possible and dominant magnetic fluctuations (see section \ref{section:electronicStructure_theory}).

\begin{figure}[h]
\begin{center}
\includegraphics[width=0.9\columnwidth]{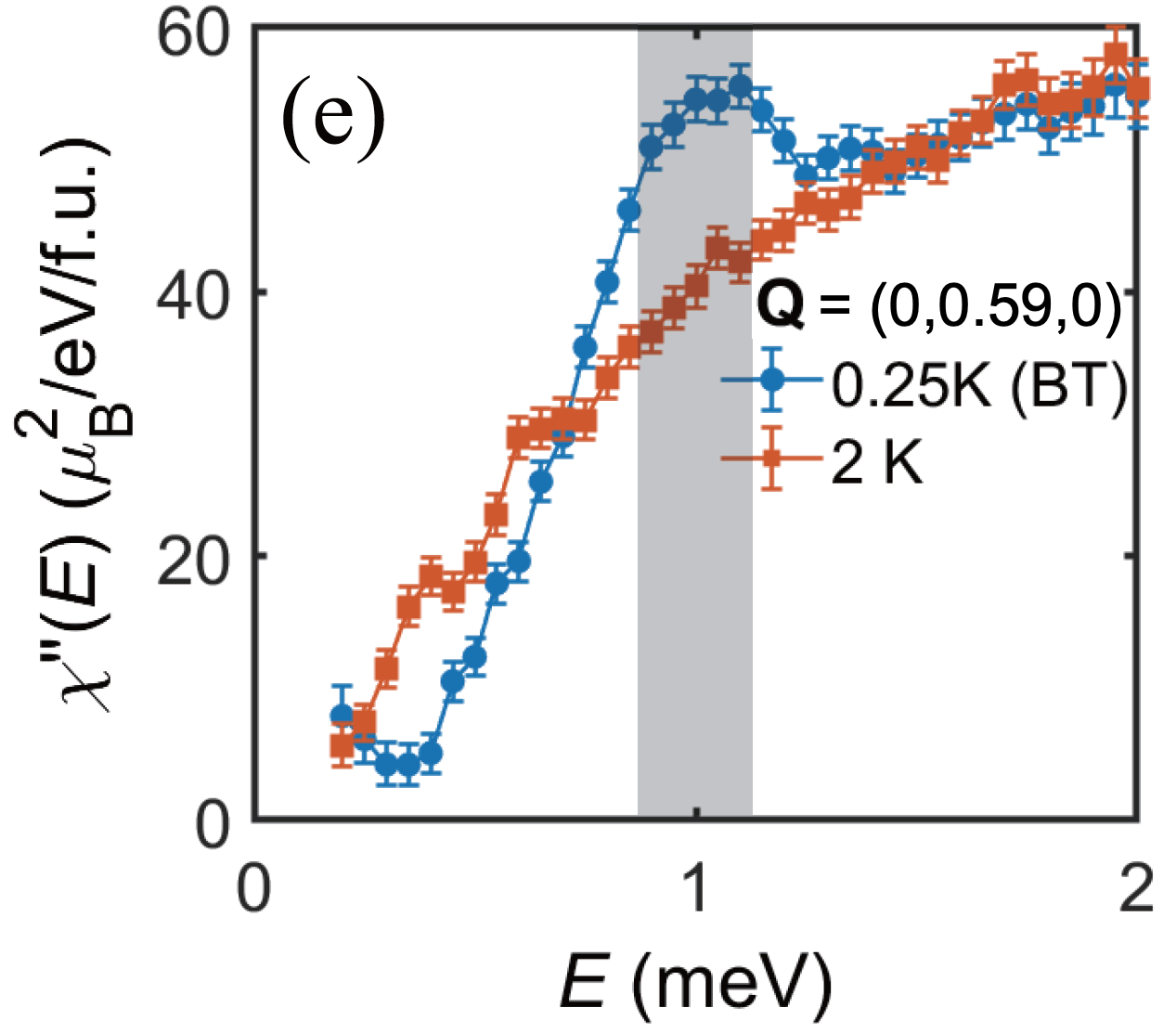}
\caption{ Comparison of energy scans below (0.25K) and above (2K) $Tsc$ in \UTe, including Bose factor corrections, at the antiferromagnetic wave-vector $\mathbf{Q}=(0,0.59,0)$. Figure from \protect{\cite{DuanArXiv2021}}.
}
\label{figResonanceNeutrons}
\end{center}
\end{figure}

Recently, inelastic neutron scattering measurements \cite{DuanArXiv2021} have revealed a resonance appearing below $\Tsc$ (see Fig. \ref{figResonanceNeutrons}) at the wave-vector where antiferromagnetic fluctuations are strongest (section \ref{section_neutron}), on 27 aligned single crystals.
This result has been confirmed even more recently by triple axis neutron experiments on a much smaller sample but with a 
10 times smaller wave-vector integration and using a unique single crystal \cite{Raymond2021}.

This is again a difference with the ferromagnetic superconductors URhGe and UCoGe, where such a resonance has not been observed. Quantitatively, it is centered at a large energy $E_r \approx 7.9 k_B \Tsc$, leading to a ratio $E_r/2 \Delta \approx 2$, taking for the value of the gap $\Delta$ the estimation from the scanning tunneling microscopy (STM) measurements ($\Delta \approx 0.25$~meV \cite{Jiao2020}). 
This is far above the value $E_r/2 \Delta \approx 0.64$ reported in the unconventional spin singlet high-$T_{\rm c}$ cuprates, or in the iron pnictide superconductors \cite{Yu2009}. 
In the other heavy fermions where such a resonance has been observed \cite{Raymond2021}, $E_r/2 \Delta$ remains close or lower than $1$, except in UBe$_{13}$ with $E_r/2 \Delta \approx 1.9$. 
Note that UBe$_{13}$ is the only one, with \UTe, strongly suspected to be $p$-wave \cite{Shimizu2019}.  
Below $E_r$, a spin-gap opens in \UTe like in other spin singlet superconductors. There is yet no definite interpretation of the significance of this resonance. Following the analogy with the cuprates or iron pnictide superconductors, it might support antiferromagnetic fluctuations as the dominant pairing mechanism \cite{DuanArXiv2021}. 
However, the differences with these well-studied unconventional singlet superconductors, and notably the location of this (broad) resonance in the overdamped region above $2 \Delta$ might indicate a different origin;
for example, the spectral weight thus redistributed in the superconducting phase could arise from itinerant modes, connected to some Fermi surface nesting features not directly involved in the pairing, or even from more local modes \cite{Raymond2021}.

%
%
\subsection{Topological superconductivity}

Potential topological properties of the superconducting excitations in \UTe where put forward already in the very first publication \UTe \cite{Ran2019}, initially purely on general grounds due to the spin-triplet pairing. However, this question became rapidly more focused after a first STM study claiming to have detected a signature of chiral surface excitations \cite{Jiao2020}. This was followed by Kerr-effect measurements \cite{Hayes2021, Wei2021}, as well as London penetration depth measurements \cite{Bae2021,Ishihara2021} all claiming the detection of anomalous effects connected to a form of chiral state.

\subsubsection{Basics of Topological superconductivity}

Realization of topological superconductivity (TSC) and Majorana fermions is one of the current central issues in modern condensed matter physics. 
Stimulated by the proposal to use Majorana fermions as qubits for quantum computation~\cite{Kitaev_2001}, about two decades of intensive study have revealed the conditions for the emergence of TSC~\cite{Qi-Zhang2011,Sato-Fujimoto2016,Sato_2017}. 
Even apart from application to quantum computation, TSC in itself is an intriguing topological phase of matter, namely, a new type of unconventional superconductivity. 
Classical definition of unconventional superconductivity relies on the symmetry; non-$s$-wave superconductors are unconventional. 
Topology provides another classification; topology of the wave function can be trivial or nontrivial. 
From this respect, several platforms of TSC have been identified (see the review \cite{Sato_2017}). 

\begin{figure}[htbp]
\begin{center}
\includegraphics[width=0.8\linewidth]{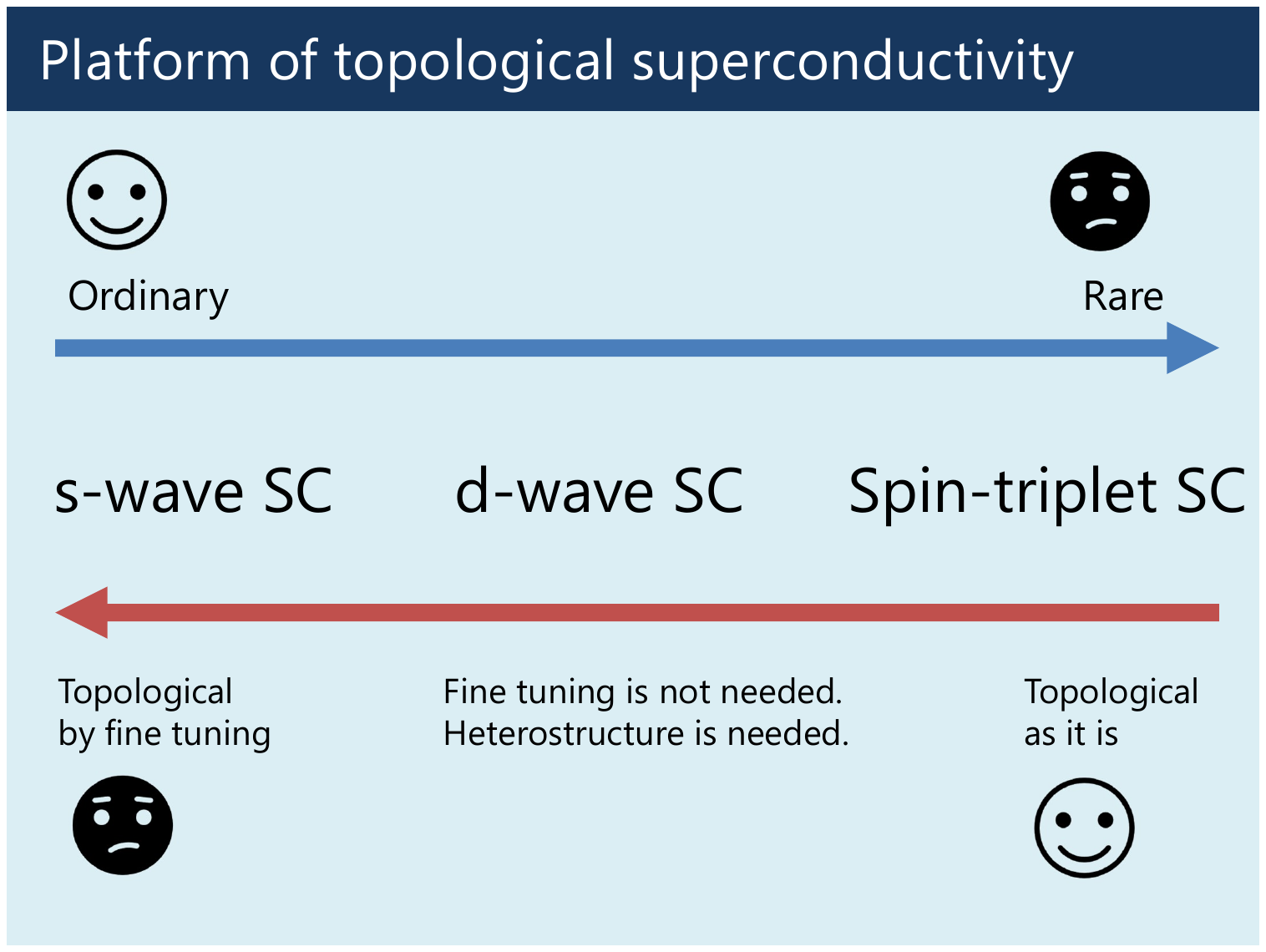}
\caption{Platform of topological superconductivity.
}
\label{fig:TSC}
\end{center}
\end{figure}

Although there is no one-to-one correspondence between symmetry and topology, they are closely linked to each other. 
Figure~\ref{fig:TSC} is a rough sketch of the platform of TSC based on the symmetry of superconductivity. 
We have a dilemma for the materials design. The $s$-wave superconductors are ubiquitous in nature. But, they are mostly trivial even from the viewpoint of topology. 
Fine tuning of parameters may enable TSC~\cite{Sato2009,Sau2010,Lutchyn2010,Oreg2010,Alicea2010,Qi2010}, and indeed tremendous efforts have been devoted for the realization of topological $s$-wave superconductivity in artificial nanowires and heterostructures. 
However, it is considered that ``evidence'' for nanowires needs further scientific verification~\cite{Zhang2020}. 

On the other hand, spin-triplet superconductors are strong candidates for topological superconductivity, as they are topologically nontrivial in most cases~\cite{Reed-Green2000,Ivanov2001,Fu-Berg2010,Sato2010}. 
But, there is a disadvantage; spin-triplet superconductors are rare in nature. 
Heavy fermion superconductors UPt$_3$, UGe$_2$, URhGe, UCoGe, and UTe$_2$ are strong candidates, as discussed in this article. 

The $d$-wave superconductors fall into an intermediate category. They are less rare than the spin-triplet superconductors, and fine tuning of parameters is not necessary for realizing TSC~\cite{Daido2016,Takasan2017,Can2021}. 
However, heterostructure~\cite{Daido2016,Can2021} or intense laser light~\cite{Takasan2017} is required in the present proposals.

\subsubsection{Possible topological superconducting states in \UTe}

Topological phases protected by crystalline symmetry, named topological crystalline superconductors or higher-order topological superconductors, are not considered here. 
Indeed, existing proposals for UTe$_2$ are for the TSC in the canonical sense, which may be protected by the local symmetries, namely, the time-reversal symmetry, particle-hole symmetry, and chiral symmetry. 
Such topological phases are classified based on the topological periodic table, so-called ten-fold way. 
A relevant part of the table for superconductors is shown in Table~\ref{tab:periodic_table}.

\begin{table}[htb]
\begin{center}
\caption{\label{tab:periodic_table} A part of the topological periodic table relevant for superconductors. Topological classification is determined by the Altland-Zirnbauer symmetry class (AS class) and the spatial dimension $d$. Time-reversal symmetric superconductors belong to the DIII class, while time-reversal symmetry breaking superconductors belong to the D class.} 
\vspace{5mm}
\begingroup
\renewcommand{\arraystretch}{1.2}
  \begin{tabular}{c|ccc}\hline \hline 
AZ class &  $d=1$ & $d=2$ & $d=3$ \\ \hline
D & $\mathbb{Z}_2$ & $\mathbb{Z}$ & 0 \\
DIII & $\mathbb{Z}_2$ & $\mathbb{Z}_2$ & $\mathbb{Z}$ \\
\hline \hline
  \end{tabular}
\endgroup
\end{center}
\end{table}

The topological periodic table tells us that three-dimensional (3D) superconductors are characterized by an integer topological invariant ($\mathbb{Z}$), which is known as the 3D winding number $\omega$, when the time-reversal symmetry is preserved (DIII class). 
In other words, 3D superconductors can be topological superconductors specified by the winding number.
On the other hand, all the 3D superconductors are topologically trivial when the time-reversal symmetry is broken (``$0$'' in the D class). 
However, even when the 3D invariant is trivial or absent, superconducting states can be distinguished from completely trivial states by the low-dimensional topological invariants.
For instance, the $k_z=0$ plane of the Brillouin zone can be regarded as an effective two-dimensional (2D) system, which can be specified by the 2D topological invariant, namely, the $\mathbb{Z}_2$ invariant $\nu_i$ ($i=1,2,3$) in the DIII class or the Chern number in the D class. 
If the 2D invariant is nontrivial, Majorana fermion may appear on surfaces preserving the quantum number $k_z$. 

Up to now, the possibility of TSC in UTe$_2$ has been discussed mainly in relation with the issue of time-reversal symmetry breaking. 
If the time-reversal symmetry is broken, the 3D odd-parity superconductors are likely to be Weyl superconductors~\cite{Yanase2016}, which are gapless superconductors with Weyl nodes protected by topology. 
The topological protection is ensured by the Chern number. 
Conversely, we have a chance to see TSC in the strong sense, when the odd-parity superconductors preserve the time-reversal symmetry. 
The 2D and 3D topological invariants in the DIII class have been investigated based on the band structure calculations for UTe$_2$~\cite{Ishizuka2019}. 
The results in Table~\ref{tab:TSC12} predict the TSC when the Fermi surfaces are as Fig.~\ref{GGA+U-FS}(b) or Fig.~\ref{GGA+U-FS}(c). 
All the odd-parity superconducting states, namely, the $A_u$, $B_{1u}$, $B_{2u}$, and $B_{3u}$ states, are specified by at least one $\mathbb{Z}_2$ invariant. 
The $A_u$ state is also a 3D topological superconductor with an odd winding number, $\omega \in 2\mathbb{Z}+1$.

\begin{table*}[tbp]
 \centering
 \caption{Gap structures, nontrivial topological indices, and surfaces hosting stable Majorana states for the Fermi surfaces in Figs.~\ref{GGA+U-FS}(b) and (c).   
}
 \label{tab:TSC12}
 \begin{tabular*}{1.0\textwidth}{@{\extracolsep{\fill}}lccc}
   \hline\hline
   FSs (i) or (ii) & & & \\
   IR & Gap structure & Topological index & Surfaces  \\\hline
   $A_u$ & Full gap & $(\omega,\nu_1,\nu_2,\nu_3)$ & (100), (010), (001) \\
   $B_{1u}$ & Point node ($\Lambda$) & $\nu_3$ & (100), (010) \\
   $B_{2u}$ & Point node ($\Delta$) & $\nu_2$ & (100), (001) \\
   $B_{3u}$ & Point node ($\Sigma, F)$ & $\nu_1$ & (010), (001) \\
   \hline\hline
 \end{tabular*}
\end{table*}

\begin{table*}[tbp]
 \centering
 \caption{Gap structures, nontrivial topological indices, and surfaces hosting stable Majorana states for the Fermi surfaces in Fig.~\ref{GGA+U-FS}(d).   
}
 \label{tab:TSC3}
 \begin{tabular*}{1.0\textwidth}{@{\extracolsep{\fill}}lccc}
   \hline\hline
   FSs (iii) & & & \\
   IR & Gap structure & Topological index & Surfaces  \\\hline
   $A_u$ & Full gap & $\omega \in 2\mathbb{Z} $ & Unpredicted \\
   $B_{1u}$ & Full gap & Trivial & None \\
   $B_{2u}$ & Point node ($\Delta, U$) & Trivial & None \\
   $B_{3u}$ & Point node ($\Sigma, F$) & Trivial & None \\
   \hline\hline
 \end{tabular*}
\end{table*}

If the Fermi surfaces are as Fig.~\ref{GGA+U-FS}(d), all the 2D $\mathbb{Z}_2$ invariants are trivial, as shown in Table~\ref{tab:TSC3}.
Thus, only the $A_u$ state can be a topological superconductor because the winding number must be zero for mirror-even superconducting states~\cite{Yoshida-Daido-Yanase2019}, namely, the $B_{1u}$, $B_{2u}$, and $B_{3u}$ states. 
We can also state that the winding number has to be an even integer, $\omega \in 2\mathbb{Z}$, in the $A_u$ state.
Estimation of $\omega$ requires detailed information of the pair potential and Fermi surfaces in this case.

In the orthorhombic systems with $D_{2h}$ point group, double superconducting transitions are needed to break the time-reversal symmetry (see section \ref{SectionSCandSymmetries}).
Thus, the time-reversal symmetric superconducting phase must exist at least near the transition temperature, even if the time-reversal symmetry is broken at zero temperature. 
In this phase, TSC can be predicted based on the topology of Fermi surfaces, as we demonstrated in Tables~\ref{tab:TSC12} and \ref{tab:TSC3}. 
Experimental verification of Fermi surfaces in UTe$_2$ is awaited also for development in the field of TSC.

\subsubsection{Double superconducting transition.}
\label{DoubleTransition}

\begin{figure}[h]
\begin{minipage}[t]{.48\textwidth}
  \centering
\includegraphics[width=0.89\columnwidth]{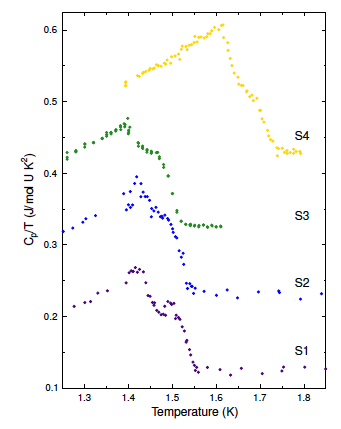}
(a)
\end{minipage}\quad
\begin{minipage}[t]{.48\textwidth}
  \centering
\includegraphics[width=0.89\columnwidth]{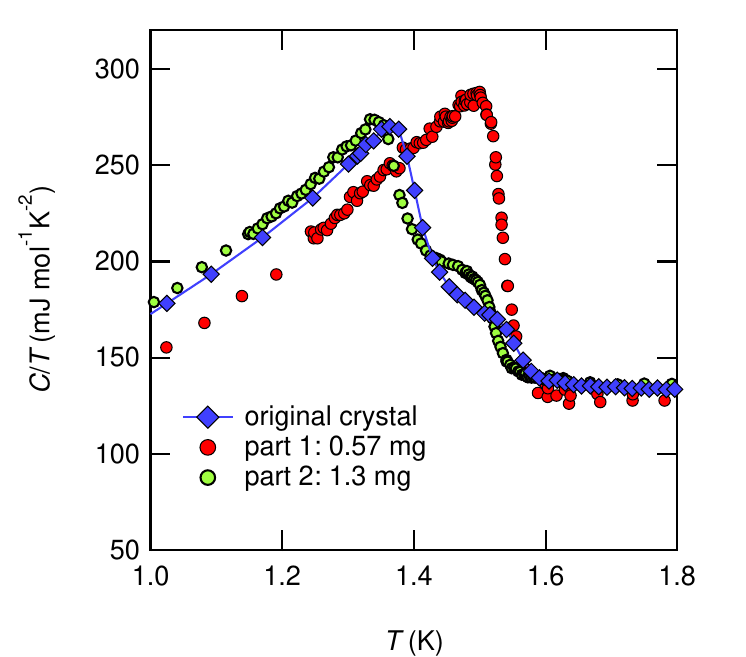}
(b)
\end{minipage}
\caption{(a) Double transition as seen on the specific heat of several \UTe samples (taken from Ref.~\cite{Hayes2021}).
(b) 
Sample with a double transition from Grenoble. After cutting in two parts, only the higher one, quite sharp, appears in one of the two pieces. See also Ref.~\cite{Thomas2021a}.
}
\label{figDoubleTransition}
\end{figure}

A main experimental debate on the superconducting state of \UTe is whether or not the double transition observed at ambient pressure (see Fig \ref{figDoubleTransition}-(a)) is an intrinsic feature or a simple problem of sample inhomogeneities.
The importance of this issue arises from the requirement of such an intrinsic double transition in order to break time-reversal symmetry in a system having only 1D IR: see section above and \ref{SectionSCandSymmetries}, as well as the discussion on \cite{Hayes2021}.
Such a double transition implies that the two (mixed) irreducible representations of the low temperature phase are nearly degenerate, something considered as rather unlikely \cite{Hayes2021}.

Of course, another possibility would be that the superconducting transition is first order, as this removes the necessity for the two irreducible representations of being completely degenerate.
The mechanism leading to a first order superconducting transition might imply for example a simultaneous ferromagnetic order (with a jump of magnetisation), or a volume change.
However, as already mentioned, no sign of a ferromagnetic order as been observed below $\Tsc$ \cite{Paulsen2021} and thermal expansion measurements only detected a jump of the thermal dilatation, not a discontinuity of the volume \cite{Thomas2021}.
Moreover, no hysteresis at $\Tsc$ has been detected.
Weakly first order transitions can be difficult to detect, but the main experimental efforts are directed these days to confirm or disclaim the intrinsic nature of the double transition rather than explore the possibility of a (hidden) first order transition.

To test whether the double transition is intrinsic or not requires either to be able to show that the transition emerge from different parts of the crystal, or at the opposite to show that the two transitions behave differently under an external tuning parameter like magnetic field or pressure. 
At first sight, the mere fact that  sharp single transitions are observed on the vast majority of \UTe single crystals favours an extrinsic origin; 
for some crystals $\frac{\Delta \Tsc}{\Tsc}$ can be as small as $0.01$ ($1 \%$), far sharper than usually observed in even very high quality crystals of the reference heavy-fermion superconductor CeCoIn$_5$ with $\frac{\Delta \Tsc}{\Tsc} \approx 0.03$.
Another unfavourable point for an intrinsic origin is that, when the double transition is observed, the peak heights and temperature splitting are not reproducible, at the opposite of UPt$_3$ for example;
in UPt$_3$, either broad transitions, or reproducible double transitions (with improved sample quality) were observed \cite{Fisher1989}.
By contrast in \UTe, a single crystal grown in Grenoble and showing two distinct superconducting transitions at $T_{\rm sc1}$=1.4~K and $T_{\rm sc2}$=1.6~K was clearly inhomogeneous;
after being cut in two pieces, one piece shows still a double transition, while the second piece had only one sharp transition at $T_{\rm sc2}$ (see Fig. \ref{figDoubleTransition}(b)).
The situation of the double transition in \UTe seems closer to the case of PrOs$_4$Sb$_{12}$ \cite{Seyfarth2006} or URu$_2$Si$_2$ \cite{Ramirez1991}, where single transitions could be isolated in parts of crystals showing initially double transitions.

Conversely, in the first paper claiming an intrinsic double transition \cite{Hayes2021}, it was mentioned that the two transitions behaved differently under field  $H\parallel c$-axis.
However, the large broadening of the transitions in this field direction, which does not correspond to an extrema of the angular dependence of $\Hc$, makes it difficult to be fully convinced that the difference is real. Remarkably, the difference between the lower and higher transition temperature is field independent for all directions in the measured field and temperature range \cite{Hayes2021}.  

Another attempt was made by studying the evolution under pressure of both transitions \cite{Thomas2020}.
However, the weakness of the specific heat jump at the upper transition above a few kbars renders the identification of the second transition quite uncertain toward zero pressure, and the proposed evolution of the two transitions will not convince a sceptic.
The most recent work done by the same group \cite{Thomas2021} explores the possibility of determining the pressure evolution of both transitions from the thermal dilatation measurement at zero pressure, using Ehrenfest relations:
\begin{equation*}
\frac{dT_{\rm sc}}{dp} = \frac{\Delta \alpha V}{\Delta C_{\rm p}/T}
\end{equation*}
Experimentally, a difficulty is to measure precisely the volume thermal expansion jump $\Delta \alpha$ at $\Tsc$ (requiring a measurement in the three crystallographic directions, with possible opposite signs, and sample dimensions smaller along the $c$-axis) and the specific heat jump, on samples where a double transition is usually associated with a rather strong broadening of the transitions.
\replaced[id=JP]{If it was initially claimed in \cite{Thomas2021} that an opposite sign was detected for the pressure evolution of both transitions, a second version of the work corrected this view, evidencing that the thermal dilation set-up was sensitive local spatial variation of the transition temperature inside the crystal. Hence, the last version of this paper now supports an extrinsic origin for the double transition, arising purely from sample inhomogeneities. It also announces an erratum to the specific heat work of Ref.\cite{Thomas2020}, and present new data showing a parallel pressure dependence of both transitions in three samples, from 0~GPa up to 1.5~GPa.}{It has been claimed in \cite{Thomas2021} that an opposite sign was detected for the pressure evolution of both transitions, which would be reliable, even if the absolute values of the slopes is not so reliable (owing to the experimental difficulties just mentioned above).
However, like for the direct pressure of the field measurements, a serious improvement on the sample quality is required to confirm this claim, and have convincing support for the intrinsic nature of this double transition.
In the course of writing this review, a new version of this work has been put on arXiv \cite{Thomas2021a}, with more systematic measurements on samples from different batches, and is now claiming that the double transition at ambient pressure in their samples is arising purely from sample inhomogeneities.}
\added[id=JP]{By contrast, this work does support again an intrinsic origin for the double transition appearing under pressure above 0.3~GPa}.

\replaced[id=JP]{Certainly, this issue underlines that there is still a requirement of sample improvement. }{Essentially, the experimental difficulties are the broad transitions on samples with double transitions, and/or the weak amplitude of one of the two transitions. 
And certainly, there is roomfor sample improvement.} 
As explained in \cite{Thomas2021, Cairns2020}, the differences in the samples is expected to arise from a difference in sample stoichiometry, which is indeed  most difficult to control reproducibly in chemical vapor transport synthesis of systems which are not line compounds (UPt$_3$ has a congruent melting).
This is certainly a most urgent problem to be solved, together with the availability of high quality crystals with large enough dimensions along the $c$-axis.

%
%
\subsubsection{Chiral superconductivity}
\label{section:chirality}

\begin{figure}[h]
\begin{center}
\includegraphics[width=0.9\columnwidth]{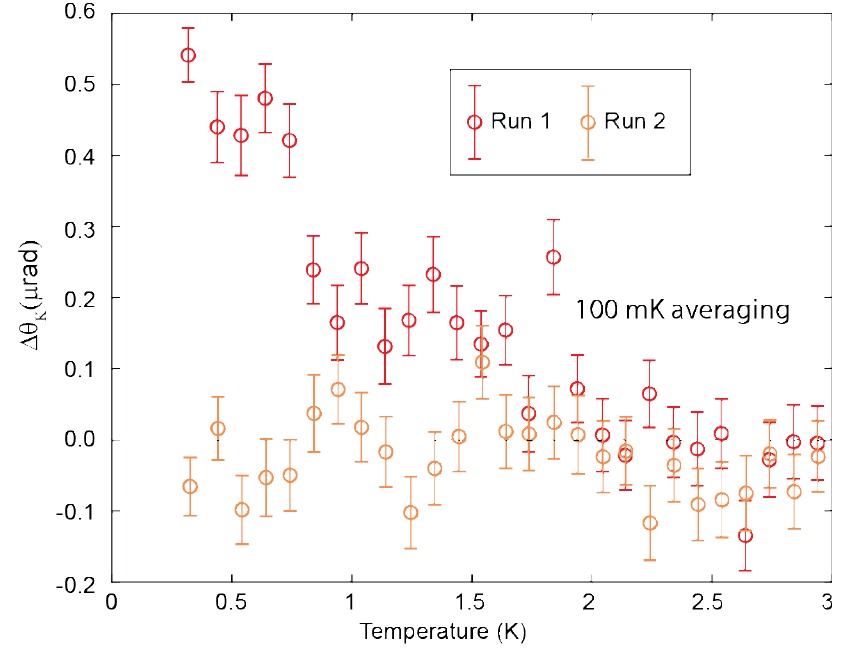}
\caption{ Measurement of the Kerr angle in zero field, when cooling below the superconducting transtition (at $\Tsc=1.6K$). On the same \UTe sample, some runs (like run 1) show a growth of the Kerr angle below $\Tsc$, while other do not (like run2). It has been argued that this difference is due to the formation of domains in the zero field experiment. Figure taken from Ref.~ \cite{Hayes2021}.
}
\label{KerrHayes}
\end{center}
\end{figure}

Early attempts to detect a tiny magnetic signature coming from a potential time reversal symmetry breaking in the superconducting state by muon spectroscopy failed, due to the very strong relaxation induced by local low energy magnetic fluctuations \cite{Sundar2019}.
So the polar Kerr effect is up to now the only measurement pointing directly to a time reversal symmetry breaking of the superconducting state (see Fig.~\ref{KerrHayes}) \cite{Hayes2021}. 
Indeed, when cooling the sample in zero field through $\Tsc$, a sizeable polar Kerr angle of maximum amplitude 500~nrad is detected in some runs. 
This Kerr effect can be ``trained'' by cooling in a very small field ($\pm$ 2.5~mT, $H \parallel c$-axis), which is considered as resulting from a field-control of the formation of chiral domains.
Moreover, the field training along $c$-axis imposes a symmetry of type $B_{3u} \pm  i B_{2u}$ or $B_{1u} \pm  i A_{u}$ for odd-parity order parameters \cite{Hayes2021}.

\begin{figure}[h]
\begin{center}
\includegraphics[width=0.9\columnwidth]{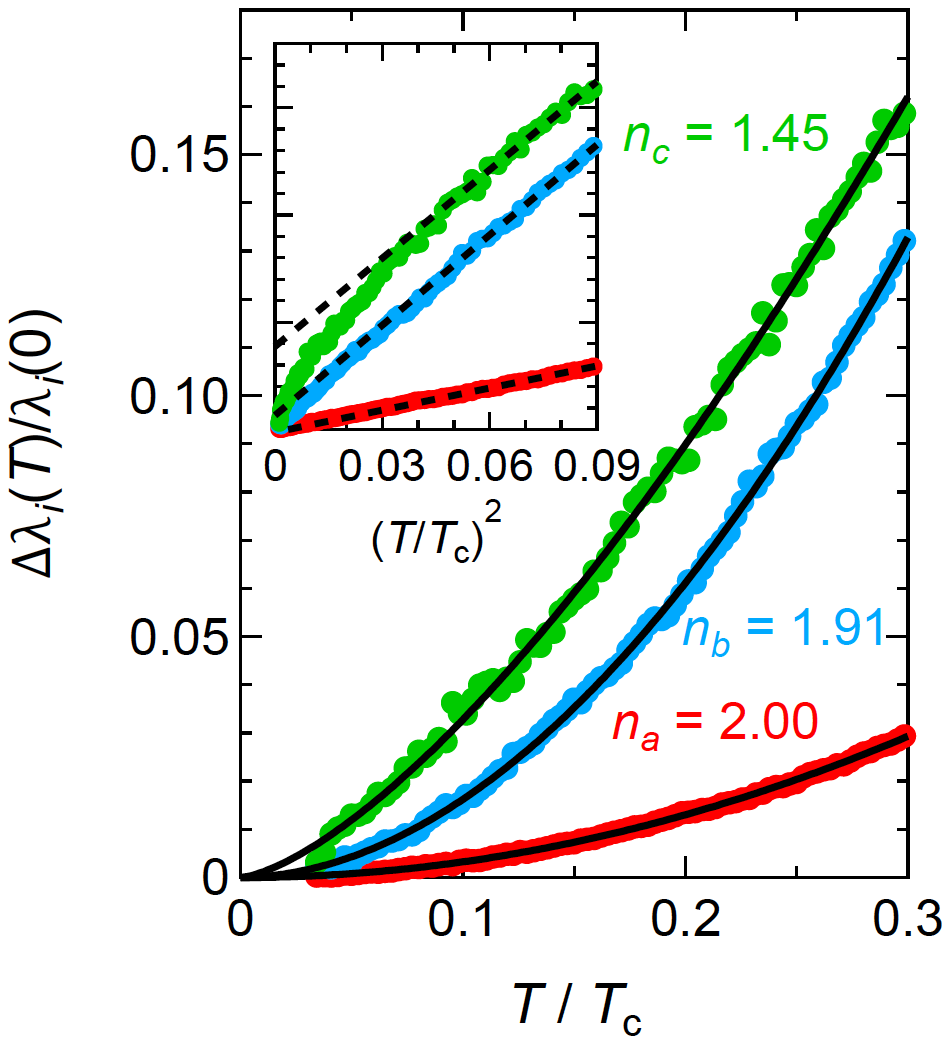}
\caption{Measurement of the temperature dependence of the change of the London penetration depth in \UTe, along the $a$ (red), $b$ (blue) and $c$ (green) axes. The anomalous power law along the $c$ axis, smaller than 2, is analyzed as arising form pairs of Weyl nodes of the gap, supporting a chiral superconducting state in \UTe. From \protect{\cite{Ishihara2021}}.
}
\label{ShibauchiLambda}
\end{center}
\end{figure}

For such a mixture of different irreducible representations, no symmetry-imposed nodes of the gap can survive. 
However, all the chiral states proposed for \UTe happen to be also non unitary states, with an order parameter of the form $\bm{d_1}$ + i $\bm{d_2}$, where $\bm{d_1}$ and $\bm{d_2}$ are two (real) ${\bm d}$-vectors belonging to two different IR.
Hence, the resulting non unitary states have two branches in their excitation spectrum, with a gap proportional to:
\begin{equation}
\Delta_{\rm \pm} \propto \sqrt{\vert \bm{d_1} \vert^2 +  \vert \bm{d_2} \vert^2 \pm 2 \vert \bm{d_1} \land \bm{d_2} \vert}
\end{equation}
Then, if for some wave-vector of the Fermi surface the conditions $\vert \bm{d_1} \vert = \vert \bm{d_2} \vert$ and $ \bm{d_1}\perp \bm{d_2}$ are met, there is a gap node in the branch $\Delta_{\rm -}$ at this point \cite{Hayes2021}, called a Weyl node due to the chirality of the excitations.
For both the  $B_{3u} \pm  i B_{2u}$ or $B_{1u} \pm  i A_{u}$, four such nodes (at least) should exist in the mirror planes $k_x=0$ and $k_y=0$, provided the Fermi surface exist at these locations \cite{Hayes2021}. 

A quite complete picture of the possible nodes and their location according to the relative weight of the 2 irreducible representations is given in the supplemental part of \cite{Ishihara2021}.
Regarding comparison with experiments, the constraint on the mixed irreducible representations from the Kerr effect arises from symmetry considerations on the training effect. 
It remains to be confirmed that the signal becomes non-zero only below the second transition, which is not so clear from the raw data, and if the change in chiral domains is the correct explanation for the absence or presence of a non-zero Kerr angle depending on the runs.

The proposals from the London penetration depth measurements come from anomalous responses. 
All measurements \cite{Metz2019,Bae2021,Ishihara2021} find a $T^2$ dependence for the low temperature variation of $\lambda_L$ when the field is along the $c$-axis, probing ``$a-b$'' plane electrodynamics.
However, in \cite{Bae2021} an anomalous response is found on the dissipative part, attributed to a surface ``normal fluid'' of chiral excitations. Quantitatively, this phenomenological explanation is however puzzling as it assigns to these excitations a coherence length with a Fermi velocity of very light bands, around 100 times larger than those deduced from the upper critical field measurements \cite{Paulsen2021}.

The work in \cite{Ishihara2021} is the first which has measured the response in all three directions and endeavours to extract the temperature dependence of the superfluid density in the three crystallographic directions.
It requires careful analysis of the sample geometries and the possible values for the London penetration depth at zero temperature, and high precision in the measurements and corrections to extract the values along the c-axis.
If a $T^2$ dependence of $\Delta \lambda_L$ is found indeed at low temperature for  $\lambda_L$ along the $a$ and $b$ axis, a weaker power law ($\approx T^{1.4}$) is found along the c-axis (see Fig. \ref{ShibauchiLambda}).
This exponent can be reproduced by calculations with a $B_{3u}+i A_u$ order parameter, having possibly Weyl nodes in the neighbourhood of the $c$ and $b$ axes.
From the results presented in \cite{Ishihara2021}-Supplement, it remains however difficult to reconcile the power laws observed in the other directions with those obtained from the model calculation.
The very anisotropic Fermi surface of \UTe might be an explanation for this discrepancy, but clearly, more work is required to settle the existence of these Weyl nodes.

\begin{figure}[h]
\begin{center}
\includegraphics[width=0.9\columnwidth]{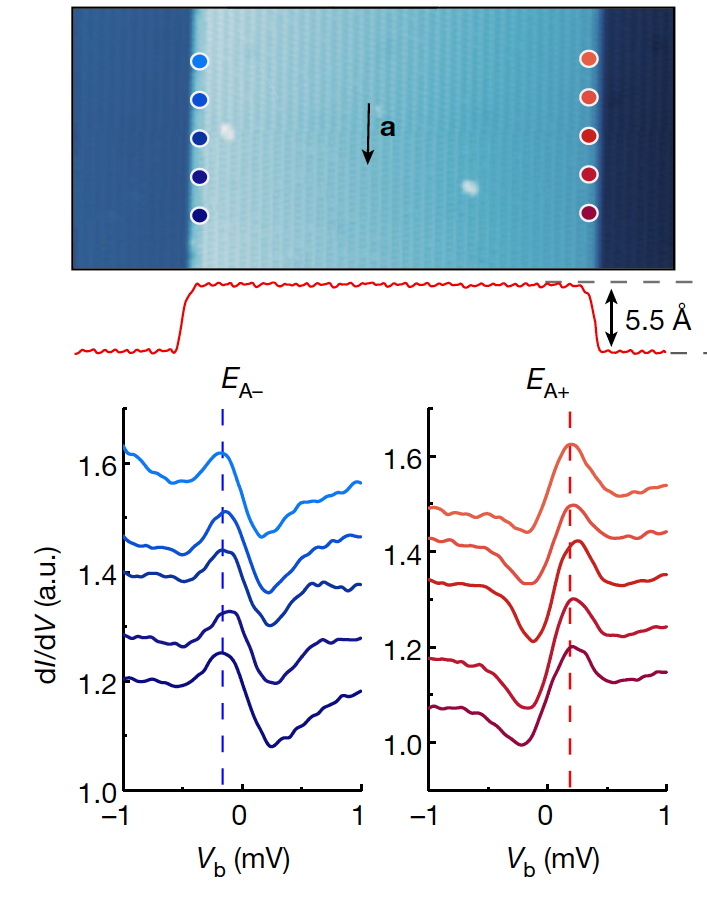}
\caption{Asymmetric spectra (bottom) measured along step edges, with colors (light to dark blue or red) corresponding to that of the dots on the top figure, showing the position where they were recorded. Measurements performed at 0.3K deep in the superconducting state. The asymmetry of the spectra is interpreted as arising from tunneling processes at non-zero in plane momentum near the step edges, probing the chiral dispersion of in-gap surface states. From \protect{\cite{Jiao2020}}.
}
\label{JiaoSTM}
\end{center}
\end{figure}

Last but not least, the STM studies \cite{Jiao2020} which were first to propose a chiral order parameter in \UTe, also detect two anomalous behaviours in their spectrum. 
First, they find a very large residual density of states at zero energy in the superconducting state (around 60 $\%$ of the normal state value \cite{Jiao2020}), which could arise from the zero energy modes associated to Weyl nodes.
Second, they find a reproducible asymmetry in the spectroscopy of the superconducting excitations near step edges on the surface (see Fig. \ref{JiaoSTM}), which they attribute in a phenomenological way to some chirality of the excitations. 
According to the surface orientation where these excitations are observed, the chiral axis ($ \bm{d_1} \land \bm{d_2}$) is required to lie along the $a$-axis, 
Hence they suggest a $B_{1u} \pm  i B_{2u}$ state, stating also that more work is required to check if this interpretation is correct.

To conclude on this issue of topological chiral superconductivity, there are presently several puzzling results and experiments, but still no converging or even coherent proposals for the mixed IR order parameter that would explain these results. The most urgent task is certainly to clarify the origin of the observed double transition in the specific heat, as it put strong constraints which will determine the framework for the possible explanations of the results. And for quantitative comparison with predictions, a much more complete knowledge of the Fermi surface is clearly mandatory.


\section{High pressure phase diagram}
\label{section_Pressure}

\subsection{Zero field measurements}

\begin{figure*}
\begin{center}
\includegraphics[angle=-90,width=0.65\columnwidth]{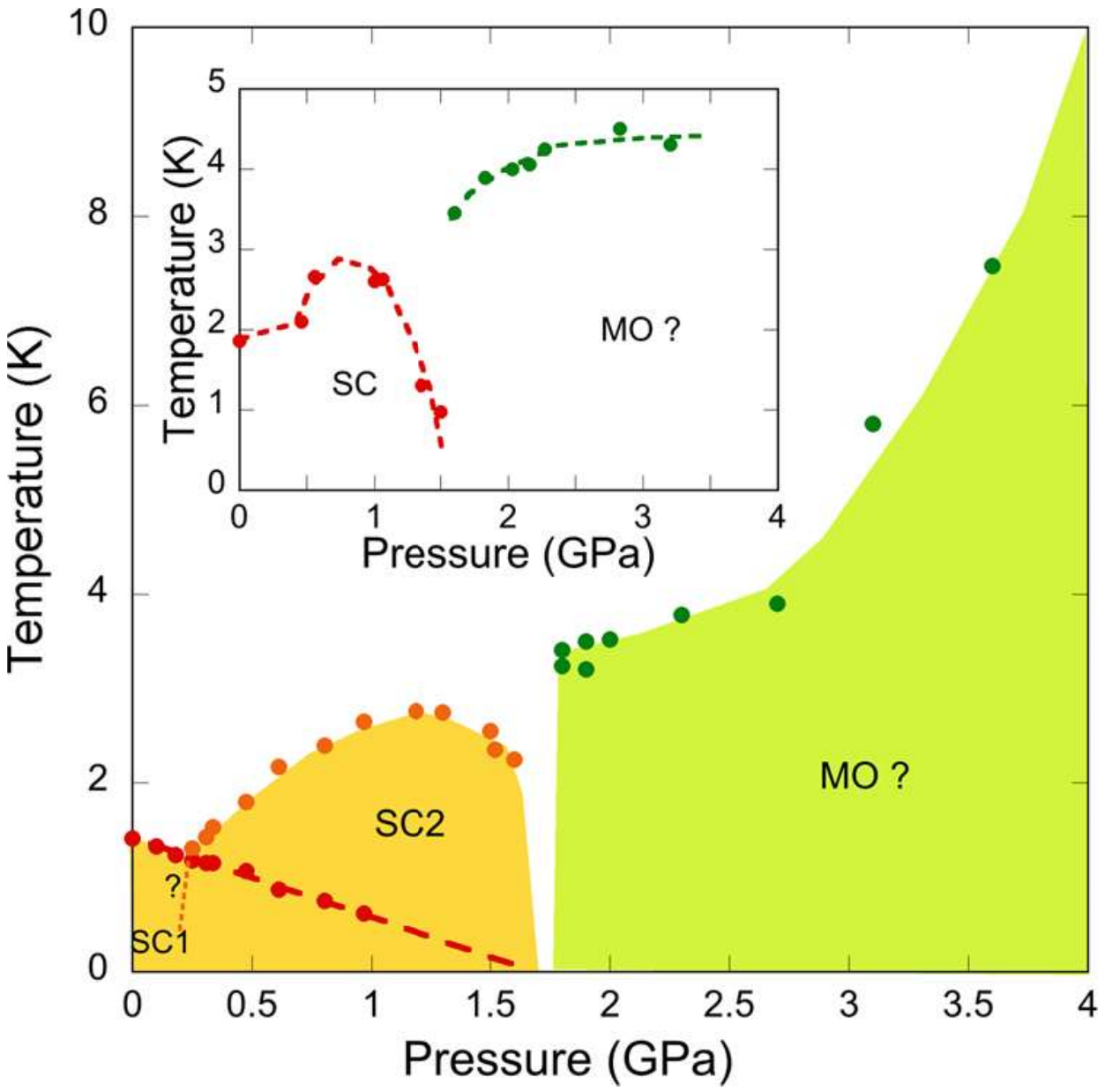}
\includegraphics[angle=-90, width=0.65\columnwidth]{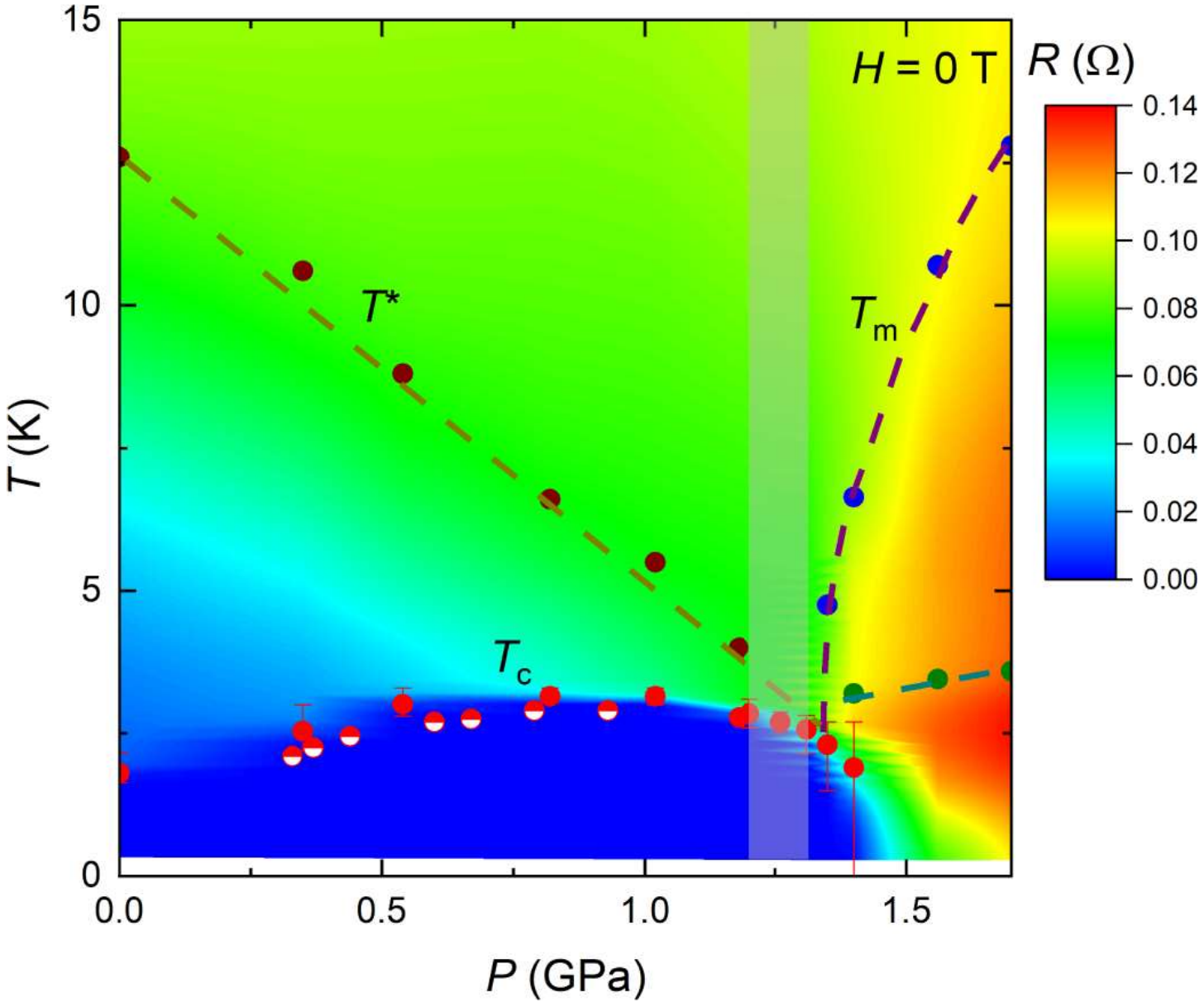}
\includegraphics[angle=-90, width=0.65\columnwidth]{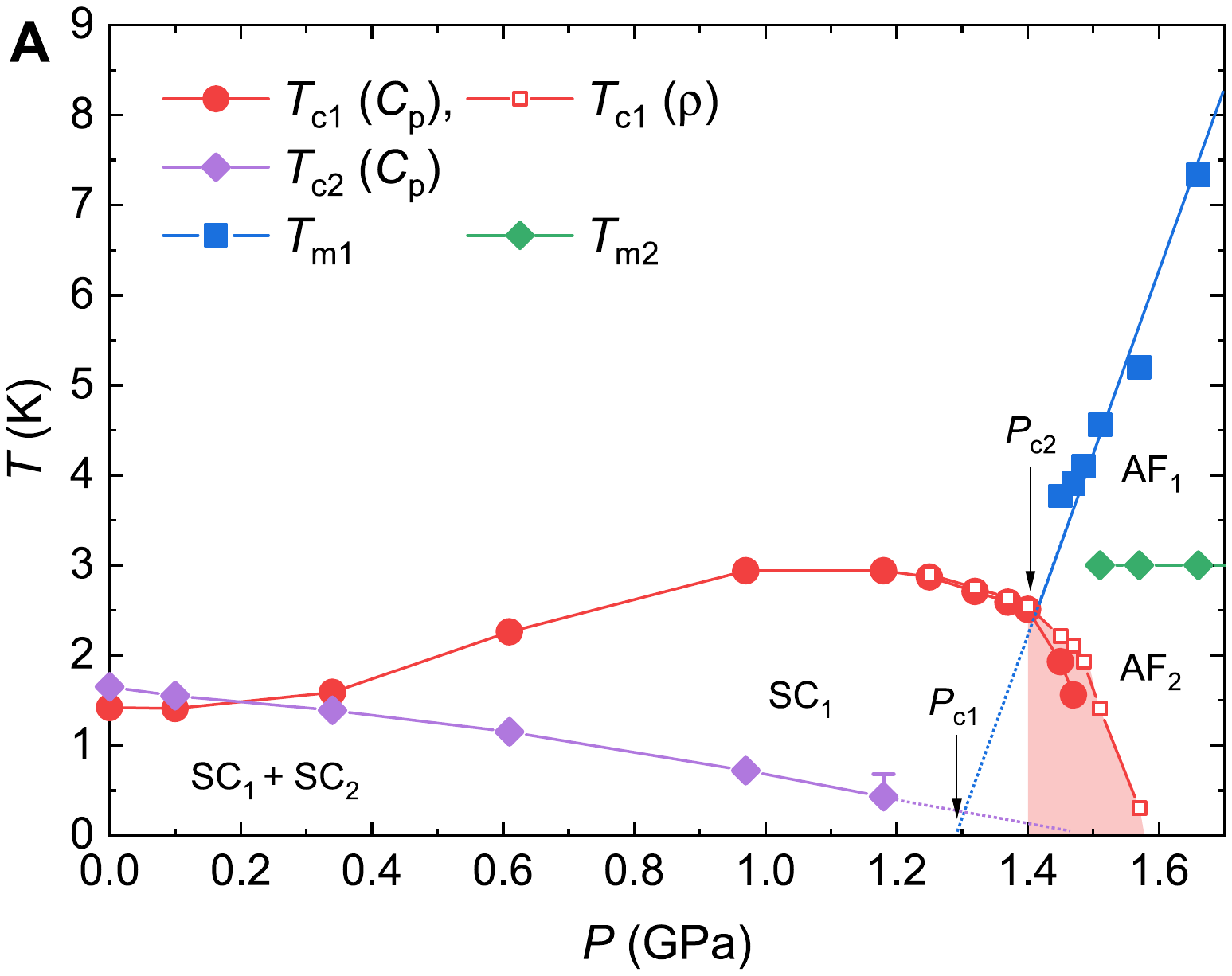}
\caption{Pressure temperature diagrams of UTe$_2$ from different groups. Left panel: From AC calorimetry in a diamond anvil cell with argon as pressure medium, At zero pressure only a single transition is observed and above 0.3~GPa two distinct superconducting transitions appear. Superconductivity is suppressed above 1.7 GPa, and a new magnetic phase is observed. Inset: From resistivity measurements in a modified Bridgman cell with Fluorinert as pressure medium (from Ref. \cite{Braithwaite2019}). Middle panel: From resistivity measurements in a piston cylinder pressure cell with Daphne oil 7373 as pressure medium. Superconductivity is suppressed above 1.5~GPa, and two magnetic transitions are observed above (from Ref. \cite{Ran2020}). Right panel: From resistivity (squares) and AC calorimetry (circles, diamonds) performed in the same piston-clamp pressure cell using Daphne oil 7373 as pressure medium. The shaded area marks the region where resistivity and specific heat show different $\Tsc$ values, indicating sample inhomogeneity due to the first order nature of the transition. In the magnetic ordered state, two different magnetic transition are reported (from Ref.   \cite{Thomas2020}).  } 
\label{pressure_PD}
\end{center}
\end{figure*}

A new surprise arises from pressure studies of UTe$_2$. Surprisingly, two superconducting phases $SC1$ and $SC2$ have been discovered for pressures above 0.3~GPa in zero magnetic field. 
These superconducting transitions are clearly observed in specific heat measurements \cite{Braithwaite2019, Thomas2020}. 
Initially $\Tsc$ decreases with pressure, and above 0.3 GPa the lower transition extrapolates linearly to zero for a pressure around 1.6~GPa. 
The high temperature superconducting phase increases in temperature up to 3~K at $p \sim 1.2$~GPa, and is rapidly suppressed for higher pressures, likely by a first order transition: 
above the critical pressure $p_{\rm c} \approx 1.6$~GPa a new magnetically ordered phase emerges.
This high pressure phase diagram, presented in Fig.~\ref{pressure_PD}, has been confirmed by experiments performed in Grenoble \cite{Braithwaite2019}, Maryland \cite{Ran2020} Los Alamos \cite{Thomas2020} and Oarai \cite{Aoki2020} (the critical pressure varies slightly due to different pressure conditions in the different experiments). 
A rather intriguing point is that the enhancement of $\Tsc$ under pressure by roughly a factor of two compared to the zero pressure value, which is similar to the field-enhancement at zero pressure of the pairing constant $\lambda$ from 1 to 2 (and the associated $\Tsc$ increase from $\approx 1.5$~K to $\approx 3$~K), from $H =0$ up to the metamagnetic transition at $\Hm$ (see section \ref{field_reinforcement} and Fig.~\ref{figCpEtCompLambda}(b)). 
This suggests that increasing pressure favors a similar pairing reinforcement than magnetic field along the $b$-axis.

All present studies of the high pressure phase diagram converge to the point that the transition from the superconducting state to the magnetically ordered state is first order, as indicated in Fig.~\ref{pressure_PD}, and there may be a small pressure range where both superconductivity and magnetic order coexist. 
Direct microscopic measurements to determine the magnetic order are not performed yet, but recent experiments seem to conclude that the ground state above $p_c$ is antiferromagnetic \cite{Thomas2020, Aoki2020, Li2021, Valiska2021}. 


The characteristic crossover temperature $T^\ast \approx 12$~K at $p =0$ (see section \ref{subsectionHpara_a}), which marks the low temperature regime in \UTe ,   decreases linearly with pressure down to $T=2$~K close to $\pc$ (see Fig.~\ref{pressure_PD}) \cite{Ran2020}  in good agreement with the thermal expansion data \cite{Willa2021}. In Ref.~\cite{Thomas2020}, evidence was given for a pressure-induced antiferromagnetic quantum critical point in intermediate valence UTe$_2$; a slight change of the valence from a dominant U$^{3+}$ to a U$^{4+}$ configuration was observed in X-ray absorption spectroscopy \cite{Thomas2020}; it was claimed to be at the origin of the change of the magnetic anisotropy under pressure on both sides of the critical pressure $\pc$ \cite{Li2021}.

\subsection{Field measurements}

\begin{figure}[bht]
\begin{center}
\includegraphics[width=0.8\columnwidth]{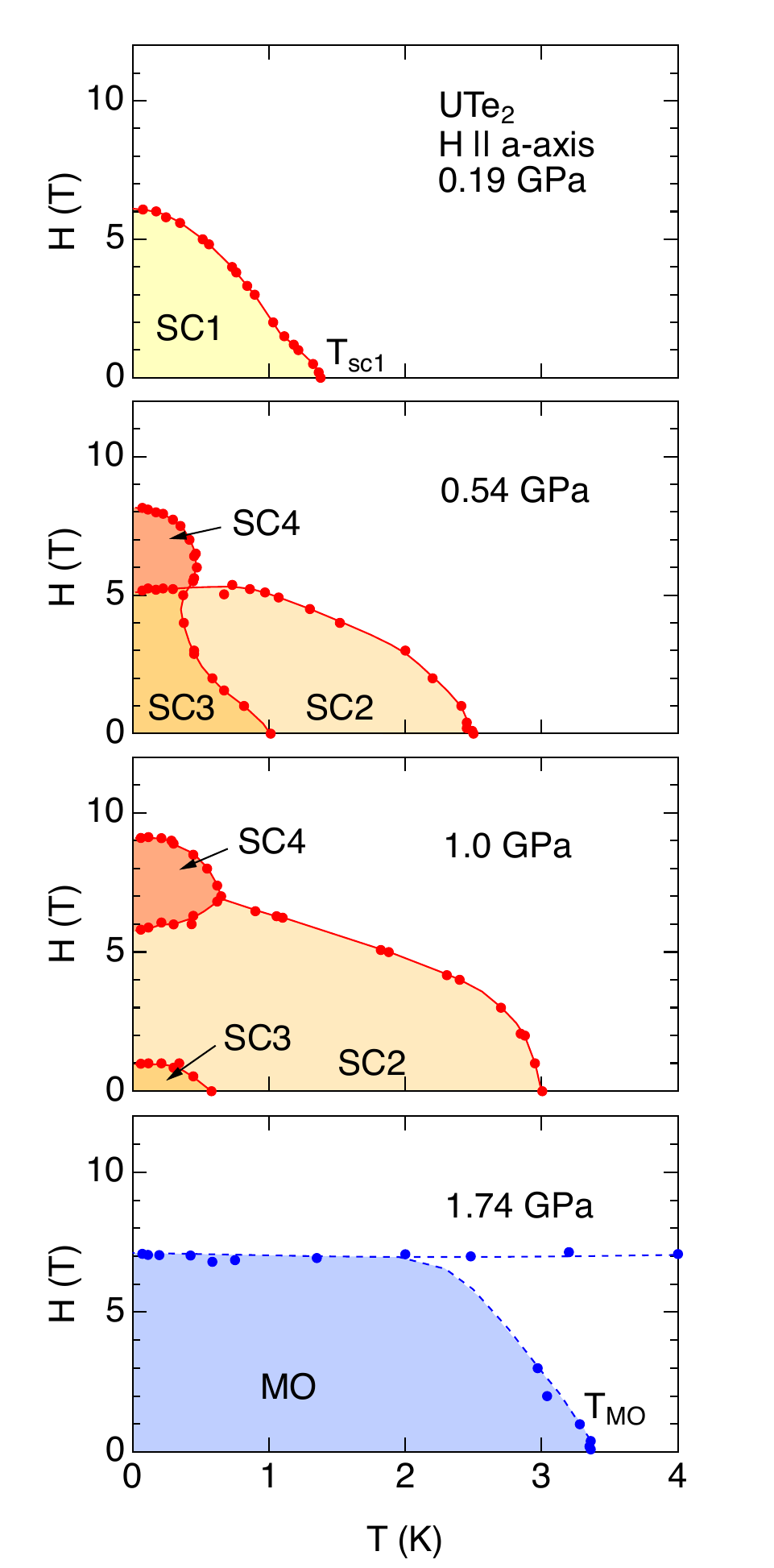}
\caption{Magnetic field $H$ versus temperature $T$ phase diagrams of UTe$_2$ for field applied along the $a$ axis at different pressure from ac calorimetry measurements. Line are guides to the eyes. (taken from Ref.\cite{Aoki2020}.}
\label{Aoki_ac_calorimetry}
\end{center}
\end{figure}

Careful studies by resistivity and ac calorimetry under pressure demonstrate the occurrence of multiple phase transitions induced by a magnetic field,  identified as superconducting transitions \cite{Knebel2020, Aoki2020}. Presently, these field-induced transitions between superconducting phases have been explored only for $H \parallel a$.
In this field direction, a pronounced enhancement of $\Hc^a$ has been observed by resistivity at low temperatures at 0.5~GPa \cite{Knebel2020}.
This was rapidly confirmed by ac calorimetry, uncovering moreover the occurence of multiple superconducting phases below $\Hc$, as shown in Fig.~\ref{Aoki_ac_calorimetry}.

\begin{figure}
\begin{center}
\includegraphics[width=0.9\columnwidth]{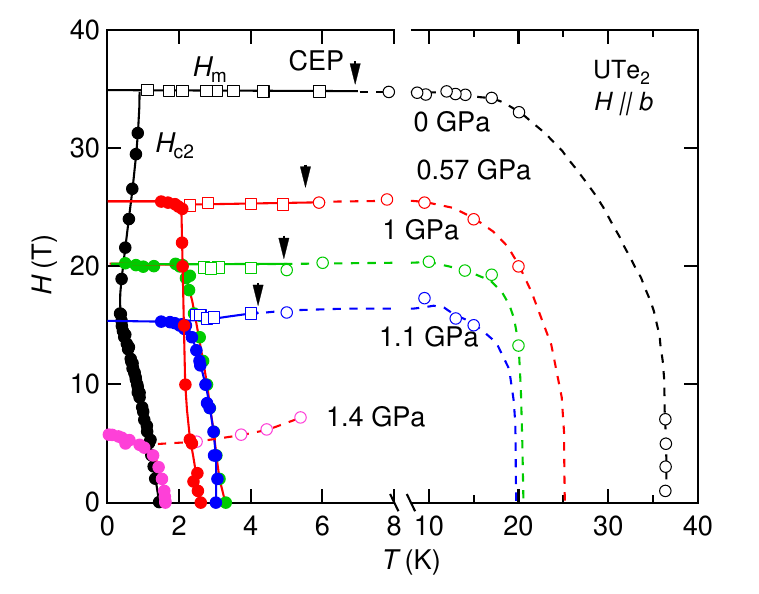}
\caption{Magnetic field versus temperature phase diagram for different pressures. Full symbols indicate the upper critical field, open squares to the metamagnetic field $\Hm$ and open circles to a crossover above the critical end point, which is marked by the arrows $\Hm$. Lines are guides to the eyes. (taken from Ref. \cite{Knebel2020}.}
\label{Knebel_Hc2_pressure}
\end{center}
\end{figure}

For $H \parallel b$ under pressure, resistivity measurements show that $\Hc^b$ remains limited by $\Hm$, and that the pressure dependence  of $\Hm$ follows that of $T_{\rm max}$ (see Fig.~\ref{Knebel_Hc2_pressure}) \cite{Knebel2020}. It has been also observed that the singular field enhancement of superconductivity vanishes above 1~GPa. By tunnel diode oscillator experiments a transition from a low field superconducting phase to a high field phase (the reinforced superconductivity) has been evidenced and it has been stressed that the low field phase disappears near 1~GPa, and for higher pressures up to $\pc$ only the (at $p=0$) high - field phase is observed.\cite{Lin2020} However, this demands still confirmation by thermodynamic measurements under high pressure. 

For $H \parallel c$, the upper critical field $\Hc^c$ has been determined by magnetoresistivity and also by ac calorimetry. 
On approaching the critical pressure $\pc$, a strong increase and an almost vertical slope of $\partial \Hc/\partial T$ is observed \cite{Braithwaite2019, Knebel2020}. 
Close to the critical pressure at $\pc = 1.45$~GPa, long range antiferromagnetic order (MO) seems to restrict the superconducting domain to a small region at high magnetic fields and low temperatures. 
At $p = 1.55$~GPa no superconductivity is observed in zero field, however field-reentrant superconductivity seems to survive just above the critical field $H_{\rm MO}$ of the magnetic order. Finally, at 1.61~GPa, no trace of superconductivity is observed anymore \cite{Aoki2021}. 

\begin{figure}
\begin{center}
\includegraphics[width=0.85\columnwidth]{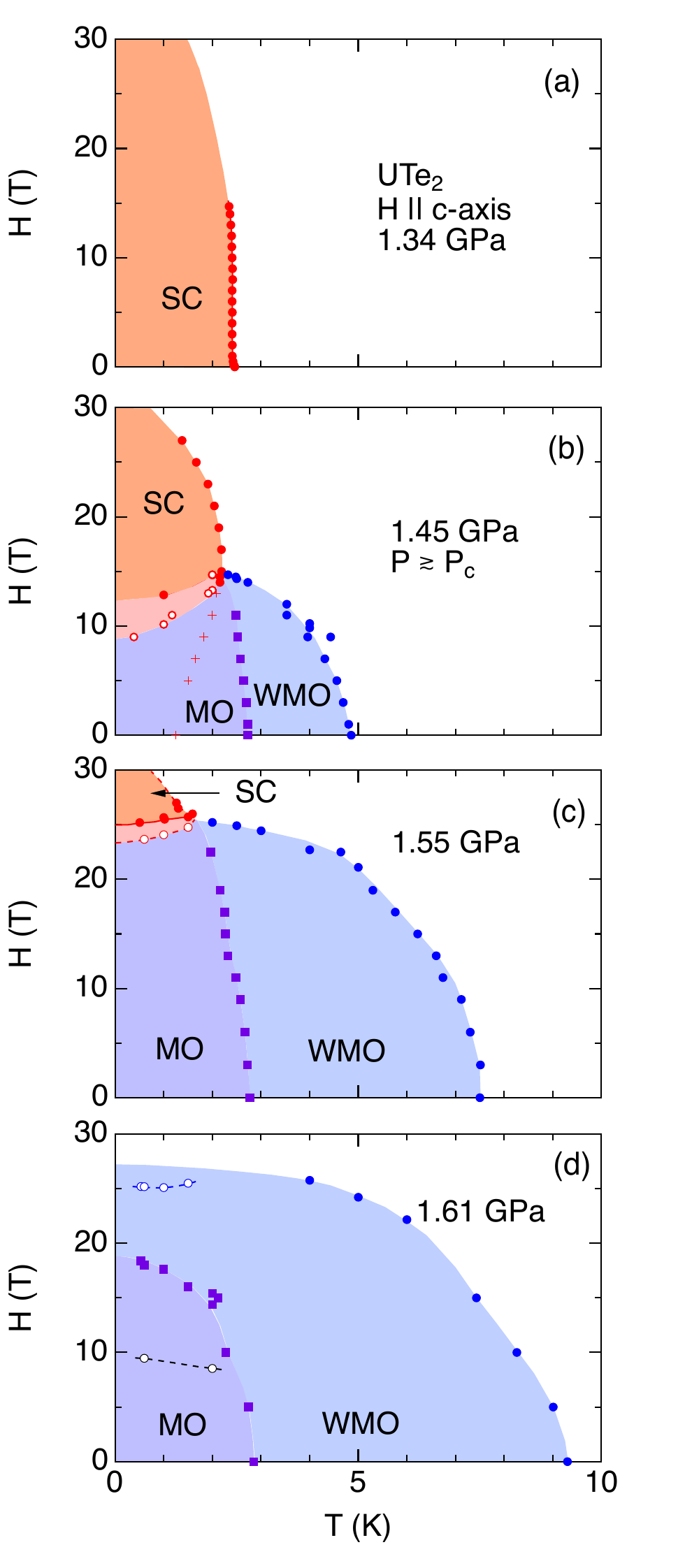}
\caption{Phase diagrams for different pressures close to the critical pressure for a magnetic field applied along the $c$ axis. (taken from Ref. \cite{Aoki2021}.}
\label{Aoki_magnetoresistance}
\end{center}
\end{figure}

The phase diagrams for different pressures on both sides of $\pc$ is shown in Fig.~\ref{Aoki_magnetoresistance}. 
From a maximum of the resistivity data for field along the three axis, another line can be drawn at higher fields or temperatures than the MO phase, delimiting a ``weak magnetic ordered phase'' (WMO).
This WMO phase is found by the different groups and different techniques used to probe \UTe under pressure.
Notably the specific heat experiment of Ref.~\cite{Thomas2020}, have well demonstrated that if the MO phase is characterized by a sharp specific heat jump, as expected for true long range magnetic ordered phases, only a rather large specific heat anomaly occurs at the WMO line (called AFM1 in this work).
The WMO phase could be a phase with only short-ranged magnetic correlations.

As regards the origin of the different superconducting phases, there are still competing explanations, and the most prominent are discussed in the theoretical section \ref{theory:superconductivity}. 
However, there are also experimental challenges. 
A first one arises from the zero field phase diagram (Fig.~\ref{pressure_PD}), and the appearance above $0.3$~GPa of a higher $\Tsc$ phase. 
The specific heat anomaly at the highest $\Tsc$ transition is very small, and grows with pressure, keeping a rough balance with the lower one: 
the proportion is opposite when the lowest $\Tsc$ transition is suppressed. 
A first natural explanation is a spatially inhomogeneous transition, with a sample volume split between the two superconducting phases, and a repartition evolving with pressure. 
This would explain why the lower transition keeps the same slope $\frac{d\Tsc}{dp}$ below and above 0.3GPa.
Explaining why such a ``phase separation'' would happen in \UTe is not easy, notably because the mechanism should be ``intrinsic'', owing to the reproducibility of the experiments by the different groups.
A more appealing alternative is that these two phases are coexisting homogeneously. 
Thermodynamically, the crossing of the two lines in the phase diagram of Fig. \ref{pressure_PD} without a change of slope of $\frac{d\Tsc}{dp}$ is possible in an homogeneous sample if the specific heat jump of the upper transition is vanishing at the crossing \cite{Braithwaite2019}, which is indeed the case experimentally within error bars.
This coincidence of $\Tsc$ and crossing of the lines, imposed by thermodynamics for homogeneous transitions, has a priori no reason to exist in the inhomogeneous scenario.

Regarding the field dependence of these two phases as measured along the a-axis (see Fig.~\ref{Aoki_magnetoresistance}), the change of behaviour of $\Hc$ at low temperatures is quite remarkable.
In the higher $\Tsc$ phase, the very strong negative curvature of $\Hc$ suggests a dominant paramagnetic limit for $H \parallel a$, close to the value expected for weak coupling singlet superconductors. 
By contrast, at low temperatures, the larger $\Hc$ of the low $\Tsc$ phase indicates a suppression of the paramagnetic limit, a feature probably very difficult to understand if the low $\Tsc$ phase is not spin triplet.
Hence, even though many questions remain for a detailed explanation of these pressure-field-temperature superconducting phase diagrams, these $\Hc$ measurements under pressure for $H\parallel a$ are probably one of the strongest support for a spin triplet superconducting state in \UTe. 

During the writing this review, a pressure study of the reentrant high-field superconducting phase for field angles between 25$^{\circ}$ and 45$^{\circ}$ in the $b-c$ plane has appeared \cite{Ran2021}. 
Depending on the angle, between 0.8 and 1~GPa, the field reinforced superconducting phase reaches $\Hm$.
But for higher pressures (1.3~GPa), remarkably, the superconducting phase above $\Hm$ appears clearly detached from the metamagnetic field, suggesting that a different mechanism is required for both phases below and above $\Hm$, as that above $\Hm$ clearly requires high magnetic fields (and not only a strong magnetic polarization).
Additionally, one may wonder if the high field phases measured close to $\pc$ along the $a$ or $c$ axes are related to this high field phase above $\Hm$.
Following the angular dependence of the reentrant superconducting phase above $\Hm$ under pressure is certainly a good experimental challenge.

The combination of pressure and magnetic fields has lead to discover an extremely rich multiphase-superconducting phase diagram, which is related to competing interactions occurring already in the normal phase, as previously discussed. 
This is also related to another unusual property of \UTe.
Indeed, as opposed to most other heavy-fermion superconductors, in \UTe, the electronic Gr\"{u}neisen parameter  of the normal phase $\Gamma_{\rm e} \approx -30$ has the same sign as the Gr\"{u}neisen parameter $\Gamma_{\rm sc}\approx - \frac{\partial \log T_{\rm sc}}{\partial \log V}$ of the superconducting phase, and with quite similar values \cite{Thomas2021}. 
In the other U-based superconductors like UPt$_3$, URu$_2$Si$_2$, and UBe$_{13}$ \cite{Flouquet1991} or ferromagnetic superconductors like URhGe \cite{Hardy_PhD, Sakarya2003} or UCoGe \cite{Gasparini2010}, $\Gamma_{\rm e}$ and $\Gamma_{\rm sc}$ have opposite signs.
This is natural as an increase of the effective mass is associated with a decrease of $T^\ast$, and with an increase of $\Tsc$ if it originates from the pressure dependence of the pairing mechanism.
The singularity points out that superconductivity in \UTe is the result of different competing interactions.

The rather weak value of $\Gamma_{\rm e} \approx -30$ may appear quite antagonistic to the location of UTe$_2$ near a ferromagnetic or antiferromagnetic second order magnetic instability. 
This value is close to those found in intermediate valence compounds \cite{Flouquet1991}. 
However, in the case of a ferromagnetic instability, experiments as well as theory, generally points out that the instability should be first order.
Large Gr\"{u}neisen parameters can be expected only if the first order transition is weak. 
As reported below, recent experiments performed in order to follow the pressure dependence of the magnetization confirm the key role of valence fluctuations under pressure \cite{Li2021}. 

\begin{figure}
\begin{center}
\includegraphics[width=0.9\columnwidth]{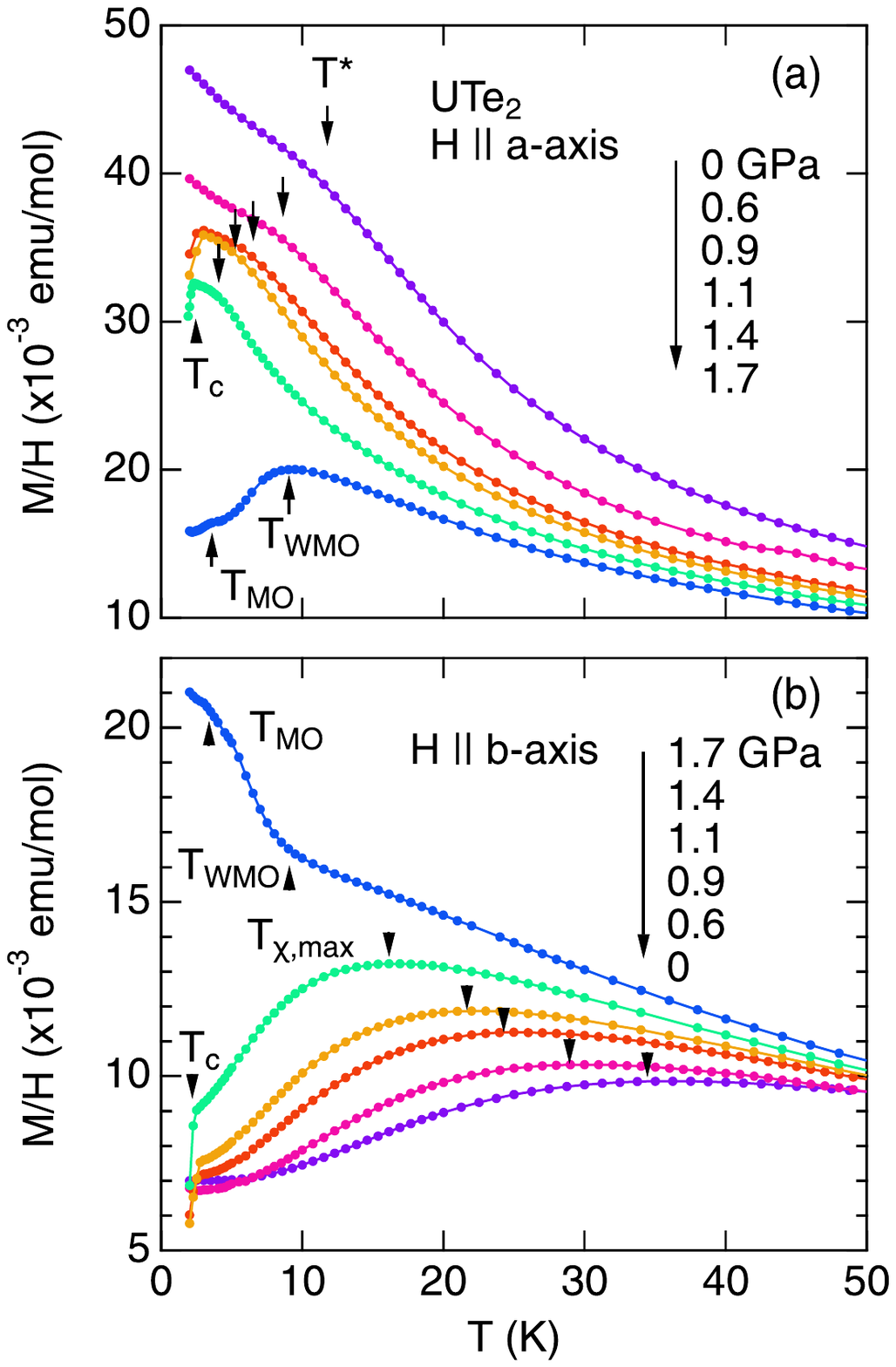}
\includegraphics[angle=-90, width=0.9\columnwidth]{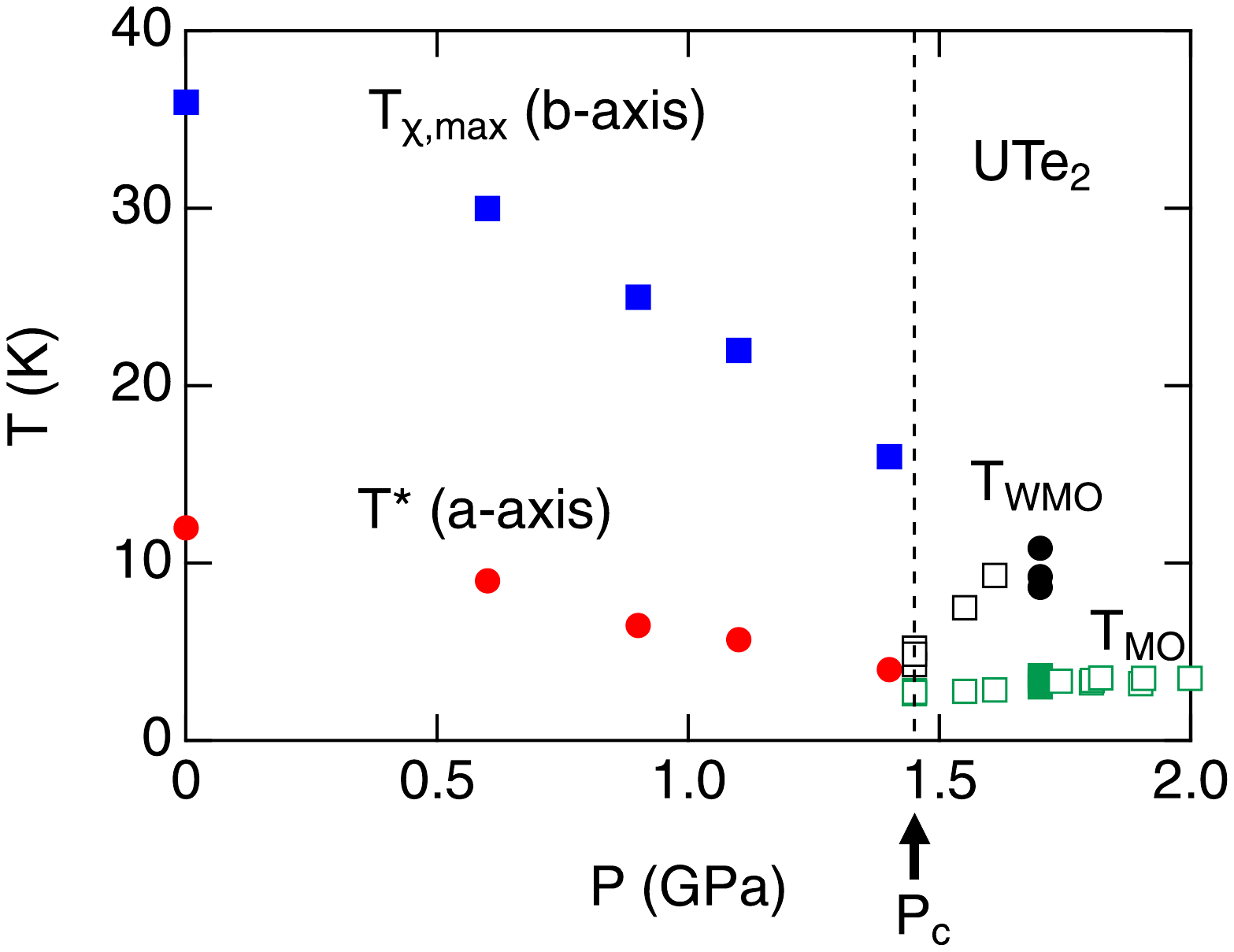}
\caption{(Upper panel) $M/H$ measured a a magnetic field of 1~T applied along the $a$ axis for different pressures. (Middle panel) $M/H$ for a field of 1~T along the $b$ axis. (lower panel) Pressure dependence of the maximum of the $b$ axis susceptibility and the characteristic temperature  $T^\ast$ of the $a$ axis susceptibility. (taken from Reference \cite{Li2021}}
\label{Li_mag_pressure}
\end{center}
\end{figure}

The compressibility of UTe$_2$ is rather high with a bulk modulus $G = 57$~GPa \cite{Honda2021}.
Applying a pressure of $p \sim \pc \sim 1.5$~GPa leads to a volume shrinking of 3\%, i.~e.~roughly the same as measured on the melting curve of $^3$He going from the liquid to the solid phase \cite{Li2021}. 
For such volume shrinking, a drastic change can occur in the  electronic properties. 
In the specific case of UTe$_2$, the major phenomenon seems to be a switch from a low pressure phase, where the magnetic properties are initially driven by the dominant U$^{3+}$ configuration to a high pressure phase  at $p > \pc$, where the starting dominant configuration will be that of U$^{4+}$. 
Tiny pressure increase of the volume can modify the magnetic anisotropy (see Ref. \cite{Derr2006} for Tm chalcogenide).
The local consequence is a drastic change in the anisotropy of the measured susceptibility: 
below $\pc$ at low temperature $\chi_a$ remains higher than $\chi_b$, while above $\pc$ the anisotropy is reversed. 
In addition, similar to Kondo lattice insulators such as SmS and SmB$_6$ a complete change of the Fermi surface can be expected \cite{Li2021}.
According to LDA calculations under pressure, the energy gap at the Fermi energy closes around $2.8\,{\rm GPa}$; the position of the 5$f$ level shifts upward under pressure \cite{Harima_private}.
Figure \ref{Li_mag_pressure} shows for $H\parallel a$ and $b$ the temperature variation of $M/H$ measured at 1~T for different pressures. 
At 2~K the magnetic hierarchy for a field of 5~T is $M_a \sim 3.5 M_c \sim 6 M_b$, while above $\pc$ at $p=1.7$~GPa it becomes $M_b \sim 1.2 M_a \sim 3 M_c$.   
For $p = 1.7$~GPa, on cooling $\chi_a$ shows a maximum at $T_{\rm WMO}$ but decreases strongly at the magnetic ordering temperature as observed for antiferromagnetic materials along their axis where the magnetic moments will be aligned. 
The location of UTe$_2$ at ambient pressure close to its U$^{3+}$ $5f^3$ configuration is in excellent agreement with core level spectroscopy \cite{Fujimori2021} but differs from angular dependent photoemission spectroscopy (ARPES) analysis, 
where a valence 
close to $5f^2$ configuration has been claimed \cite{Miao2020} (see section \ref{Electronic_state_exp}). 
The Ising anisotropy observed at ambient pressure is in good agreement with the crystal field splitting of the 5$f$ level with $\pm J_z$ doublet ground state. 
Switching to the $5f^2$ configuration, a drastic change in the crystal field scheme will appear; 
the $5f^2$ configuration is quite analogous to Pr$^{3+}$ situation, with various crystal field singlets. 

Independent from the precise definition of the crossover temperature $T^\ast$ at zero pressure (maxima of $C/T$, $\rho_c$ or $\partial \rho_a/\partial T$, minima of $\alpha$ or $   \partial M_a/\partial T$), $T^\ast$ collapses just above $\pc$ (see Fig.~\ref{Li_mag_pressure}).
The finite value of $T^\ast$ at $\pc$ suggests that the change from the 
U$^{3+}$ to U$^{4+}$ configuration could be first order. 
The verification of the first order nature of this valence change needs experimental detection, like a volume discontinuity at $\pc$. 
The situation is complex because pressure modifies the relative strength of the correlations and at the same time increases of the width of the electronic bands. 
Moreover, the instability at $\pc$ can be driven by a Fermi surface Lifshitz transition. 
An interesting theoretical objective would be to precise quantitatively how the anisotropy of the interactions is suppressed under pressure and what is its role in the hierarchy of the different pairing channels.

\section{Theoretical perspectives}

\label{section_theoretical_perspectives}

\subsection{Magnetism}

\begin{figure}[htbp]
\begin{center}
\includegraphics[width=0.8\columnwidth]{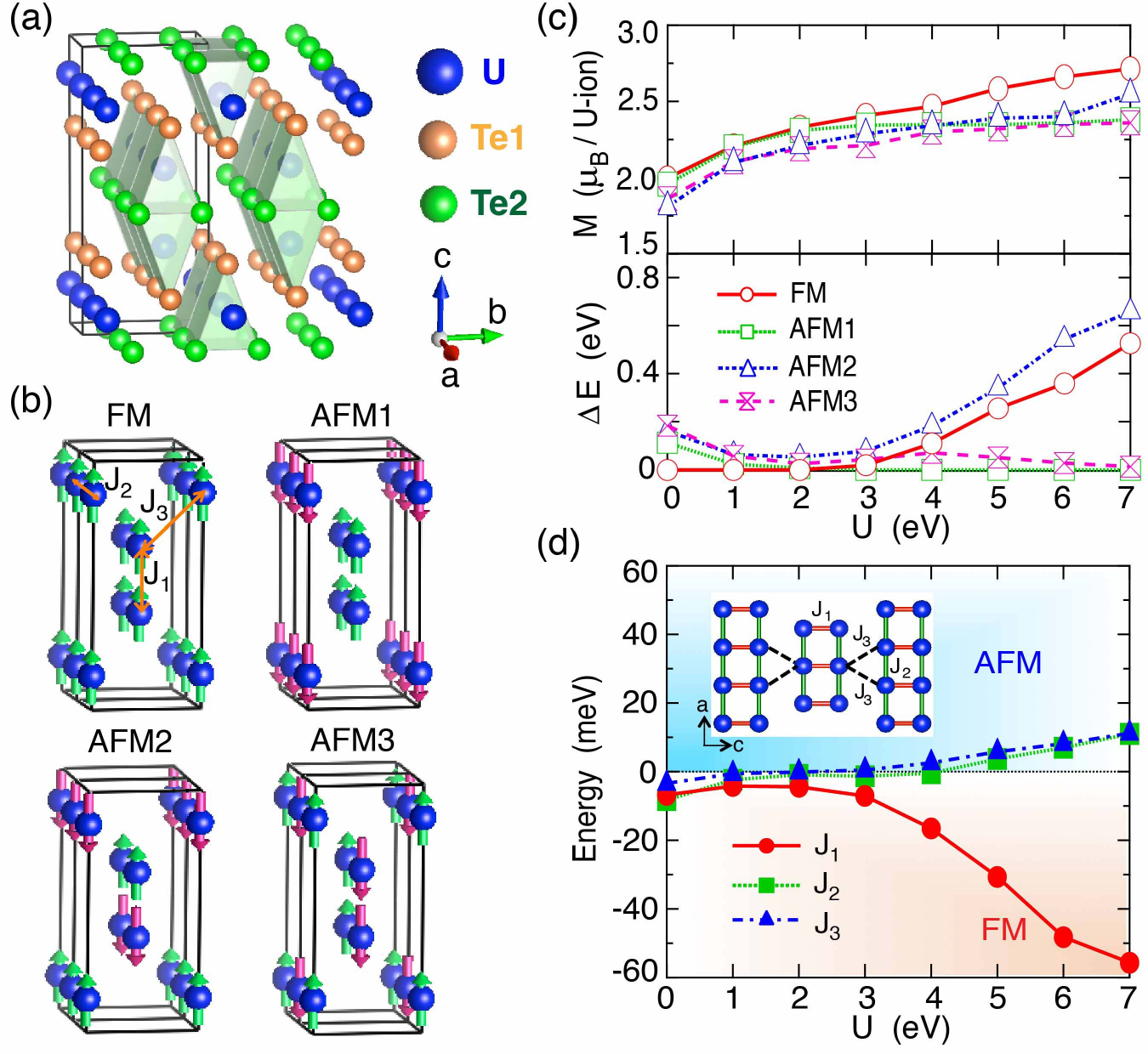}
\caption{Magnetic structures proposed in Ref.~\cite{Xu_UTe2}.
}
\label{Magnetism_localized}
\end{center}
\end{figure}

\begin{figure*}[tbp]
\includegraphics[width=0.9\linewidth]{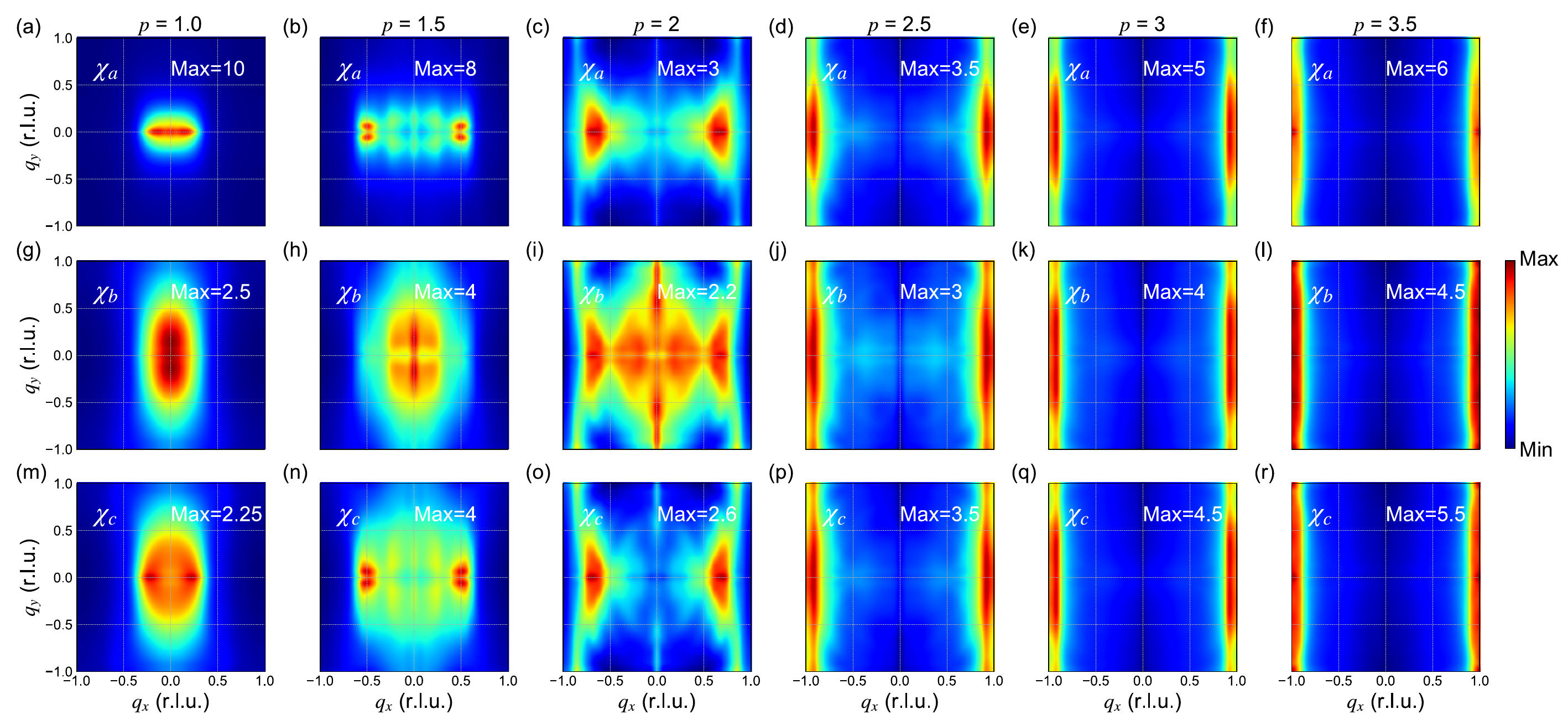}
\centering
\caption{Magnetic susceptibility calculated based on the 24-band periodic Anderson model~\cite{Ishizuka2021}. 
(a-f) $\chi_a(\bm q, 0)$, (g-l) $\chi_b(\bm q, 0)$, and (m-r) $\chi_c(\bm q, 0)$. Momentum dependence on the $q_x$-$q_y$ plane at $q_z = 0$ is drawn for various parameters $p$ corresponding to the applied pressure. 
Figures taken from Ref.~\cite{Ishizuka2021}.}
\label{Magnetism_itinerant}
\end{figure*}

Despite the claim of nearly ferromagnetic magnetic correlations in the first paper~\cite{Ran2019}, it is now expected that the magnetic correlations in UTe$_2$ are not limited to a single spin fluctuation component. 
The magnetic interactions and fluctuations in UTe$_2$ have been theoretically investigated based on the localized and itinerant picture of $f$-electrons. 
The magnetic interactions for localized $f$-electrons were calculated from first-principles~\cite{Xu_UTe2}, and various magnetic structures in Fig.~\ref{Magnetism_localized} including ferromagnetic and antiferromagnetic states were predicted. 
The magnetic fluctuations due to the itinerant $f$-electrons were calculated based on the model reproducing the band structure for an intermediate Coulomb interaction (Fig.~\ref{GGA+U-FS}(c)). In this calculation~\cite{Ishizuka2021}, a 24-band periodic Anderson model for the Te$5p$-, U$6d$-, and U$5f$-electrons has been adopted, and an effective parameter $p$ multiplying the $f$-$f$ and $c$-$f$ hybridization is taken into account. Because the hybridization in the heavy fermion states is enhanced by the applied pressure, the parameter $p$ is considered as an indication of the pressure effect. Fig.~\ref{Magnetism_itinerant} shows the momentum dependence of the magnetic susceptibility deduced from these calculations.  

Both Figs.~\ref{Magnetism_localized} and \ref{Magnetism_itinerant} show an instability between ferromagnetic and antiferromagnetic order on increasing respectively the Hubbard $U$ or the hybridization. Thus, competing or coexisting ferromagnetic and antiferromagnetic fluctuations are expected in UTe$_2$. The dominant magnetic fluctuations may change as a function of the magnetic field and applied pressure. This is at least roughly consistent with the experimental indications of antiferromagnetic correlation~\cite{Knafo2021,Duan2020,Thomas2020,Aoki2020} as well as ferromagnetic one~\cite{Ran2019,Aoki2019,Tokunaga2019,Sundar2019}. 
From Fig.~\ref{Magnetism_itinerant} it is predicted that the dominantly ferromagnetic correlations change to antiferromagnetic correlations on increasing the pressure. This appears consistent with experimental observations~\cite{Ran2019,Aoki2019,Tokunaga2019,Sundar2019,Thomas2020,Aoki2020}.

For the parameter at ambient pressure $p=1.0$, the ferromagnetic fluctuations show a moderately Ising anisotropy, $\chi_a > \chi_b \simeq \chi_c$ and $\chi_a/\chi_b \sim 4$. In a large pressure region $p>2$, the antiferromagnetic fluctuations are dominant, and a weak anisotropy stabilizes the antiferromagnetic order with magnetic moment along the {\it a}-axis. These calculated results on magnetic anisotropy are surprisingly consistent with the experiments although the degeneracy of the $j=5/2$ multiplet is not taken into account in the 24-band periodic Anderson model. In this model, the magnetic anisotropy arises from the local Rashba spin-orbit coupling which may appear in locally noncentrosymmetric crystals~\cite{Fischer2011,Maruyama2012}. Indeed, UTe$_2$ is a case, where the global point group is centrosymmetric $D_{2h}$, but the local symmetry of U sites is noncentrosymmetric $C_{2v}$~\cite{Ishizuka2021,Shishidou2021}. 
Note that the magnetic anisotropy plays an essential role in the following results on superconductivity.  
It should also be noticed that the predicted magnetic wave-vector of antiferromagnetic fluctuations is not equivalent to the experiment at ambient pressure~\cite{Duan2020}. It is ${\bm Q} \parallel$[100] in Fig.~\ref{Magnetism_itinerant}, while ${\bm Q} \parallel$[010] in Ref.~\cite{Duan2020}. This discrepancy may be resolved by taking into account the $j=5/2$ multiplet and/or tuning the hybridization parameters.


\subsection{Theories for superconductivity in UTe$_2$}
\label{theory:superconductivity}

To clarify the superconducting phase of UTe$_2$, microscopic theoretical studies have been conducted based on the 24-band periodic Anderson model~\cite{Ishizuka2021} and on the model with exchange interactions~\cite{Shishidou2021}. 
In Ref.~\cite{Ishizuka2021}, the magnetic-fluctuation-mediated superconductivity has been studied with employing the same model as for Fig.~\ref{Magnetism_itinerant}, and the pressure dependence of the superconducting instability has been predicted. 
In this model, at ambient pressure, the ferromagnetic fluctuations stabilize odd-parity spin-triplet superconductivity, and the moderately Ising-type magnetic anisotropy favors a ${\bm d}$-vector perpendicular to the {\it a}-axis, consistent with an intuitive expectation. 
On the other hand, in the high pressure region, antiferromagnetic fluctuations dominate and rather stabilize an even-parity spin-singlet superconducting state.

\begin{figure}[htbp]
\begin{center}
\includegraphics[width=0.95\columnwidth]{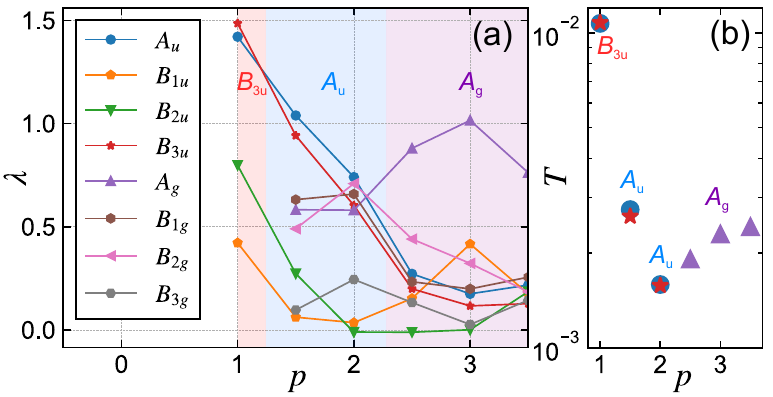}
\caption{(a) Eigenvalues $\lambda$ of the Eliashberg equation for various IRs of $D_{2h}$ point group. The parameter $p \geq 1$ indicates applied pressure. The Coulomb interaction is set so that the Stoner factor is $\alpha_{\rm sf} = 0.98$. (b) Transition temperatures of the $A_u$, $B_{3u}$, and $A_g$ superconducting states for a fixed Coulomb interaction $U = 1.9$. Figures taken from Ref.~\cite{Ishizuka2021}.
}
\label{Eliashberg_lambda}
\end{center}
\end{figure}

Such a pressure dependence of superconductivity is theoretically captured by the eigenvalues of the linearized Eliashberg equation~\cite{Yanase2003}. The maximum eigenvalue determines the stable 
superconducting state.
Figure~\ref{Eliashberg_lambda} shows the $p$ dependence of the maximum eigenvalues for several IRs of the $D_{2h}$ point group, revealing stable odd-parity superconductivity at low pressures $p \leq 2$ as well as stable even-parity superconductivity at high pressures $p>2$.
For the odd-parity superconductivity, the $B_{3u}$ state and $A_{u}$ state are almost degenerate. This is probably a consequence of the Ising-type ferromagnetic fluctuations with $\chi_a > \chi_b \sim \chi_c$. Because the dominant component of the pair potential is ${\bm d}({\bm k}) = k_b \hat{c}$ in the $B_{3u}$ state and ${\bm d}({\bm k}) = k_b \hat{b}$ in the $A_{u}$ state, these states are approximately related by a rotation of ${\bm d}$-vector in the $b$-$c$ plane. When the magnetic fluctuations are nearly isotropic in this plane, the $B_{3u}$ state and $A_{u}$ state are nearly degenerate. The other odd-parity superconducting states of different IRs, namely, the $B_{1u}$ and $B_{2u}$ states, do not contain a pair potential with orbital component $k_b$ and spin component perpendicular to the $a$-axis, and therefore, these states are not stable.
The even-parity superconducting states are similarly classified into the $A_g$, $B_{1g}$, $B_{2g}$, and $B_{3g}$ states (Table~\ref{tab:classification}). Figure~\ref{Eliashberg_lambda} shows that the $A_g$ state is stable in the high pressure region. Although this symmetry is equivalent to the conventional $s$-wave superconductors, the order parameter changes sign in the Brillouin zone like in the $s_\pm$-wave superconducting state of iron-based superconductors. Thus, the $A_g$ state is gapless (line node)~\cite{Ishizuka2021}. By contrast, the $A_u$ state is nearly point-nodal but fully gapped, while in the $B_{3u}$ state the quasiparticle spectrum has point nodes. 
As we discussed in section \ref{sectionGapNodes}, existing experimental results indicate point nodes in the superconducting gap at ambient pressure. This is consistent with the odd-parity $A_u$ and $B_{3u}$ states rather than with the even-parity $A_g$ state. 

After recent theoretical work~\cite{gap_node}, it has been recognized that the naive expectation for the superconducting gap structure based on the point group symmetry and basis functions~\cite{Sigrist-Ueda1991} (see Table~\ref{tab:classification} for example) breaks down in some cases. 
However, an exhaustive classification theory remains consistent with the classical theory~\cite{sumita2021topological} for the orthorhombic and symmorphic UTe$_2$ system~\cite{Ishizuka2019}. 
We show the nodal structure of UTe$_2$ predicted based on the band structure calculation and symmetry analysis in Tables~\ref{tab:TSC12} and \ref{tab:TSC3}.
When time-reversal symmetry is broken, odd-parity superconducting states are likely to possess Weyl nodes~\cite{Yanase2016}, which are topologically-protected point nodes.
Prediction of Weyl superconducting states is compatible with several experiments claiming to have revealed point nodes of the gap (see e.g. \cite{Ishihara2021}). 
However, for more accurate comparison with predictions, we need detailed information on the Fermi surface and pair potential, particularly for Weyl nodes. 
Indeed, the Weyl nodes are not constrained by symmetries, so they can appear at any point of the Fermi surface, depending on the amplitude of the different components of the order parameter (see Ref.~\cite{Hayes2021} and section \ref{section:chirality}).
This is indeed an intriguing property of the Weyl nodes: Weyl points in superconductors are not protected by crystalline symmetry in most cases, although they are stable due to the topological protection.

\begin{figure}[htbp]
\begin{center}
\includegraphics[width=0.9\linewidth]{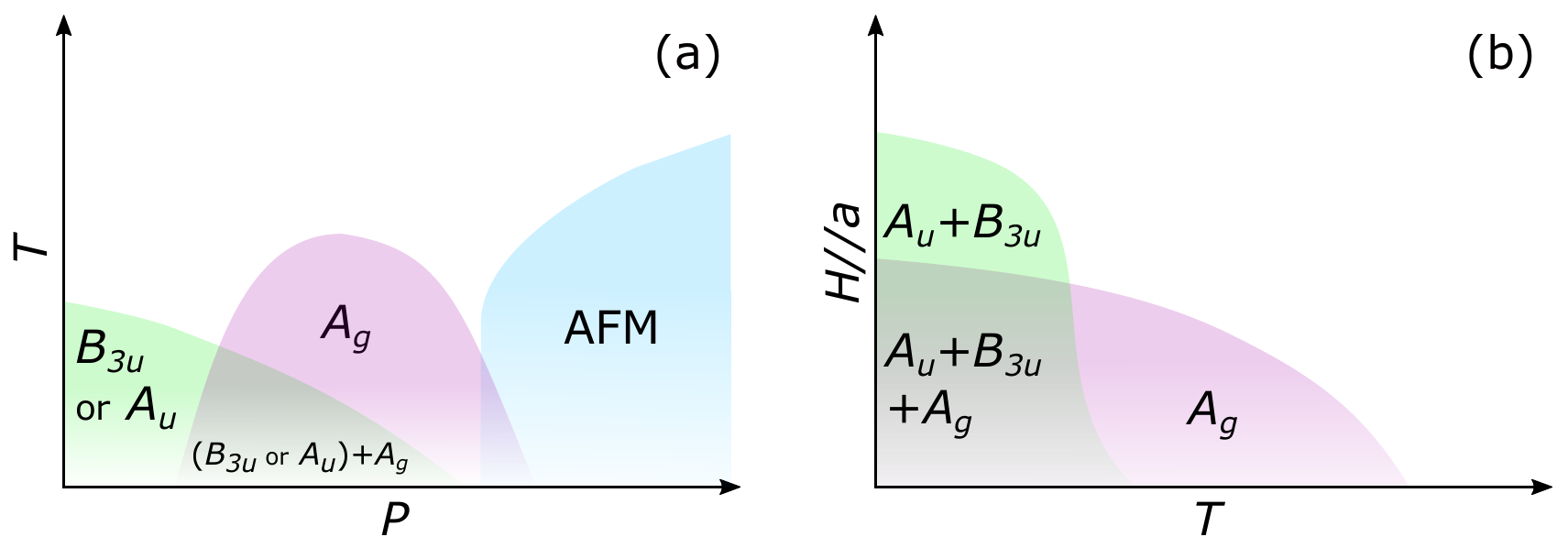}
\caption{
A theoretical interpretation of phase diagrams (a) in the $P$-$T$ plane and (b) in the $T$-$H_a$ plane. Figures taken from Ref.~\cite{Ishizuka2021}.}
\label{phase}
\end{center}
\end{figure}

The analysis of the periodic Anderson model (Fig.~\ref{Eliashberg_lambda}), has allowed to propose a first microscopic interpretation of the experimental superconducting phase diagrams in the pressure-temperature ($P$-$T$) plane and the temperature-magnetic-field ($T$-$H_a$) plane.
This interpretation is presented on Fig.~\ref{phase}. 
A superconducting phase transition from the odd-parity $B_{3u}$ or $A_u$ state to the even-parity $A_g$ state occurs under the applied pressure, coinciding with the crossover in magnetic fluctuations from ferromagnetic to antiferromagnetic. 
The coexisting mixed-parity superconducting state would be stabilized in the intermediate pressure region (Fig.~\ref{phase}(a)).
In the mixed-parity superconducting state, either the $B_{3u}$($A_u$) + $A_g$ state or the $B_{3u}$($A_u$) + $i A_g$ state are expected \cite{note2,WangFu2017,kanasugi2021anapole}, and space inversion symmetry is spontaneously broken. 
The former is similar to states from noncentrosymmetric superconductors, where the crystalline inversion symmetry is broken~\cite{Bauer2012NCS,Smidman2017}. 
The latter is more exotic and breaks both inversion and time-reversal symmetries. 
These states are distinguished by the preserved symmetry; the time-reversal symmetry is preserved in the $B_{3u}$($A_u$) + $A_g$ state, while the combination of space inversion and time-reversal operations, called $PT$ symmetry, is preserved in the $B_{3u}$($A_u$) + $i A_g$ state. 
From the viewpoint of symmetry, the former corresponds to the odd-parity electric multipole state, while the latter corresponds to the odd-parity magnetic multipole state~\cite{Watanabe2018}. 
When the conventional BCS theory for intraband pairing applies, 
the latter is stable~\cite{WangFu2017}, and thus, the $B_{3u}$($A_u$) + $i A_g$ state is the most likely in UTe$_2$. 

Such a spontaneously parity-mixed pairing state was referred to in a review article for $^3$He published forty years ago with the comment ``there seems at present no experimental evidence which require this hypothesis''~\cite{Leggett1975}. 
Even now, such experimental evidence has not been reported. 
Hence, it is desirable to examine whether UTe$_2$ may be the first material where such a state would exist. 
Even apart from issues of UTe$_2$, smoking gun experiments for the inversion symmetry breaking in superconductors are awaited. 
At present, no experiments have detected an inversion symmetry breaking in the bulk, unlike for the time-reversal symmetry breaking and rotation symmetry breaking. 
The polar Kerr effect and $\mu$SR measurements have been widely used as probes of broken time-reversal symmetry in superconductors~\cite{Kallin_2016}. 
The rotation symmetry breaking in superconductors, named nematic superconductivity, can be studied by various bulk measurements~\cite{Yonezawa2019}. 
On the other hand, theoretical exploration of the spontaneously mixed-parity superconducting state is now in progress~\cite{kanasugi2021anapole,Watanabe2021a,Watanabe2021b}. 

Corresponding to Fig.~\ref{Aoki_ac_calorimetry}, the $T$-$H_a$ phase diagram under pressure is drawn in Fig.~\ref{phase}(b). A purely odd-parity superconducting state will be stable in the high field region because the paramagnetic depairing effect is partly avoided. 
This phase is denoted as $A_u + B_{3u}$ because the $B_{3u}$ and $A_u$ states can not be distinguished by symmetry under a magnetic field $H \parallel a$ (see Table~\ref{tab:mixing}). 
Compared with Fig.~\ref{phase}(b) we see rich phases in the experimental phase diagram (Fig.~\ref{Aoki_ac_calorimetry}). 
This may be due to the degeneracy of the $B_{3u}$ and $A_u$ states. 
For instance, the $B_{3u}$ dominant state and the $A_u$ dominant state may appear in the phase diagram as thermodynamically distinct phases.

In this section, we have discussed a scenario based on the ferromagnetic-antiferromagnetic crossover.
However, the symmetry of superconductivity and the interpretation of the multiple phases in UTe$_2$ are far from settled. 
Even at ambient pressure, experimental results have been interpreted based on various scenarios, as shown in Table~\ref{tab:Nodes}. 
Accordingly, theories proposed various superconducting states. Most of them assume purely odd-parity spin-triplet superconductivity, such as the $B_{3u} + i B_{2u}$ state~\cite{Shishidou2021} and the $B_{3u} + i B_{1u}$ state~\cite{Nevidomskyy_UTe2}. 
Assuming that long-lived ferromagnetic fluctuation plays the same role as the ferromagnetic long-range order and ignoring the spin-orbit coupling, a spin-triplet pairing state with ${\bm d}({\bm k}) = (k_b + i k_c) (\hat{b} + i \hat{c})$ was also proposed~\cite{Kittaka2020,Machida2020,Machida2021}. 
This state is classified into a $B_{3u} + i A_{u}$ state in the terminology of IR. 
Firm theoretical grounds for the assumptions have not been obtained up to now. 
To settle these issues, a close comparison between theories and experiments, which is lacking at present, is highly desired. 
In particular, a smoking gun experiment revealing the symmetry breaking is awaited. 

As detailed discussions have already exposed (sections \ref{NMR_normal}, \ref{section_neutron}), the nature of magnetic correlations in UTe$_2$ and its pressure and magnetic field dependence are still open issues. 
In particular, only the antiferromagnetic fluctuation have been observed in the neutron scattering~\cite{Duan2020,Knafo2021}, despite various indications for the ferromagnetic correlation and spin-triplet superconductivity. 
This issue is essential for the mechanism of superconductivity. 
Comparing the theoretical results in Fig.~\ref{Magnetism_itinerant} and Fig.~\ref{Eliashberg_lambda}, we notice that the spin-triplet superconductivity is stable in the ferromagnetic-antiferromagnetic crossover region around $p=2$. 
Thus, strictly ferromagnetic correlations are not necessary for spin-triplet superconductivity.

\section{Conclusions}

\UTe\ appears to be an exciting new paramagnetic heavy fermion superconductor. 
Regarding the superconducting phase, the number of experimental and theoretical investigations triggered by the first publication \cite{Ran2019} three years ago is certainly a record in the field of heavy-fermion superconductors. 
This enthusiasm  has been supported by the heady prospects of topological spin triplet superconductivity,  the stunning discoveries of high-field reinforced and induced superconducting states, as well as pressure-induced exotic superconducting phase diagrams. 
Naturally, for such a ``young'' system, most results and hypothesis require confirmation. 

Progress has already been done on the crystals, displaying now superconducting transitions larger than $1.8$~K, smaller residual specific heat terms, but still not good enough purity which would allow for the observation of quantum oscillations. Hence, the low temperature bulk Fermi surface, and the experimental determination of the electronic structure are still controversial.

\replaced[id=JP]{The key issue of the topological character of the superconductivity of \UTe has been mainly focused up to now on a possible chiral state. However, with the growing evidence for an extrinsic character of the double transition, this trail should probably be bent toward other features, related more deeply to the highly probable spin-triplet superconducting ground state.}{The difficult subject of topological superconductivity and possibly Weyl nodes requires first of all a clarification of the origin of the double superconducting transition, observed in some samples, a target probably accessible with improvements of the crystals, as well as the pressure and thermal dilatation measurements already performed.} 

Determining the role of the competing magnetic fluctuations in the pairing mechanism for the different regions of temperature, pressure and magnetic field is probably the greatest challenge, like in all unconventional superconductors. It  requires a precise experimental knowledge and theoretical description of the normal phase of \UTe. Large progress has already been achieved by neutron scattering and magnetic resonance experiments, but the role of short range ferromagnetic fluctuation has still to be clarified.  
The rapid progress of theoretical approaches, starting from realistic band calculations of these complex materials to propose predictions on the magnetic and superconducting properties brings a new hope. 
\UTe is certainly an excellent case study for these new methods, owing to the richness of its phase diagrams, and stimulated by the amazing  possibility of mixed singlet-triplet states due to spontaneous space inversion symmetry breaking in the superconducting state.

\section*{Acknowledgments}

We thank our colleagues D.Braithwaite, W.~Knafo, G.~Lapertot, A.~Miyake, Q. Niu, C.~Paulsen, A.~Pourret, A. Rosuel, S.~Raymond, G.~Seyfahrt, I.~Sheikin, M. Vali\v{s}ka, S. Rousseau, C. Marcenat, T. Helm, M. Kimata, F. Honda, D. Li, Y. Homma, Y. J. Sato, Y. Shimizu, A. Nakamura, G. Nakamine, K. Kinjo, S. Kitagawa, H. Sakai, and S. Kambe for collaborations in the various experiments. We also thank J. Ishizuka, S. Sumita, and A. Daido for collaborations. We acknowledge very helpful critical reading of the manuscript by M.~Houzet, J.-P.~Sanchez, and H.~Suderow.
We further thank for fruitful discussions K.~Asayama, H.~Harima, K.~Miyake, K. Machida, M. Houzet, M. Zhitomirsky, V. Mineev, S. Fujimoto, S. Fujimori, K. Izawa, S. Kittaka, T. Sakakibara, S. Imajo, Y. Kohama, K. Willa, F. Hardy, C. Meingast, and D.~F.~Agterberg. 

We have got financial support from the Cross-Disciplinary Program on Instrumentation of CEA, the French Alternative Energies and the Atomic Energy Commission, the French National Agency for Research ANR within the project FRESCO No. ANR-20-CE30-0020 and FETTOM ANR-19-CE30-0037, the CEA Exploratory program TOPOHALL, 
JSPS KAKENHI (JP18H05227, JP18H01178, JP20H05159, JP19H00646, JP20K20889, JP20H00130, JP20KK0061), SPIRITS 2020 of Kyoto University, and ICC-IMR.
We acknowledge support of the LNCMI-CNRS, member the European
Magnetic Field Laboratory (EMFL).


\section{References}

\bibliographystyle{iopart-num_GK}
\bibliography{main}

\begin{thebibliography}{100}
\bibitem{Hill_note}
 The Hill limit gives a rational criterion for the overlap of the $5f$ wave
  functions. For next neighbour U-U distances smaller than the Hill limit, the
  $5f$ wave functions overlap and are forming narrow a narrow band and have
  itinerant character. For larger U-U distances there is no direct overlap of
  the $5f$ wave function and the $5f$ states are quasi-localized.
\bibitem{note1}
 In the presence of extra degrees of freedom such as orbital and sublattice,
  the relation between symmetry and the internal structure of Cooper pairs
  becomes rich and complicated. Manifestation of exotic superconducting state
  due to the locally noncentrosymmetric multi-sublattice structure has recently
  been reported in CeRh$_2$As$_2$.
\bibitem{note2}
 In principle, states with phase difference other than $0$, $\pi$ and $\pm
  \pi/2$ are possible, where time-reversal, inversion, and $PT$ symmetry are
  broken. However, such states are not stable in the BCS theory.
\end{thebibliography}


\end{document}